\pdfoutput=1

\documentclass[12pt,a4paper]{article}

\usepackage{ifthen} 
\newboolean{pdflatex}
\setboolean{pdflatex}{true} 

\newboolean{articletitles}
\setboolean{articletitles}{true} 

\newboolean{uprightparticles}
\setboolean{uprightparticles}{false} 

\def\paperauthors{LHCb collaboration} 
\def\paperasciititle{Search for CP violation in Xib- -> pK-K- decays} 
\def\papertitle{Search for \CP violation\\ in \boldmath\XibToKKp decays} 
\def\paperkeywords{{High Energy Physics}, {LHCb}} 
\def\papercopyright{\the\year\ CERN for the benefit of the LHCb collaboration} 
\def\paperlicence{CC BY 4.0 licence}
\def\paperlicenceurl{https://creativecommons.org/licenses/by/4.0/}

\usepackage[top=1in, bottom=1.25in, left=1in, right=1in]{geometry}

\usepackage{microtype}
\usepackage{lineno}  
\usepackage{xspace} 
\usepackage{caption} 

\usepackage{graphicx}  
\usepackage{color}
\usepackage{colortbl}
\graphicspath{{./figs/}} 

\usepackage{amsmath} 
\usepackage{amssymb}
\usepackage{amsfonts}
\usepackage{upgreek} 

\usepackage{arydshln} 

\newcommand*\patchAmsMathEnvironmentForLineno[1]{%
\expandafter\let\csname old#1\expandafter\endcsname\csname #1\endcsname
\expandafter\let\csname oldend#1\expandafter\endcsname\csname
end#1\endcsname
 \renewenvironment{#1}%
   {\linenomath\csname old#1\endcsname}%
   {\csname oldend#1\endcsname\endlinenomath}%
}
\newcommand*\patchBothAmsMathEnvironmentsForLineno[1]{%
  \patchAmsMathEnvironmentForLineno{#1}%
  \patchAmsMathEnvironmentForLineno{#1*}%
}
\AtBeginDocument{%
\patchBothAmsMathEnvironmentsForLineno{equation}%
\patchBothAmsMathEnvironmentsForLineno{align}%
\patchBothAmsMathEnvironmentsForLineno{flalign}%
\patchBothAmsMathEnvironmentsForLineno{alignat}%
\patchBothAmsMathEnvironmentsForLineno{gather}%
\patchBothAmsMathEnvironmentsForLineno{multline}%
\patchBothAmsMathEnvironmentsForLineno{eqnarray}%
}

\usepackage{hyperxmp}

\usepackage[pdftex,
            pdfauthor={\paperauthors},
            pdftitle={\paperasciititle},
            pdfkeywords={\paperkeywords},
            pdfcopyright={Copyright (C) \papercopyright},
            pdflicenseurl={\paperlicenceurl}]{hyperref}

\usepackage[colorinlistoftodos,textsize=scriptsize]{todonotes}

\usepackage[all]{hypcap} 

\usepackage{xspace} 
\usepackage{upgreek}

\def\lhcb   {\mbox{LHCb}\xspace}

\def\MagUp {\mbox{\em Mag\kern -0.05em Up}\xspace}

\ifthenelse{\boolean{uprightparticles}}%
{

 \def\Pmu         {\ensuremath{\upmu}\xspace}

 \def\Ppi         {\ensuremath{\uppi}\xspace}

 \def\Ppsi        {\ensuremath{\uppsi}\xspace}

 \def\PDelta      {\ensuremath{\Delta}\xspace}                 
 \def\PXi         {\ensuremath{\Xi}\xspace}                 
 \def\PLambda     {\ensuremath{\Lambda}\xspace}                 
 \def\PSigma      {\ensuremath{\Sigma}\xspace}                 
 \def\POmega      {\ensuremath{\Omega}\xspace}                 
 \def\PUpsilon    {\ensuremath{\Upsilon}\xspace}

 \def\PB      {\ensuremath{\mathrm{B}}\xspace}                 
                  
 \def\PD      {\ensuremath{\mathrm{D}}\xspace}

 \def\PJ      {\ensuremath{\mathrm{J}}\xspace}                 
 \def\PK      {\ensuremath{\mathrm{K}}\xspace}

 \def\PX      {\ensuremath{\mathrm{X}}\xspace}

 \def\Pb      {\ensuremath{\mathrm{b}}\xspace}                 
 \def\Pc      {\ensuremath{\mathrm{c}}\xspace}

 \def\Pi      {\ensuremath{\mathrm{i}}\xspace}

 \def\Pp      {\ensuremath{\mathrm{p}}\xspace}

 \def\Ps      {\ensuremath{\mathrm{s}}\xspace}

 \def\thebaroffset{0.0em}
}
{

 \def\Pmu         {\ensuremath{\mu}\xspace}

 \def\Ppi         {\ensuremath{\pi}\xspace}

 \def\Ppsi        {\ensuremath{\psi}\xspace}                 
                  
 \mathchardef\PDelta="7101
 \mathchardef\PXi="7104
 \mathchardef\PLambda="7103
 \mathchardef\PSigma="7106
 \mathchardef\POmega="710A
 \mathchardef\PUpsilon="7107
                  
 \def\PB      {\ensuremath{B}\xspace}                 
                  
 \def\PD      {\ensuremath{D}\xspace}

 \def\PJ      {\ensuremath{J}\xspace}                 
 \def\PK      {\ensuremath{K}\xspace}

 \def\PX      {\ensuremath{X}\xspace}

 \def\Pb      {\ensuremath{b}\xspace}                 
 \def\Pc      {\ensuremath{c}\xspace}

 \def\Pi      {\ensuremath{i}\xspace}

 \def\Pp      {\ensuremath{p}\xspace}

 \def\Ps      {\ensuremath{s}\xspace}

 \def\thebaroffset{0.18em}
}
\newcommand{\offsetoverline}[2][\thebaroffset]{\kern #1\overline{\kern -#1 #2}}%

\makeatletter
\ifcase \@ptsize \relax
  \newcommand{\miniscule}{\@setfontsize\miniscule{4}{5}}
\or
  \newcommand{\miniscule}{\@setfontsize\miniscule{5}{6}}
\or
  \newcommand{\miniscule}{\@setfontsize\miniscule{5}{6}}
\fi
\makeatother

\DeclareRobustCommand{\optbar}[1]{\shortstack{{\miniscule (\rule[.5ex]{1.25em}{.18mm})}
  \\ [-.7ex] $#1$}}

\def\mumu       {{\ensuremath{\Pmu^+\Pmu^-}}\xspace}

\def\squark    {{\ensuremath{\Ps}}\xspace}

\def\cquark    {{\ensuremath{\Pc}}\xspace}

\def\bquark    {{\ensuremath{\Pb}}\xspace}
\def\bquarkbar {{\ensuremath{\overline \bquark}}\xspace}

\def\pion   {{\ensuremath{\Ppi}}\xspace}
\def\piz    {{\ensuremath{\pion^0}}\xspace}
\def\pip    {{\ensuremath{\pion^+}}\xspace}
\def\pim    {{\ensuremath{\pion^-}}\xspace}

\def\kaon    {{\ensuremath{\PK}}\xspace}
\def\Kbar    {{\ensuremath{\offsetoverline{\PK}}}\xspace}

\def\KorKbar {\kern \thebaroffset\optbar{\kern -\thebaroffset \PK}{}\xspace}

\def\Kp      {{\ensuremath{\kaon^+}}\xspace}
\def\Km      {{\ensuremath{\kaon^-}}\xspace}

\def\KS      {{\ensuremath{\kaon^0_{\mathrm{S}}}}\xspace}

\def\Kstarm  {{\ensuremath{\kaon^{*-}}}\xspace}

\def\D       {{\ensuremath{\PD}}\xspace}

\def\DorDbar {\kern \thebaroffset\optbar{\kern -\thebaroffset \PD}\xspace}
\def\Dz      {{\ensuremath{\D^0}}\xspace}

\def\Dp      {{\ensuremath{\D^+}}\xspace}
\def\Dm      {{\ensuremath{\D^-}}\xspace}

\def\DpDm    {\ensuremath{\Dp {\kern -0.16em \Dm}}\xspace}

\def\Dstarp  {{\ensuremath{\D^{*+}}}\xspace}

\def\Ds      {{\ensuremath{\D^+_\squark}}\xspace}

\def\B       {{\ensuremath{\PB}}\xspace}
\def\Bbar    {{\ensuremath{\offsetoverline{\PB}}}\xspace}

\def\BorBbar {\kern \thebaroffset\optbar{\kern -\thebaroffset \PB}\xspace}

\def\Bzb     {{\ensuremath{\Bbar{}^0}}\xspace}
\def\Bd      {{\ensuremath{\B^0}}\xspace}

\def\BdorBdbar {\kern \thebaroffset\optbar{\kern -\thebaroffset \Bd}\xspace}
\def\Bu      {{\ensuremath{\B^+}}\xspace}
\def\Bub     {{\ensuremath{\B^-}}\xspace}

\def\Bm      {{\ensuremath{\Bub}}\xspace}

\def\Bs      {{\ensuremath{\B^0_\squark}}\xspace}
\def\Bsb     {{\ensuremath{\Bbar{}^0_\squark}}\xspace}
\def\BsorBsbar {\kern \thebaroffset\optbar{\kern -\thebaroffset \Bs}\xspace}

\def\jpsi     {{\ensuremath{{\PJ\mskip -3mu/\mskip -2mu\Ppsi}}}\xspace}

\def\Y#1S{\ensuremath{\PUpsilon{(#1S)}}\xspace}

\def\proton      {{\ensuremath{\Pp}}\xspace}
\def\antiproton  {{\ensuremath{\overline \proton}}\xspace}

\def\Lz          {{\ensuremath{\PLambda}}\xspace}

\def\LorLbar     {\kern \thebaroffset\optbar{\kern -\thebaroffset \PLambda}\xspace}

\def\Sigmares    {{\ensuremath{\PSigma}}\xspace}

\def\Xires       {{\ensuremath{\PXi}}\xspace}

\def\Xiresbar       {{\ensuremath{\offsetoverline{\Xires}}}\xspace}

\def\Omegares    {{\ensuremath{\POmega}}\xspace}

\def\Lc          {{\ensuremath{\Lz^+_\cquark}}\xspace}

\def\Lb           {{\ensuremath{\Lz^0_\bquark}}\xspace}

\def\Xib          {{\ensuremath{\Xires_\bquark}}\xspace}
\def\Xibz         {{\ensuremath{\Xires^0_\bquark}}\xspace}
\def\Xibm         {{\ensuremath{\Xires^-_\bquark}}\xspace}

\def\Xibbarp      {{\ensuremath{\Xiresbar{}_\bquark^+}}\xspace}
\def\Omegab       {{\ensuremath{\Omegares^-_\bquark}}\xspace}

\newcommand{\decay}[2]{\ensuremath{#1\!\to #2}\xspace} 

\def\to                 {\ensuremath{\rightarrow}\xspace}

\def\CP                {{\ensuremath{C\!P}}\xspace}

\def\AT#1     {\ensuremath{A_{\mathrm{T}}^{#1}}\xspace}           

\def\C#1      {\ensuremath{\mathcal{C}_{#1}}\xspace}                       
\def\Cp#1     {\ensuremath{\mathcal{C}_{#1}^{'}}\xspace}                    
\def\Ceff#1   {\ensuremath{\mathcal{C}_{#1}^{\mathrm{(eff)}}}\xspace}        
\def\Cpeff#1  {\ensuremath{\mathcal{C}_{#1}^{'\mathrm{(eff)}}}\xspace}       
\def\Ope#1    {\ensuremath{\mathcal{O}_{#1}}\xspace}                       
\def\Opep#1   {\ensuremath{\mathcal{O}_{#1}^{'}}\xspace}                    

\newcommand{\ket}[1]{\ensuremath{|#1\rangle}}              

\newcommand{\nospaceunit}[1]{\ensuremath{\text{#1}}}       
\newcommand{\aunit}[1]{\ensuremath{\text{\,#1}}}       

\newcommand{\tev}{\aunit{Te\kern -0.1em V}\xspace}
\newcommand{\gev}{\aunit{Ge\kern -0.1em V}\xspace}
\newcommand{\mev}{\aunit{Me\kern -0.1em V}\xspace}
\newcommand{\kev}{\aunit{ke\kern -0.1em V}\xspace}
\newcommand{\ev}{\aunit{e\kern -0.1em V}\xspace}
 
\newcommand{\mevc}{\ensuremath{\aunit{Me\kern -0.1em V\!/}c}\xspace}
\newcommand{\gevc}{\ensuremath{\aunit{Ge\kern -0.1em V\!/}c}\xspace}
\newcommand{\mevcc}{\ensuremath{\aunit{Me\kern -0.1em V\!/}c^2}\xspace}
\newcommand{\gevcc}{\ensuremath{\aunit{Ge\kern -0.1em V\!/}c^2}\xspace}

\def\mum  {\ensuremath{\,\upmu\nospaceunit{m}}\xspace}

\def\fb   {\ensuremath{\aunit{fb}}\xspace}
\def\invfb   {\ensuremath{\fb^{-1}}\xspace}

\newcommand{\stat}{\aunit{(stat)}\xspace}
\newcommand{\syst}{\aunit{(syst)}\xspace}

\newcommand{\chisq}{\ensuremath{\chi^2}\xspace}

\newcommand{\chisqip}{\ensuremath{\chi^2_{\text{IP}}}\xspace}

\def\gsim{{~\raise.15em\hbox{$>$}\kern-.85em
          \lower.35em\hbox{$\sim$}~}\xspace}
\def\lsim{{~\raise.15em\hbox{$<$}\kern-.85em
          \lower.35em\hbox{$\sim$}~}\xspace}

\def\pt         {\ensuremath{p_{\mathrm{T}}}\xspace}

\def\ptot       {\ensuremath{p}\xspace}

\def\evtgen     {\mbox{\textsc{EvtGen}}\xspace}

\def\geant      {\mbox{\textsc{Geant4}}\xspace}

\def\photos     {\mbox{\textsc{Photos}}\xspace}

\def\pythia     {\mbox{\textsc{Pythia}}\xspace}

\def\tell1  {TELL1\xspace}
\def\ukl1   {UKL1\xspace}

\newcommand{\ie}{\mbox{\itshape i.e.}\xspace}

\def\Xibp    {{\ensuremath{\Xiresbar{}^+_\bquark}}\xspace}
\def\KKp 	  	{\proton \Km \Km}
\def\KKK 	  	{\Kp \Km \Km}

\def\xb 	  	{\X_b}
\def\BuToKKK 	  	{\decay{\Bm}{\KKK}} 

\def\XibToKKp 	  	{\decay{\Xibm}{\KKp}}
\def\XibToKpip 	  	{\decay{\Xibm}{\proton \Km \pim}}

\def\OmegabToKKp  	{\decay{\Omegab}{\KKp}}

\def\N     {{\ensuremath{N}}\xspace}

\def\XibToNKK		{\decay{\Xibm}{\N(\proton \piz) \Km \Km}}

\def\XibToKstKp		{\decay{\Xibm}{\Kstarm(\Km \piz) \Km \proton}}

\def\OmegabToKstKp	{\decay{\Omegab}{\Kstarm(\Km \piz) \Km \proton}}

\def\xb                       {\ensuremath{\PX^-_\bquark}\xspace}

\def\pK                       {\ensuremath{\Pp\PK^-}\xspace}
\def\Sigmapi                  {\ensuremath{\PSigma^+\Ppi^-}\xspace}

\def\Y                        {\ensuremath{\mathcal{N}}\xspace}

\def\klow                     {\ensuremath{\kaon^-_{\rm low}}\xspace}
\def\khigh                    {\ensuremath{\kaon^-_{\rm high}}\xspace}

\def\mpklowsq                 {\ensuremath{m_{\rm low}^{2}}\xspace}
\def\mpkhighsq                {\ensuremath{m_{\rm high}^{2}}\xspace}
\def\mkksq                    {\ensuremath{m_{\kaon\kaon}^{2}}\xspace}
\def\mpklow                   {\ensuremath{m_{\rm low}}\xspace}
\def\mpkhigh                  {\ensuremath{m_{\rm high}}\xspace}

\def\NRthreehalfpos             {\ensuremath{{\rm NR}(\frac{3}{2}^+)}\xspace}

\newcommand{\bei}{\begin{itemize}}
\newcommand{\eei}{\end{itemize}}
\newcommand{\beqn}{\begin{eqnarray}}
\newcommand{\eeqn}{\end{eqnarray}}
\newcommand{\beqns}{\begin{eqnarray*}}
\newcommand{\eeqns}{\end{eqnarray*}}
\newcommand{\beq}{\begin{equation}}
\newcommand{\eeq}{\end{equation}}
\newcommand{\beqa}{\begin{eqnarray}}
\newcommand{\eeqa}{\end{eqnarray}}

\usepackage{cite} 
\usepackage{mciteplus}
\usepackage{subfig}
\usepackage{rotating}
\usepackage{multirow}
\usepackage{afterpage}

\begin{document}

\renewcommand{\thefootnote}{\fnsymbol{footnote}}
\setcounter{footnote}{1}

\begin{titlepage}
\pagenumbering{roman}

\vspace*{-1.5cm}
\centerline{\large EUROPEAN ORGANIZATION FOR NUCLEAR RESEARCH (CERN)}
\vspace*{1.5cm}
\noindent
\begin{tabular*}{\linewidth}{lc@{\extracolsep{\fill}}r@{\extracolsep{0pt}}}
\ifthenelse{\boolean{pdflatex}}
{\vspace*{-1.5cm}\mbox{\!\!\!\includegraphics[width=.14\textwidth]{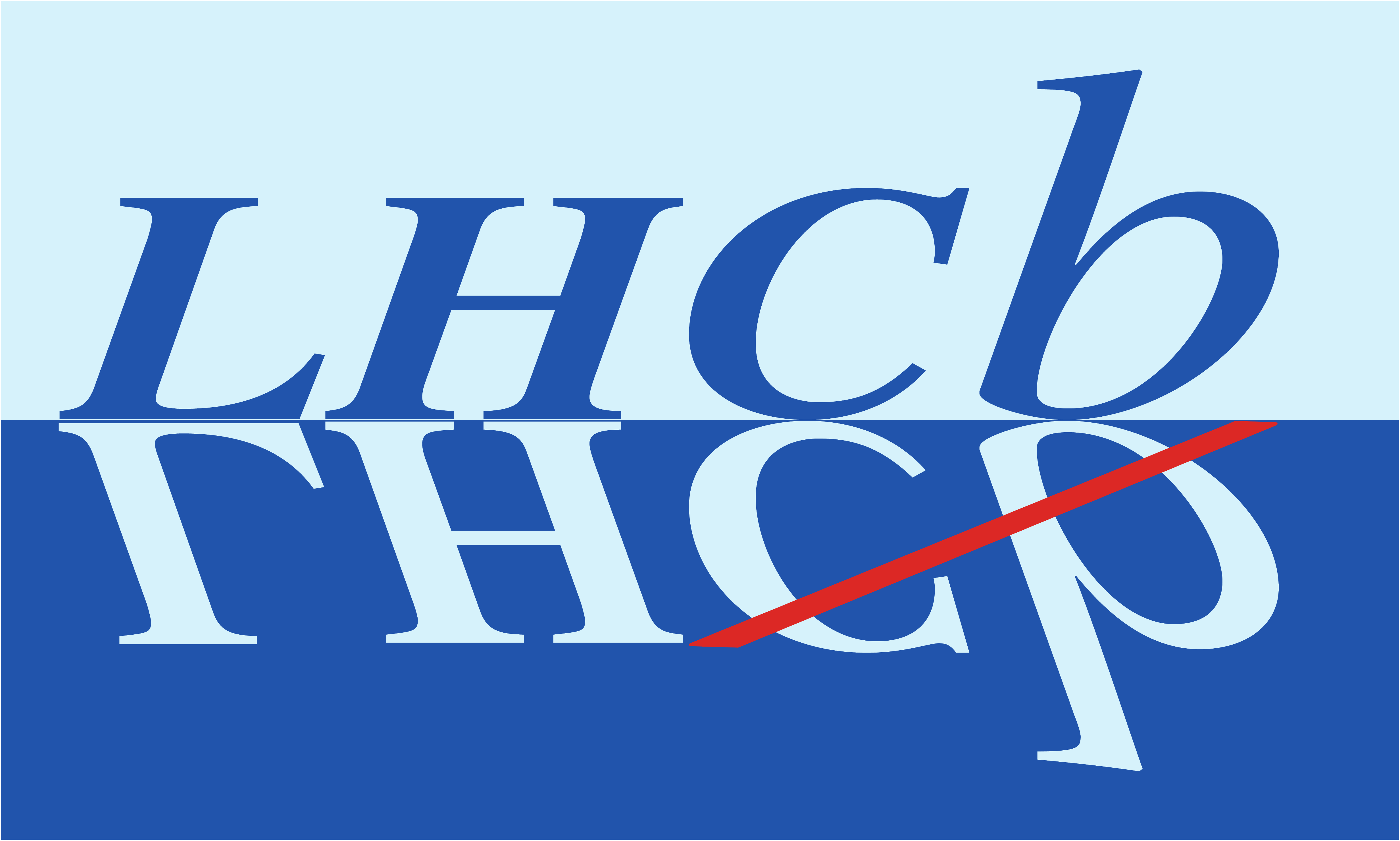}} & &}%
{\vspace*{-1.2cm}\mbox{\!\!\!\includegraphics[width=.12\textwidth]{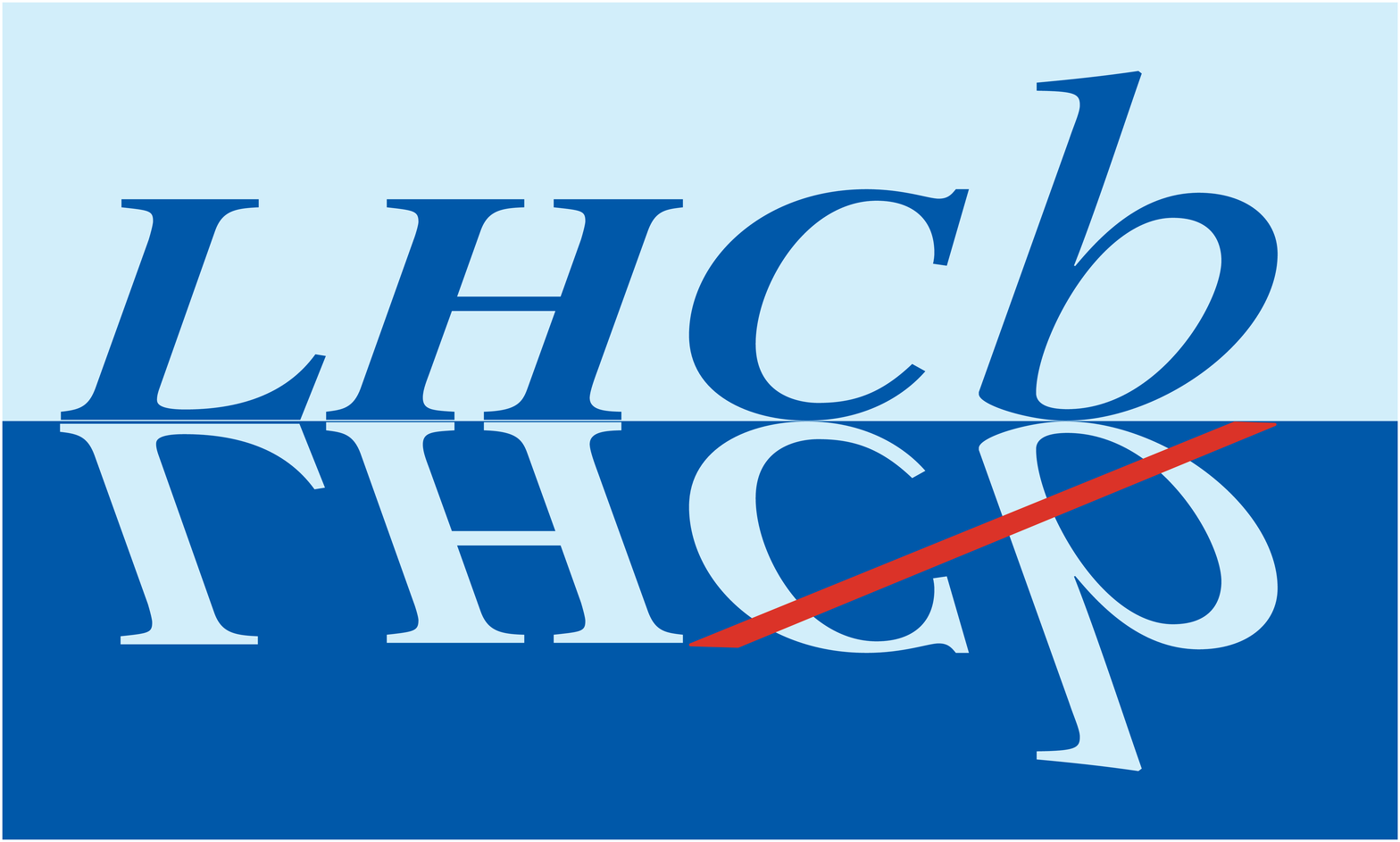}} & &}%
\\
 & & CERN-EP-2021-062 \\  
 & & LHCb-PAPER-2020-017 \\  
 & & \today \\ 
 & & \\
\end{tabular*}

\vspace*{3.0cm}

{\normalfont\bfseries\boldmath\huge
\begin{center}
  \papertitle 
\end{center}
}

\vspace*{1.5cm}

\begin{center}
\paperauthors\footnote{Authors are listed at the end of this paper.}
\end{center}

\vspace*{0.5cm}

\begin{abstract}
  \noindent
  A search for \CP\ violation in charmless three-body \XibToKKp\ decays is performed using $pp$ collision data recorded with the \lhcb detector, corresponding to integrated luminosities of $1\invfb$ at a centre-of-mass energy $\sqrt{s} = 7\tev$, $2\invfb$ at $\sqrt{s} = 8\tev$ and $2\invfb$ at $\sqrt{s} = 13\tev$.
   A good description of the phase-space distribution is obtained with an amplitude model containing contributions from $\Sigmares(1385)$, $\Lz(1405)$, $\Lz(1520)$, $\Lz(1670)$, $\Sigmares(1775)$ and $\Sigmares(1915)$ resonances.
  The model allows for \CP-violation effects, which are found to be consistent with zero.
  The branching fractions of $\decay{\Xibm}{\Sigmares(1385)\Km}$, $\decay{\Xibm}{\Lz(1405)\Km}$, $\decay{\Xibm}{\Lz(1520)\Km}$, $\decay{\Xibm}{\Lz(1670)\Km}$, $\decay{\Xibm}{\Sigmares(1775)\Km}$ and $\decay{\Xibm}{\Sigmares(1915)\Km}$ decays are also reported.
  In addition, an upper limit is placed on the product of ratios of \Omegab\ and \Xibm\ fragmentation fractions and the \OmegabToKKp and \XibToKKp\ branching fractions.

\end{abstract}

\vspace*{0.25cm}

\begin{center}
	Published in Phys. Rev. D 104, 052010
\end{center}

\vspace{\fill}

{\footnotesize 
\centerline{\copyright~\papercopyright. \href{\paperlicenceurl}{\paperlicence}.}}
\vspace*{2mm}

\end{titlepage}


\newpage
\setcounter{page}{2}
\mbox{~}
\renewcommand{\thefootnote}{\arabic{footnote}}
\setcounter{footnote}{0}
\cleardoublepage


\pagestyle{plain} 
\setcounter{page}{1}
\pagenumbering{arabic}


\section{Introduction}
\label{sec:Introduction}

In the Standard Model (SM), \CP\ violation, defined as the breaking of symmetry under the combined charge conjugation and parity operations, 
owes its origin to a single irreducible complex phase in the Cabibbo--Kobayashi--Maskawa (CKM) matrix~\cite{Cabibbo:1963yz,Kobayashi:1973fv}.
All effects of \CP\ violation in particle decays observed so far are consistent with this paradigm.
However, the degree of \CP\ violation permitted in the SM is inconsistent with the observed matter-antimatter asymmetry in the Universe~\cite{Sakharov:1967dj,Shaposhnikov:1991cu}.
This motivates further searches for sources of \CP\ violation beyond the SM.

Interference between two amplitudes with different weak and strong phases leads to \CP\ violation in decay, where weak phases are those that change sign under \CP\ conjugation while strong phases do not.
In the SM, weak phases are associated with the complex elements of the CKM matrix and strong phases are associated with hadronic final-state effects. 
Two such amplitudes are potentially present in decays of \bquark hadrons to final states that do not contain charm quarks, which therefore provide fertile ground for studies of \CP violation.
Significant asymmetries have been observed between \B and \Bbar partial widths 
in \decay{\Bzb}{\Km\pip}~\cite{Lees:2012mma,Duh:2012ie,Aaltonen:2014vra,LHCb-PAPER-2013-018,LHCb-PAPER-2018-006} 
\decay{\Bzb}{\pip\pim}~\cite{Lees:2012mma,Duh:2012ie,Adachi:2013mae} and \decay{\Bsb}{\Kp\pim}~\cite{Aaltonen:2014vra,LHCb-PAPER-2013-018} decays.
Even larger \CP-violation effects have been observed in regions of the phase space of \Bm decays to $\pip\pim\pim$, $\Km\pip\pim$, $\Kp\Km\Km$ and $\Kp\Km\pim$ final states~\cite{LHCb-PAPER-2013-027,LHCb-PAPER-2013-051,LHCb-PAPER-2014-044,LHCb-PAPER-2018-051,LHCb-PAPER-2019-017,LHCb-PAPER-2019-018}.

Breaking of \CP\ symmetry has not yet been observed in the properties of any baryon.  
Tests of this symmetry have been performed through studies of \Lb baryon decays to
$\proton\pim$, $\proton\Km$~\cite{Aaltonen:2014vra,LHCb-PAPER-2018-025}, 
$\KS\proton\pim$~\cite{LHCb-PAPER-2013-061}, 
$\Lz\Kp\Km$, $\Lz\Kp\pim$~\cite{LHCb-PAPER-2016-004},
$\proton\pim\pip\pim$, $\proton\pim\Kp\Km$, $\proton\Km\pip\pim$ 
and $\proton\Km\Kp\Km$~\cite{LHCb-PAPER-2019-028,LHCb-PAPER-2018-044,LHCb-PAPER-2018-001} final states,
as well as \Xibz decays to $\proton\Km\pip\pim$ and $\proton\Km\pip\Km$~\cite{LHCb-PAPER-2018-044,LHCb-PAPER-2018-001}.
No significant evidence of \CP\ violation has been found in any of these studies, nor in measurements of the properties of charm baryon decays~\cite{PDG2020}.
In light of the large \CP-violation effects observed in three-body charmless decays of \B\ mesons, it is of great interest to extend the range of searches in \bquark-baryon decays.
In particular, the recently observed \XibToKKp\ decay~\cite{LHCb-PAPER-2016-050} provides an interesting new opportunity to search for \CP-violation effects.

In this paper, the first amplitude analysis of \XibToKKp\ decays is reported.
This is also the first amplitude analysis of any \bquark-baryon decay mode allowing for \CP-violation effects.
A search for the previously unobserved \OmegabToKKp\ decay is also presented.
The analysis reported here is performed using proton-proton ($pp$) collision data recorded with the \lhcb detector, corresponding to integrated luminosities of
$1 \invfb$ at a centre-of-mass energy of $\sqrt{s} = 7 \tev$ collected in 2011, 
$2 \invfb$ at $\sqrt{s} = 8 \tev$ in 2012 
and $2 \invfb$ at $\sqrt{s} = 13 \tev$ in 2015 and 2016.
The data-taking period of 2011 and 2012 is referred to hereafter as Run~1 and that of 2015 and 2016 as Run~2.
The inclusion of charge-conjugate processes is implied throughout the paper, except where asymmetries are discussed.

This paper is organised as follows.
Section~\ref{sec:Detector} gives a brief description of the \lhcb detector, trigger requirements and simulation software.
The signal candidate selection procedure is set out in Sec.~\ref{sec:offselection}.
In Sec.~\ref{sec:invmassfit}, the procedure for estimating the signal and background yields that enter the amplitude fit is explained.
Section~\ref{sec:amplitudeana} covers the modelling of the distribution of decays across the phase space.
Sections~\ref{sec:systematics} and~\ref{sec:results} contain a description of the systematic uncertainties associated with the analysis procedure and a presentation of the results, respectively.
A brief summary of the analysis is given in Sec.~\ref{sec:summary}.

\section{Detector, trigger and simulation}
\label{sec:Detector}

The \lhcb detector~\cite{LHCb-DP-2008-001,LHCb-DP-2014-002} is a single-arm forward
spectrometer covering the \mbox{pseudorapidity} range $2<\eta <5$,
designed for the study of particles containing \bquark or \cquark
quarks. The detector includes a high-precision tracking system
consisting of a silicon-strip vertex detector surrounding the $pp$
interaction region, a large-area silicon-strip detector located
upstream of a dipole magnet with a bending power of about
$4{\mathrm{\,Tm}}$, and three stations of silicon-strip detectors and straw
drift tubes placed downstream of the magnet.
The tracking system provides a measurement of the momentum, \ptot, of charged particles with
a relative uncertainty that varies from 0.5\% at low momentum to 1.0\% at 200\gev.\footnote{Natural units with $\hbar=c=1$ are used throughout this paper.}
The minimum distance of a track to a primary $pp$ collision vertex (PV), the impact parameter (IP), 
is measured with a resolution of $(15+29/\pt)\mum$,
where \pt is the component of the momentum transverse to the beam direction, in\,\gev.
Different types of charged hadrons are distinguished using information
from two ring-imaging Cherenkov detectors. 
Photons, electrons and hadrons are identified by a calorimeter system consisting of scintillating-pad and preshower detectors, an electromagnetic and a hadronic calorimeter.
Muons are identified by a system composed of alternating layers of iron and multiwire proportional chambers.
The magnetic field deflects oppositely charged particles in opposite directions and this can lead to detection asymmetries.
Periodically reversing the magnetic field polarity throughout the data-taking reduces this effect to a negligible level.
Approximately 60\% of 2011 data, 50\% of 2012 data, 61\% of 2015 data and 53\% of 2016 data were collected in the ``down'' polarity configuration, and the rest in the ``up'' configuration.

The online event selection is performed by a trigger, which consists of a hardware stage, based on information from the calorimeter and muon systems, followed by a software stage, which applies a full event reconstruction.
During offline analysis, reconstructed candidates are associated with trigger decisions.
Events considered in the analysis are required to have been triggered at the hardware level in one of two ways: either through one of the final-state tracks of the signal decay depositing sufficient energy in the calorimeter system, or by one of the other tracks in the event, not reconstructed as part of the signal candidate, fulfilling any hardware trigger requirement.
At the software stage, it is required that at least one charged particle associated to the \bquark-hadron candidate has high \pt and high \chisqip, where \chisqip is defined as the difference in PV fit \chisq with and without the inclusion of a specific particle. 
A multivariate algorithm~\cite{BBDT} is used to identify secondary vertices consistent with being a two- or three-track \bquark-hadron decay. 
The PVs are fitted with and without 
the tracks that comprise 
the \bquark-baryon candidate, and the PV that gives the smallest \chisqip is associated with the candidate.
Finally, the momentum scale for charged particles is calibrated using  samples of $\decay{\jpsi}{\mumu}$, $\decay{\Bu}{\jpsi\Kp}$ and $\decay{\PLambda}{\proton \pi^-}$~decays collected concurrently with the~data sample used for this analysis~\cite{LHCb-PAPER-2012-048,LHCb-PAPER-2013-011}.

Simulation samples are used to investigate background from other \bquark-hadron decays and to study the detection and reconstruction efficiency of the signal.
In the simulation, $pp$ collisions are generated using
\pythia~\cite{Sjostrand:2006za,*Sjostrand:2007gs} with a specific \lhcb
configuration~\cite{LHCb-PROC-2010-056}.  Decays of unstable particles
are described by \evtgen~\cite{Lange:2001uf}, in which final-state
radiation is generated using \photos~\cite{davidson2015photos}. 
The interaction of the generated particles with the detector, and its response, are implemented using the \geant\ toolkit~\cite{Allison:2006ve, *Agostinelli:2002hh} as described in Ref.~\cite{LHCb-PROC-2011-006}.

\section{Offline selection}
\label{sec:offselection}

The offline selection consists of an initial filtering stage followed by a requirement on the output of a multivariate algorithm (MVA).
Compared to the procedure applied to select the \XibToKKp\ channel in Ref.~\cite{LHCb-PAPER-2016-050}, improvements in both stages lead to a significant increase in efficiency.
In particular, the inclusion in the multivariate algorithm of particle identification (PID) variables that distinguish the final-state charged hadrons from misidentified particles, is found to separate signal from background effectively.

In the filtering stage, tracks are required to be of good quality, to satisfy $\ptot > 1500\mev$ and $\pt > 250\mev$, and to be displaced from all PVs. 
Tracks associated to proton candidates must, at this stage, satisfy a loose PID requirement and all tracks are required to not be associated to hits in the muon system. 
Each \bquark-hadron (henceforth denoted as \xb) candidate must form a good-quality decay vertex that is separated significantly from any PV, and must be consistent with originating from its associated PV. 
Only \xb\ candidates with $\pt > 3500\mev$ and invariant mass $5545 < m(\KKp) < 6470\mev$ are retained for further analysis.

In the selected $m(\KKp)$ range there are three main categories of background that contribute: 
combinatorial background that results from random association of unrelated tracks;
partially reconstructed background due to \bquark-hadron decays into final states similar to the signal, but with additional soft particles that are not reconstructed;
and cross-feed background that results from misidentification of one or more final-state particles. 
The MVA classifier is designed primarily to reduce combinatorial background while retaining high signal efficiency, but also has some discriminating power against the other background sources.
It is trained with a signal sample comprised of simulated \XibToKKp\ decays generated uniformly across the phase space, and a background sample obtained from candidates in data in the sideband regions $5545.0 < m(\KKp) < 5634.4 \mev$ and $6209.0 < m(\KKp) < 6470.0 \mev$.
The latter of these regions is dominated by combinatorial background, as its lower threshold excludes possible \OmegabToKKp\ decays from the sample.
The former region includes also contributions from sources of partially reconstructed background such as \XibToNKK\ or \XibToKstKp\ decays.
Potential cross-feed background from \BuToKKK\ decays is removed by assigning the proton candidate the kaon mass and vetoing the  $m(\KKK)$ region within $\pm 45\mev$ around the known \Bm\ mass~\cite{PDG2020}.
This veto corresponds to approximately $\pm 3$ times the invariant mass resolution for \BuToKKK\ decays. 

Variables that exhibit good discriminating power between the signal and background samples are chosen as inputs to the MVA.
These are:
the angle between the \xb\ candidate's momentum vector and the line connecting its decay vertex to its associated PV;
the scalar sum of the \pt\ of all final-state tracks; 
the \chisqip\ of the highest \pt\ final-state track and of the \xb\ candidate;
the square of the significance of the distance between the \xb\ decay vertex and its associated PV; 
the vertex fit \chisq\ per degree of freedom of the \xb candidate; 
the minimum change in the \xb\ candidate vertex fit \chisq\ when including an additional track;
variables that characterise the PID information of the proton and kaon candidates; 
and a variable that quantifies the isolation of the \xb\ candidate.
The last of these is defined as the \pt\ asymmetry between the \xb candidate and the tracks within a circle, centred on the \xb candidate (but excluding its decay products), with a radius $\sqrt{\delta\eta^2 + \delta\phi^2} < 1.7$ in the space of pseudorapidity $\eta$ and azimuthal angle $\phi$ (in radians) around the beam direction~\cite{LHCb-PAPER-2012-001}.

To describe accurately the proton and kaon PID variables, the quantities in simulation are resampled according to values obtained from data calibration samples of \decay{\Lb}{\Lc\pim}, \decay{\Ds}{\phi\pip} and \decay{\Dstarp}{\Dz\pip} decays~\cite{LHCb-DP-2018-001}.
The procedure accounts for correlations between the variables associated to a particular track, as well as the dependence of the PID response on \pt, $\eta$, and the multiplicity of tracks in the event.
All other MVA input variables show good agreement between simulation and data, as validated with a control sample of \decay{\Bm}{\proton\antiproton\Km} decays.
The MVA input variables are also found to not be correlated strongly either with \xb\ candidate mass or with position in the phase space of the decay.

Several types of MVA classifier are investigated, with a gradient boosted decision tree algorithm giving the best performance~\cite{xgboost}.
Four classifiers are trained separately with samples separated by data-taking period (Run~1 or Run~2) and by even or odd event numbers. 
The event number identifies the proton-proton bunch crossing, from which the \xb\ candidate was recorded, in a certain operational period of the experiment.
To avoid possible MVA overtraining, for each data-taking period the classifier trained on the sample with even event numbers is validated and employed on the sample with odd event numbers, and vice versa.

A threshold on the output of the MVA is chosen to maximise ${\cal N}_S/\sqrt{{\cal N}_S+{\cal N}_B}$, where ${\cal N}_S$ and ${\cal N}_B$ represent the estimated numbers of \XibToKKp\ signal and combinatorial background candidates, respectively, within a signal region of $\pm 40\mev$ around the \Xibm\ mass from Ref.~\cite{PDG2014}.
This range corresponds to approximately $\pm 2.5$ times the \XibToKKp\ invariant mass resolution.
The value of ${\cal N}_S$ is estimated using the signal efficiency evaluated from simulation, multiplied by the \XibToKKp\ branching fraction, the \Xibm\ fragmentation fraction, the $\bquark\bquarkbar$ production cross-section~\cite{LHCb-PAPER-2016-031} and the integrated luminosity for the relevant data-taking period.
The product of the \XibToKKp\ branching fraction and the \Xibm\ fragmentation fraction is obtained from the results of Ref.~\cite{LHCb-PAPER-2016-050}, where the \BuToKKK\ channel is used for normalisation, by multiplying the \Bm\ fragmentation fraction in the relevant kinematic range~\cite{LHCb-PAPER-2013-004} and the \BuToKKK branching fraction~\cite{PDG2020}.
The value of ${\cal N}_B$ is estimated from data by fitting the region $6125 < m(\KKp) < 6470\mev$ with a linear function and extrapolating the result into the signal region.
The MVA output requirements have efficiencies of about $52\%$ and $61\%$ for Run~1 and Run~2, respectively, with combinatorial background rejection of about $98\%$ for both data-taking periods.
The choice of ${\cal N}_S/\sqrt{{\cal N}_S+{\cal N}_B}$ as the figure of merit is intended to obtain a sufficiently large data sample to make an amplitude analysis viable. 
After all offline selection requirements are applied, each selected event contains a single \xb\ candidate.

The variables describing the phase space of the decay, which are used in the amplitude analysis, are calculated following a kinematic fit in which the \xb\ candidate mass is fixed to the \Xibm\ mass from Ref.~\cite{PDG2014}.
This procedure improves resolution of these variables and ensures that all decays remain within the phase-space boundary. 
The difference between the \Xibm\ mass value used in this fit and recent more precise results~\cite{LHCb-PAPER-2014-048,LHCb-PAPER-2016-053,LHCb-PAPER-2018-047} has negligible impact on the analysis.
The experimental resolution of the $m(\proton\Km)$ invariant mass, in the region with the narrowest resonance considered in this analysis, the $\PLambda(1520)$ state, is expected to be around $1.5\mev$.
This is smaller than the $\PLambda(1520)$ width, and therefore effects related to finite resolution in the phase-space variables are not considered further.

The expected \XibToKKp\ signal efficiency, assuming uniform distribution of decays across the phase space and taking into account the \lhcb\ detector acceptance, reconstruction and both online and offline selection criteria, is 
$\left(1.159 \pm 0.005\right)\%$ for Run~1 and $\left(1.748 \pm 0.006\right)\%$ for Run~2.
The corresponding \OmegabToKKp\ signal efficiencies are $\left(1.257 \pm 0.005\right)\%$ and $\left(1.921 \pm 0.006\right)\%$.
The quoted uncertainties are due to the limited size of the simulation samples only.

\section{$\boldsymbol{\xb}$ candidate mass fit}
\label{sec:invmassfit}

\begin{figure}[!tb]
\centering
\includegraphics*[width=0.49\textwidth]{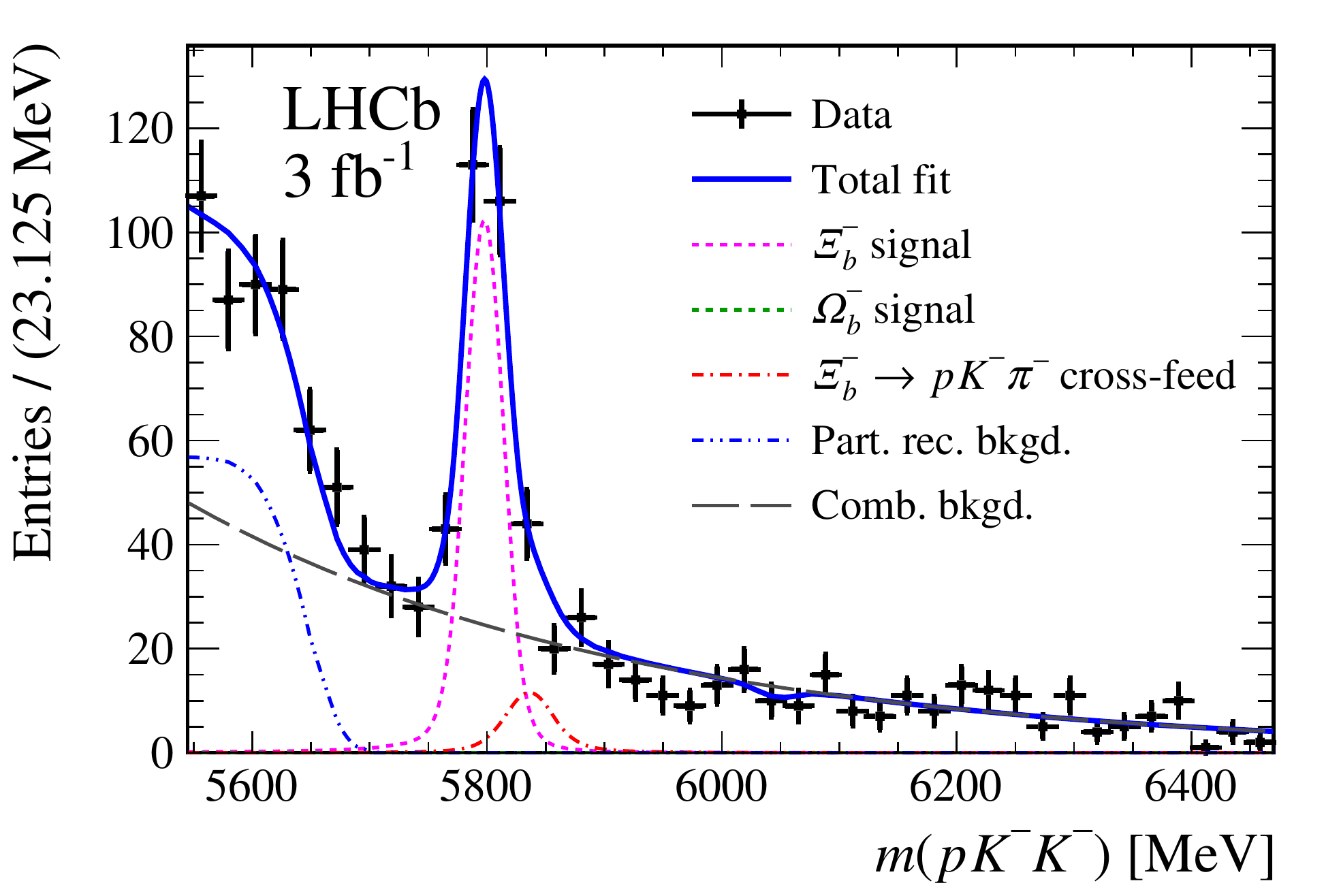}
\includegraphics*[width=0.49\textwidth]{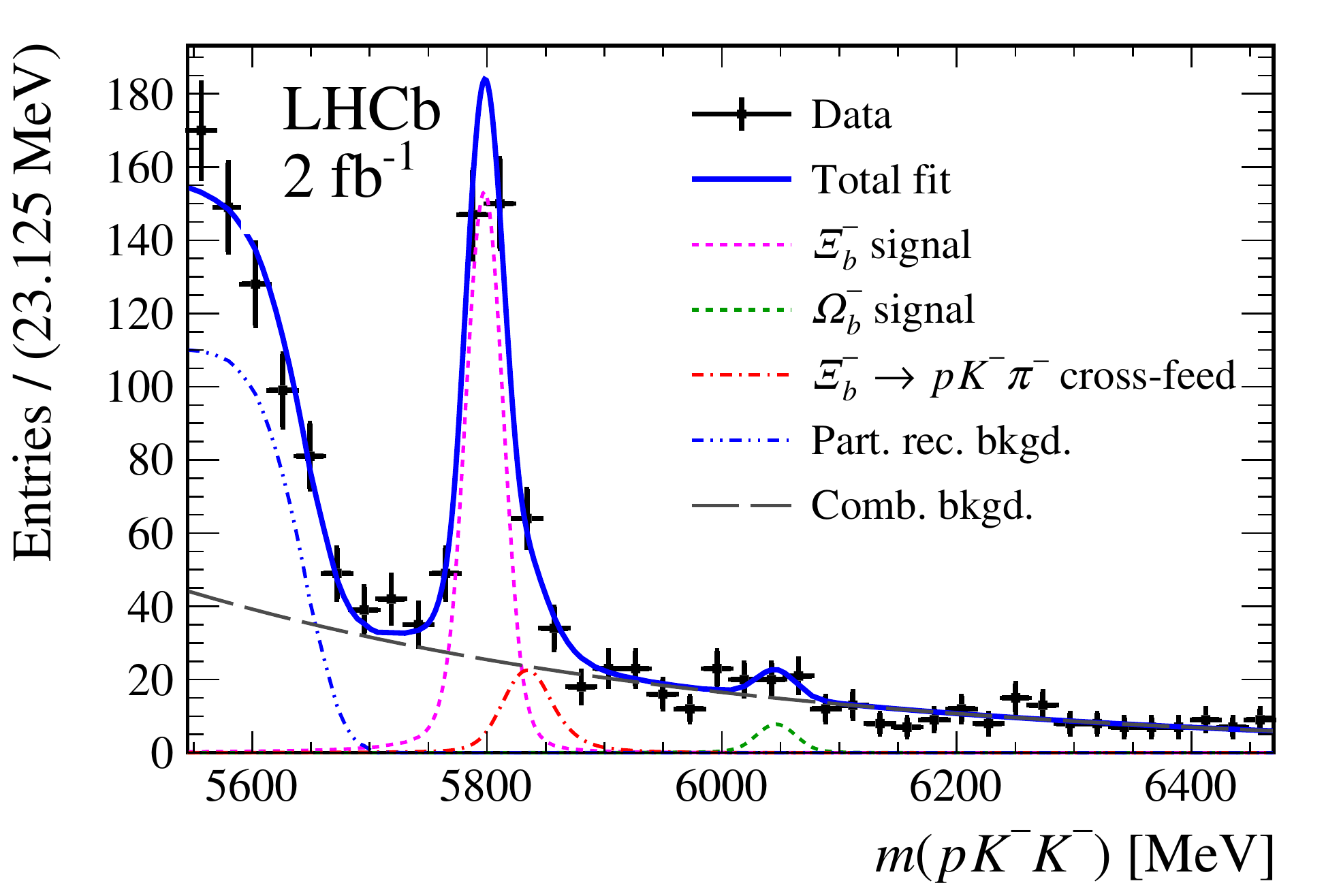}
\caption{\small 
  Distributions of $\KKp$ invariant mass for \xb\ candidates in (left)~Run~1 and (right)~Run~2 data with results of the unbinned extended maximum-likelihood fits superimposed.
  The total fit result is shown as the blue solid curve, with individual components shown as indicated in the legend.
}
\label{fig:fitdata}
\end{figure}

Distributions of $m(\KKp)$ for selected \xb\ candidates are shown in Fig.~\ref{fig:fitdata} for Run~1 and Run~2 separately.
The signal yields are obtained from unbinned extended maximum-likelihood fits to these distributions. 
The fit model is composed of signal and background components whose shape parameters are mostly obtained from fits to the corresponding simulation samples, after imposing the same selection requirements as on the data.
One exception is the combinatorial background component, which is modelled by an exponential function with slope parameter allowed to vary freely in the fit to data.

Signal \XibToKKp\ and \OmegabToKKp\ components are each modelled with the sum of two Crystal Ball (CB) functions~\cite{Skwarnicki:1986xj} where the core width and peak position are shared and with independent power-law tails on both sides.
The tail parameters and the relative normalisation of the CB functions are determined from simulation.
The peak positions are fixed to the \Xibm\ mass from Ref.~\cite{LHCb-PAPER-2014-048} and the known \Omegab\ mass~\cite{PDG2020}, and a scale factor relating the width in data to that in simulation is introduced.

A possible cross-feed background contribution from \XibToKpip\ decays~\cite{LHCb-PAPER-2016-050}, where the pion is misidentified as a kaon, is modelled with the sum of two CB functions.
All shape parameters of this function are fixed according to the values obtained from a fit to simulation but the width is scaled by the same factor as the signal components.
The phase-space distribution of these decays is not known, and the simulation sample is weighted according to a model, inspired by the $m(\proton\Km)$ and $m(\proton\pim)$ mass spectra observed in $\Lb \to \jpsi \proton \Km$~\cite{LHCb-PAPER-2015-029} and $\Lb \to \jpsi \proton\pim$~\cite{LHCb-PAPER-2016-015} decays, which consists of the $\Lz(1405)$, $\Lz(1520)$, $\Lz(1690)$, $N(1440)$, $N(1520)$, $N(1535)$ and $N(1650)$ resonances.
The yield of the \XibToKpip\ cross-feed component is expressed relative to the \XibToKKp signal yield and constrained within uncertainty according to the previous branching fraction ratio measurement~\cite{LHCb-PAPER-2016-050} and relative selection efficiency.
The expected relative yields are $0.15 \pm 0.06$ and $0.14 \pm 0.03$ for Run~1 and Run~2, respectively.

Partially reconstructed and combinatorial background contributions are also included in the fit model.
It is found that the $m(\KKp)$ distributions of various potential sources of partially reconstructed background, such as \XibToNKK\ or \XibToKstKp\ decays, are very similar~\cite{LHCb-PAPER-2015-029}.
Therefore, the baseline fit model includes a single partially reconstructed background component, which is modelled from simulated \XibToKstKp\ decays with an ARGUS function~\cite{Albrecht:1990am} convolved with a Gaussian function.
The threshold of the ARGUS function is fixed to the known value of $m_{\Xibm}-m_{\piz}$~\cite{PDG2020,LHCb-PAPER-2014-048}, and the width parameter of the Gaussian function is taken from the fit to simulation and scaled by the same factor as the signal components. 
Negligible contributions are expected from partially reconstructed \Omegab\ decays, such as \OmegabToKstKp. 

The results of the fits to Run~1 and Run~2 data are shown in Table~\ref{table:fitdata} and Fig.~\ref{fig:fitdata}.
The free parameters of each fit are the two signal yields, the partially reconstructed and combinatorial background yields, the width scale factor and the exponential shape parameter of the combinatorial background, while the cross-feed background yield is constrained to its expectation relative to the \XibToKKp\ signal yield.

\begin{table}[!tb] 
\caption{\small
  Yields obtained from fits to the $m(\KKp)$ distributions, in the full invariant mass range.
  The quoted uncertainties are statistical only.
}
\label{table:fitdata}
\centering
\renewcommand{\arraystretch}{1.1}
\begin{tabular}{lr@{$\,\pm\,$}lr@{$\,\pm\,$}l} 
\hline 
Parameter & \multicolumn{2}{c}{Run~1} &  \multicolumn{2}{c}{Run~2} \\
\hline
\XibToKKp\ yield  		& 193 & 21	& 297 & 23 	\\
\OmegabToKKp\ yield 		& $-4$ & 6 	&  15 &  9 	\\
Partially reconstructed background yield & 231 & 34 	& 442 & 36 	\\
Combinatorial background yield 		& 721 & 50 	& 775 & 51 	\\
\hline
\end{tabular} 
\end{table}

Only candidates in the $m(\KKp)$ signal region of $\pm 40\mev$ around the \Xibm\ mass from Ref.~\cite{PDG2014} are retained for the amplitude analysis.
In this region, the yields of the signal, cross-feed and combinatorial components are $N_{\rm sig} = 181 \pm 20$, $N_{\rm cf}=16 \pm 7$ and $N_{\rm comb} = 90 \pm 6$ for Run~1, and $N_{\rm sig} = 278 \pm 21$, $N_{\rm cf}=25 \pm 6$ and $N_{\rm comb} = 95 \pm 6$ for Run~2, where the quoted uncertainties are statistical only.
These correspond to signal purities of $(63\pm 3)\%$ and $(70 \pm 2)\%$ for Run~1 and Run~2, respectively.
The contribution from partially reconstructed background in the signal region is negligible.

No significant signal from the \OmegabToKKp\ decay is observed.
The results of the fits are used to set limits on the product of its branching fraction with the fragmentation fraction for \Omegab\ production, normalised to the corresponding quantities for \XibToKKp\ decay, \ie
\begin{equation}
\label{eq:bfr}
\mathcal{R} = \frac{f_{\Omegab}}{f_{\Xibm}} \times \frac{{\cal B}(\OmegabToKKp)}{{\cal B}(\XibToKKp)} = \frac{\epsilon(\XibToKKp)}{\epsilon(\OmegabToKKp)} \times \frac{N(\OmegabToKKp)}{N(\XibToKKp)} \,,
\end{equation}
where $N$ and $\epsilon$ denote yield and efficiency, respectively, for the indicated mode, while $f_{\Xibm}$ and $f_{\Omegab}$ are the \Xibm\ and \Omegab\ fragmentation fractions.
Results for the ratio $\mathcal{R}$ are reported, both for Run~1 and Run~2 separately and combined, in Sec.~\ref{sec:results}.

\section{Amplitude analysis}
\label{sec:amplitudeana}

\begin{figure}[!tb]
\centering
\includegraphics[width=0.49\textwidth]{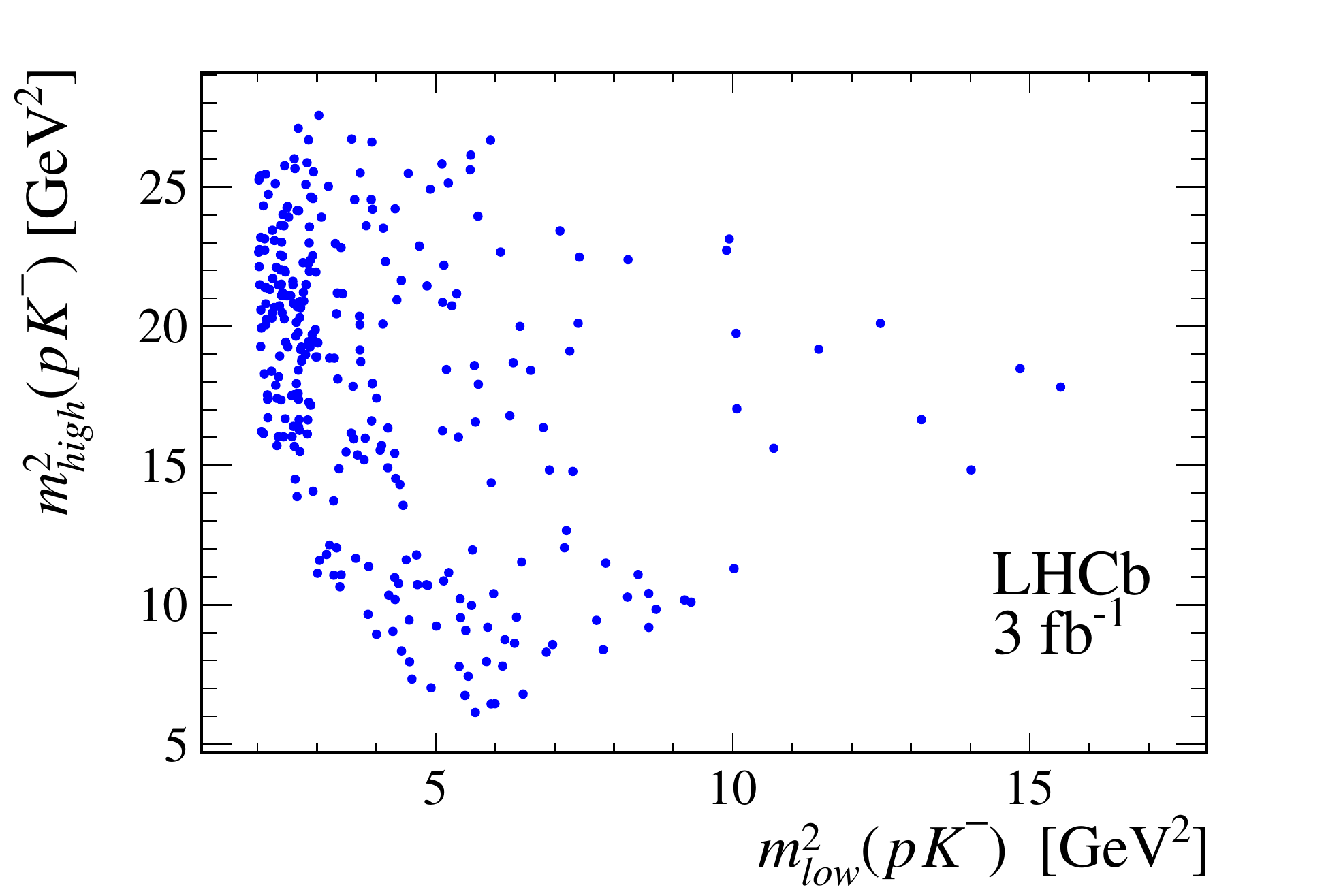} 
\includegraphics[width=0.49\textwidth]{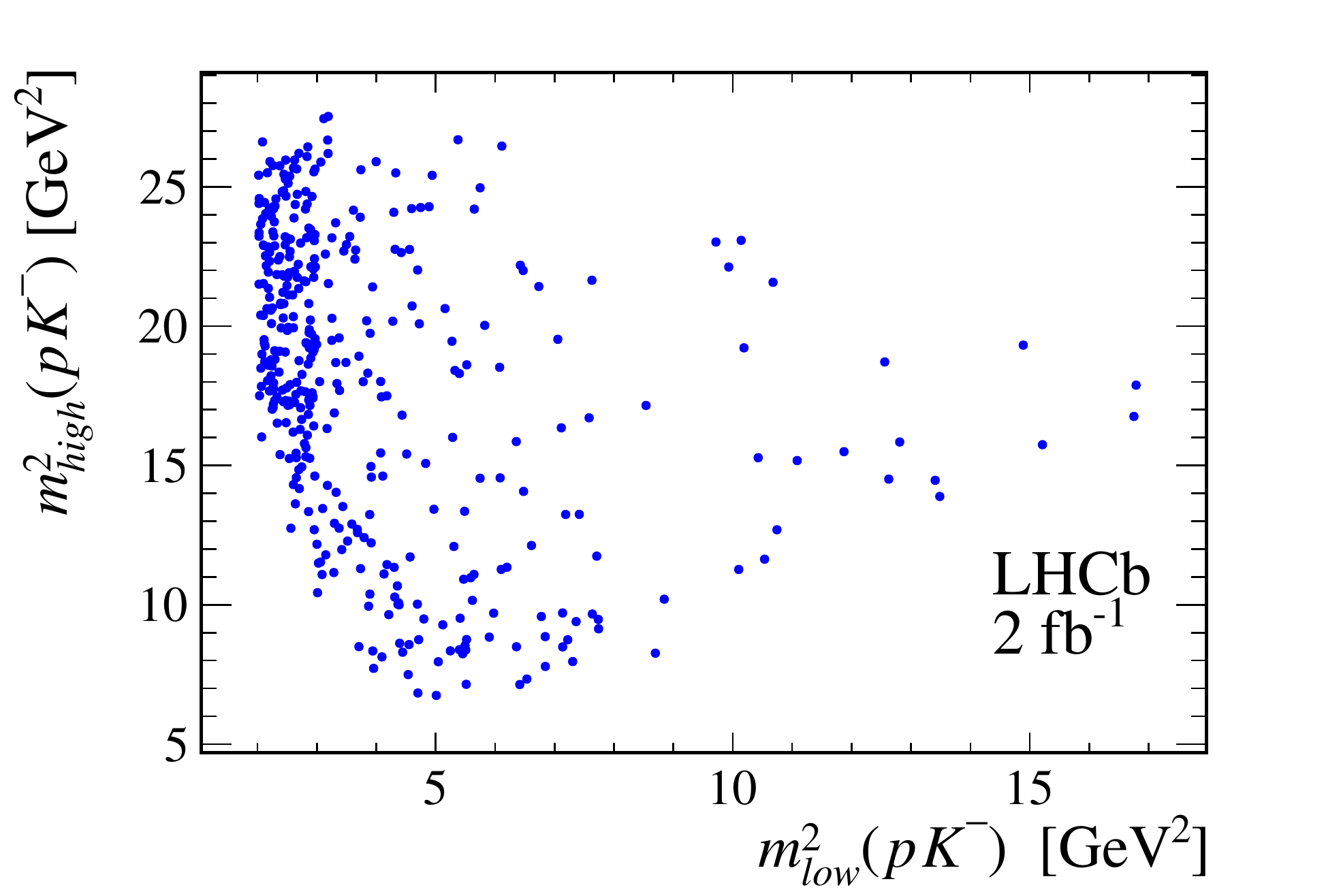} \\
\includegraphics[width=0.49\textwidth]{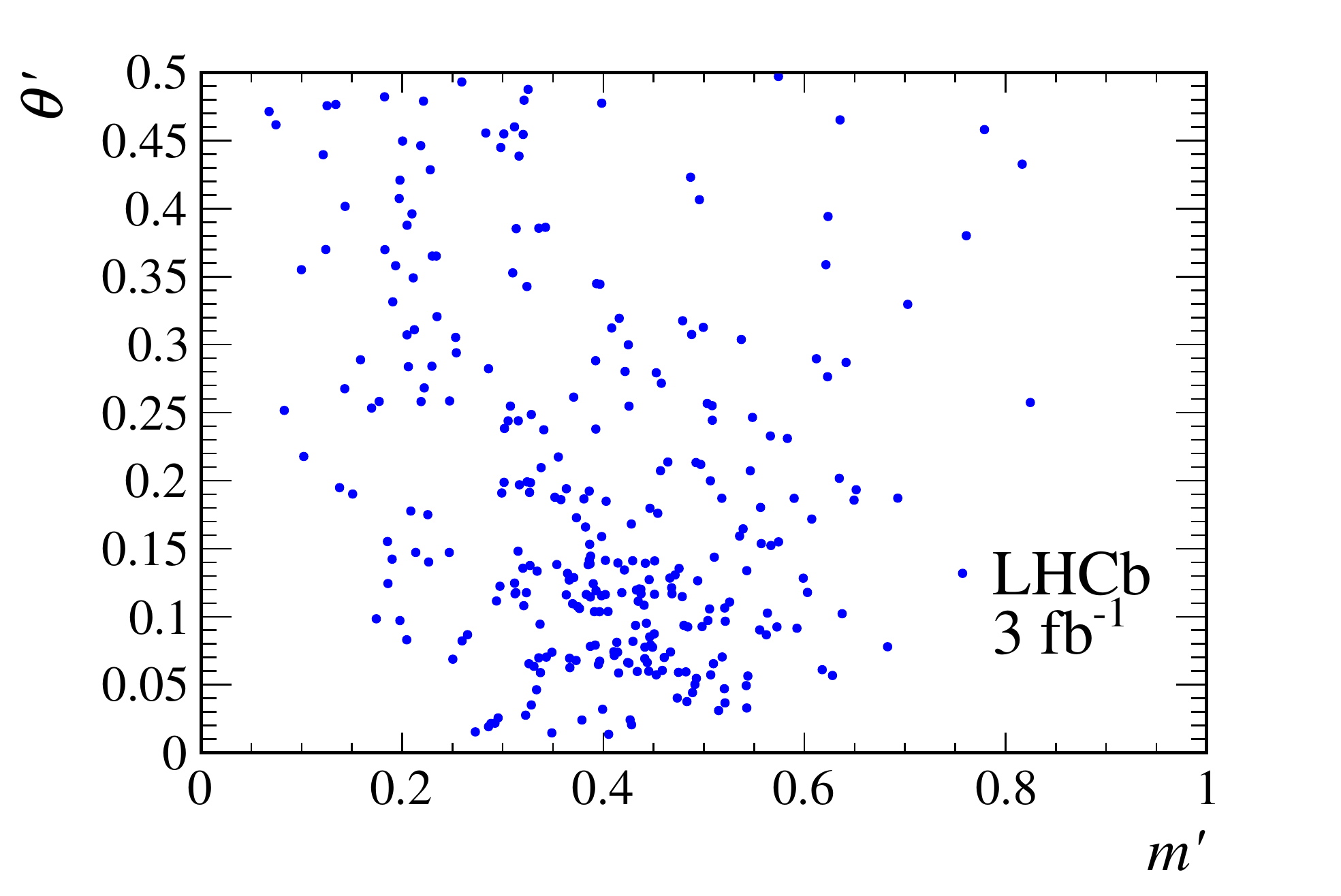}
\includegraphics[width=0.49\textwidth]{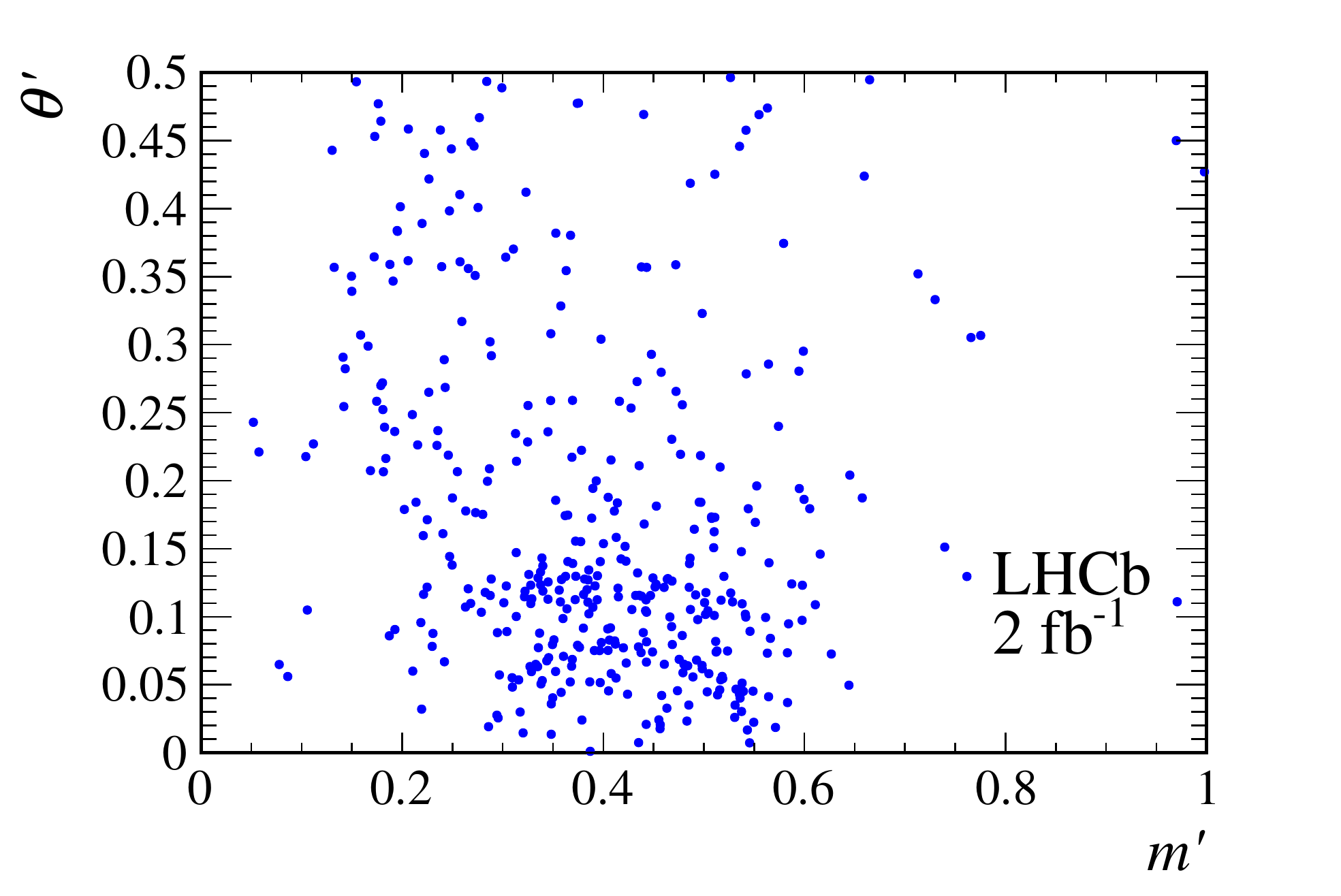}
\caption{\small
  Distributions of selected candidates from (left)~Run~1 and (right)~Run~2 data in the (top)~Dalitz-plot and (bottom)~square Dalitz-plot representations of the phase space.
}
\label{fig:data_run}
\end{figure}

The phase space of the three-body decay of a, potentially polarised, \bquark baryon has five degrees of freedom.
A baseline assumption is made that \Xibm\ baryons produced in $\proton\proton$ collisions within the LHCb acceptance have negligible polarisation, as observed for \Lb\ baryons~\cite{LHCb-PAPER-2012-057,LHCb-PAPER-2020-005}.
As a result, the phase space of the $\decay{\Xibm}{\proton\kaon^-_1\kaon^-_2}$ decay is characterised by two independent kinematic variables (subscripts here distinguish the two kaons in the final state).
Since no resonances are expected to decay to $\kaon^-_1\kaon^-_2$, these variables chosen are the squared invariant masses $m^2(\proton\kaon^-_1)$ and $m^2(\proton\kaon^-_2)$.
The presence of two identical kaons in the final state imposes a Bose symmetry that the decay amplitudes must be invariant under the exchange of these two particles. 
As a result, the two-dimensional distribution of $m^2(\proton\kaon^-_1)$ and $m^2(\proton\kaon^-_2)$ has a symmetry under interchange of the variables.
This motivates the use of the variables \mpklowsq\ and \mpkhighsq, which denote the lower and higher of $m^2(\proton\kaon^-_1)$ and $m^2(\proton\kaon^-_2)$, respectively, effectively removing a duplicated half of the $\left( m^2(\proton\kaon^-_1), m^2(\proton\kaon^-_2) \right)$ plane.
References hereafter to the Dalitz plot (DP) of \XibToKKp decays refer to the two-dimensional $\left( \mpklowsq, \mpkhighsq \right)$ distribution.
The DP distributions of selected candidates in Run~1 and Run~2 are shown in the top row of Fig.~\ref{fig:data_run}.

It is common practice in amplitude analysis to use the so-called ``square'' Dalitz plot (SDP) variables~\cite{Aubert:2005sk,LHCb-PAPER-2019-017}, which in this case are defined as
\begin{equation}
\label{eq:sqdp}
m^\prime 	= \frac{1}{\pi} \arccos{\left(2\frac{m(\Km \Km) - m_{\rm \text{min}}(\Km \Km)}{m_{\rm \text{max}}(\Km \Km) - m_{\rm \text{min}}(\Km \Km)}\right)}\quad \text{and} \quad 
\theta^\prime 	= \frac{1}{\pi} \theta(\Km \Km)\,.
\end{equation}
Here $m_{\rm \text{min}}(\Km \Km) = 2 m_\kaon$ and $m_{\rm \text{max}}(\Km \Km) = m_\Xib - m_\proton$ represent the kinematic limits of $m(\Km\Km)$ for \XibToKKp decay and
$\theta(\Km\Km)$ is the angle between one \Km direction in the $\Km\Km$ centre-of-mass frame and the direction of the $\Km\Km$ system in the \Xibm\ centre-of-mass frame.
The symmetry of the final state requires that distributions are symmetric with respect to $\theta^\prime = 0.5$, so only the region $\theta^\prime \in \left[ 0, 0.5 \right]$ is considered.
These SDP variables provide improved granularity, when using uniform binning, in the regions close to the DP boundaries that tend to be populated most densely.
This is beneficial for example in the modelling of the signal efficiency. 
Furthermore, the mapping to a square space aligns the bin boundaries to the kinematic boundaries of the phase space. 
As such, all efficiencies and background distributions in the analysis are obtained as functions of the SDP variables.
The SDP distributions of selected candidates in Run~1 and Run~2 are shown in the bottom row of Fig.~\ref{fig:data_run}.

\subsection{Modelling of the signal component}
\label{sec:model_signal}

The probability density function (PDF) for the signal component is expressed as
\begin{equation}
\label{eq:sig_cpv}
\mathcal{P}^Q_{\rm sig}(\Omega) = 
\frac{\epsilon^Q(\Omega)}{\Gamma} \frac{d\Gamma^Q}{d\Omega}\,,
\end{equation}
where $Q = +1$ for \Xibm\ decays and $Q = -1$ for \Xibp\ decays and $\Omega$ denotes the phase space in terms of the DP variables. 
The efficiency is denoted by $\epsilon^Q(\Omega)$, and can differ for $Q = +1$ and $-1$ to accommodate efficiency asymmetries; as described in Sec.~\ref{sec:bkg_shapes}, the efficiency maps are determined using SDP coordinates, denoted $\Omega^\prime$, but at any point in the phase space $\epsilon^Q(\Omega) = \epsilon^Q(\Omega^\prime)$.
The term $d\Gamma^Q/d\Omega$ describes the differential decay densities for \Xibm\ and \Xibp\ decays, including both local and overall rate asymmetries and the normalisation factor $\Gamma$ is 
\begin{equation}
\label{eq:norm}
\Gamma = \int_{\Omega}\left(\epsilon^{Q=+1}(\Omega)\frac{d\Gamma^{Q=+1}}{d\Omega} + \epsilon^{Q=-1}(\Omega)\frac{d\Gamma^{Q=-1}}{d\Omega}\right)d\Omega\,.
\end{equation}
Equations~\eqref{eq:sig_cpv} and~\eqref{eq:norm} assume no asymmetry in the production rates of \Xibm\ and \Xibp\ baryons produced within the LHCb acceptance from high-energy $pp$ collisions, consistent with measurement~\cite{LHCb-PAPER-2018-047}.
The effect of such a production asymmetry would, in this analysis, mimic a global (\ie\ phase-space independent) difference between $\epsilon^{Q=+1}$ and $\epsilon^{Q=-1}$, and therefore the systematic uncertainty due to this assumption can be evaluated straightforwardly.

The differential decay density is expressed as 
\begin{equation}
\label{eq:decay_dens}
\frac{d\Gamma^Q}{d\Omega}=
    \frac{1}{\left(8\pi m_\Xib\right)^3} \sum_{M_\Xib,\lambda_{\proton}}\left|
	\sum_{R}
	A^{Q}_{R,M_\Xib,\lambda_{\proton}}
(\Omega)\right|^2\,,
\end{equation}
where $A^{Q}_{R,M_\Xib,\lambda_{\proton}}$ denotes the symmetrised decay amplitude for a given intermediate state $R$, \Xibm spin component along a chosen quantisation axis $M_\Xib$, and proton helicity $\lambda_{\proton}$.
The quantisation axis is chosen to be the direction opposite to the proton momentum in the \Xibm rest frame, and the proton helicity is defined in the rest frame of the \mbox{$\Km\Km$}~system to ensure explicit symmetry between the \mbox{$\proton\klow$} and \mbox{$\proton\khigh$}~decay chains.
Here $\klow$ is the kaon whose four-momentum is used in the definition of \mpklowsq and $\khigh$ denotes the other kaon.
The amplitude in Eq.~\eqref{eq:decay_dens} has been summed incoherently over the spins of the initial and final states (corresponding to an average over initial states) and 
coherently over all contributing intermediate states.

The helicity formalism is used to parametrise the decay dynamics. 
A detailed description of this formalism can be found in Refs.~\cite{LHCb-PAPER-2016-015,LHCb-PAPER-2016-019,LHCb-PAPER-2015-029,LHCb-PAPER-2016-061,Mikhasenko:2019rjf}.
In particular, the Dalitz-plot decomposition procedure~\cite{Mikhasenko:2019rjf} is followed to express the symmetrised decay amplitude as
\begin{equation}
\label{eq:amplsym}
	A^{Q}_{R,M_\Xib,\lambda_p}(\mpklowsq,\mpkhighsq) = T^{Q}_{R,M_\Xib,\lambda_p}(\mpklowsq,\mpkhighsq)+
	(-1)^{M_\Xib+\lambda_p} T^{Q}_{R,M_\Xib,\lambda_p}(\mpkhighsq,\mpklowsq)\,.
\end{equation}
The first term corresponds to the amplitude for the weak decay $\decay{\Xibm}{R\khigh}$, where $R$ decays to \mbox{$\proton\klow$} via the strong interaction.
This decay amplitude is expressed as
\begin{align} 
\label{eq:Tsymm}
\begin{split}
	T^{Q}_{R,M_\Xib,\lambda_p}(\mpklowsq,\mpkhighsq) 
	= \sum_{\lambda_R,\lambda_p'} 
	\Big(
	d_{M_\Xib,\lambda_R}^{J_\Xib}(\theta_{R}) &
	d_{\lambda_R,\lambda_p'}^{J_R}(\theta_{p})
	d_{\lambda_p',\lambda_p}^{J_p}(\zeta) \\
	& \times \eta_{\lambda_p'} 
	(-1)^{\lambda_p'-\lambda_p} 
	h^{Q}_{R,\lambda_R}
	R(\mpklowsq)
	\Big)\,,
\end{split}
\end{align}
where the amplitude is summed coherently over the allowed helicities of the intermediate state, $\lambda_R$, and of the proton, $\lambda_p'$, defined in the rest frames of the \Xibm and the \mbox{$\proton\klow$}~systems, respectively.
The \Xibm, $R$ and proton spins are denoted by $J_\Xib$, $J_R$ and $J_\proton$, respectively. 

The three functions of the form $d^{J}_{\lambda,\lambda'}$ in Eq.~\eqref{eq:Tsymm} are the small Wigner d-matrix elements~\cite{Wigner:102713} that impose angular momentum conservation giving rise to the condition $|\lambda_{R}| \leq 1/2$. 
As a result, for intermediate states with any half-integer spin, only helicities corresponding to $\lambda_{R} = \pm 1/2$ contribute to the amplitude.
The three angles $\theta_R$, $\theta_p$ and $\zeta$ are functions of the DP variables. 
The angle $\theta_R$, defined in the \Xibm rest frame, is formed between the direction opposite to the proton momentum and the combined momentum of the \mbox{$\proton\klow$}~system. 
The angle $\theta_p$ is between the direction opposite to the $\khigh$ momentum in the \Xibm rest frame and the proton momentum in the rest frame of the \mbox{$\proton\klow$}~system.
The angle $\zeta$ gives the Wigner rotation that is required to relate the proton helicity state, $\ket{\lambda_p^\prime}$, defined in the \mbox{$\proton\klow$}~rest frame to the proton helicity state, $\ket{\lambda_p}$, defined in the \mbox{$\Km\Km$}~rest frame. 
This angle, computed in the proton rest frame, is formed between the momenta of $\klow$ and of the \mbox{$\Km\Km$}~system.
Mathematical definitions of these three angles, each defined in the range $[0,\pi]$, are
\begin{align}
	\cos \theta_R &= \frac{(m_\Xib^2+m_\kaon^2-\mpklowsq)(m_\Xib^2+m_\proton^2-\mkksq) - 2m_\Xib^2(\mpkhighsq-m_\kaon^2-m_\proton^2)}{\sqrt{{\cal K}(m_\Xib^2,m_\proton^2,\mkksq)} \sqrt{{\cal K}(m_\Xib^2,\mpklowsq,m_\kaon^2)} },\label{ang:1}\\
	\cos \theta_p &= \frac{2\mpklowsq(\mpkhighsq-m_\kaon^2-m_\proton^2)-(\mpklowsq+m_\proton^2-m_\kaon^2)(m_\Xib^2-\mpklowsq-m_\kaon^2)}{\sqrt{{\cal K}(m_\Xib^2,m_\kaon^2,\mpklowsq)} \sqrt{{\cal K}(\mpklowsq,m_\proton^2,m_\kaon^2)} }\label{ang:2}\,,\\
	\cos \zeta &= \frac{2m_\proton^2(\mpkhighsq-m_\Xib^2-m_\kaon^2)+(m_\Xib^2+m_\proton^2-\mkksq)(\mpklowsq-m_\proton^2-m_\kaon^2)}{\sqrt{{\cal K}(m_\Xib^2,m_\proton^2,\mkksq)} \sqrt{{\cal K}(\mpklowsq,m_\proton^2,m_\kaon^2)} }\label{ang:3}\,,
\end{align}
where $\mkksq = m_\Xib^2+2m_\kaon^2+m_\proton^2-\mpklowsq-\mpkhighsq$, the K\"{a}ll\'en function is given by ${\cal K}(a,b,c) = a^2 + b^2 + c^2 - 2 (ab + ac + bc)$ and
the \Xibm, \kaon and \proton masses are denoted by $m_\Xib$, $m_\kaon$ and $m_\proton$, respectively. 

The second term in Eq.~\eqref{eq:amplsym} corresponds to the weak decay of $\decay{\Xibm}{R\,\klow}$ where $R$ now decays to $\decay{R}{\proton\khigh}$.
The expression for this amplitude can be obtained by interchanging $\mpklowsq\leftrightarrow\mpkhighsq$ in Eqs.~\eqref{eq:Tsymm}--\eqref{ang:3}.

The term $\eta_{\lambda_p'}$, in Eq.~\eqref{eq:Tsymm}, arises as a consequence of parity conservation in the strong decay of the intermediate state $R$.
It is defined as
\begin{gather*}
\eta_{\lambda_p'}\,=\,
\begin{cases}
1 \text{ when } \lambda_p' = 1/2 \,,\\
(-1)^{\frac{3}{2}-J_R}\eta_{R} \text{ when } \lambda_p' = -1/2\,,
\end{cases}
\end{gather*}
where $\eta_R$ is the intrinsic parity of $R$.

The complex coefficient $h^{Q}_{R,\lambda_{R}}$, in Eq.~\eqref{eq:Tsymm}, encapsulates the combined couplings of the weak decay of the initial state and the strong decay of the intermediate state.
This coefficient, subsequently referred to as the helicity coupling, can be expressed as
\begin{equation}
\label{helcity_cpv}
	h^{Q}_{R,\lambda_{R}}  = (x_{R,\lambda_{R}} + Q\,\delta x_{R,\lambda_{R}}) + i\,(y_{R,\lambda_{R}} + Q\,\delta y_{R,\lambda_{R}})\,,
\end{equation}
where $x_{R,\lambda_R}$ and $y_{R,\lambda_R}$ denote the real and imaginary components of the \CP-conserving part of the coupling, while $\delta x_{R,\lambda_R}$ and $\delta y_{R,\lambda_R}$ are \CP-violating parameters.

The term $R(\mpklowsq)$, in Eq.~\eqref{eq:Tsymm}, describes the lineshape of each resonant or nonresonant contribution. 
Resonances are parametrised with relativistic Breit--Wigner (RBW) functions, $F_{\rm RBW}$, that are modified by Blatt--Weisskopf barrier factors, $B_{L_{\Xib}}$ and $B_{L_{R}}$, and are given by
\begin{equation}
R(m^2_x) =
B_{L_{\Xib}}(p|p_0,d) \left(\frac{p}{m_{\Xib}}\right)^{L_{\Xib}} \;
F_{\rm RBW}(m^2_x|m_{0},\Gamma_{0}) \;
B_{L_{R}}(q|q_0,d) \left(\frac{q}{m_{0}}\right)^{L_{R}}.
\label{eq:resshape}
\end{equation}
Here $m^2_x$ is either $\mpklowsq$ or $\mpkhighsq$, while $p$ is the magnitude of the resonance momentum in the $\Xibm$ centre-of-mass frame and $q$ is the magnitude of the proton momentum in the resonance centre-of-mass frame.
The symbols $p_0$ and $q_0$ denote the values of these quantities at the resonance peak, \ie\ when $m_x=m_0$. 
The orbital angular momentum released in the \Xibm\ decay is denoted $L_{\Xib}$ while that in the resonance decay is denoted $L_{R}$.
Angular momentum conservation in the \Xibm\ decay imposes the condition $J_{R}-1/2 \le L_{\Xib}\le J_{R}+1/2$.
The minimal value $L_{\Xib} = J_{R} - 1/2$ is assumed when calculating $R(m^2_x)$.
Angular momentum conservation in the resonance decay limits $L_{R}$ to $J_{R}\pm\frac{1}{2}$, which is then uniquely defined by parity conservation in the decay, $\eta_{R}=(-1)^{L_{R}+1}$.

The Blatt--Weisskopf barrier functions are
\begin{eqnarray}
\label{eq:blattw}
B_{0}(k|k_0,d) &=&1 \,,\\
B_{1}(k|k_0,d) &=& \sqrt{ \frac{1+(k_0\,d)^2}{1+(k\phantom{_0}\,d)^2} } \,,\\
B_{2}(k|k_0,d) &=& \sqrt{ \frac{9+3(k_0\,d)^2+(k_0\,d)^4}{9+3(k\phantom{_0}\,d)^2+(k\phantom{_0}\,d)^4} } \,,\\
B_{3}(k|k_0,d) &=& \sqrt{ \frac{225+45(k_0\,d)^2+6(k_0\,d)^4+(k_0\,d)^6}{225+45(k\phantom{_0}\,d)^2+6(k\phantom{_0}\,d)^4+(k\phantom{_0}\,d)^6} } \,,
\end{eqnarray}
and account for suppression creating high values of the orbital angular momentum $L$, which depends on the momentum of one of the decay products, $k$, in the centre-of-mass frame of the decaying particle and on the size of the decaying particle given by the constant $d$. 
The value $d=5.0\gev^{-1}$ is used for $\Xibm$ decays while $1.5\gev^{-1}$ is used for resonances~\cite{LHCb-PAPER-2016-061}. 

The relativistic Breit--Wigner amplitude is given by
\begin{equation}   
  F_{\rm RBW}(m^2_x | m_0, \Gamma_0 ) = \frac{1}{m_0^2-m^2_x - i m_0 \Gamma(m_x)} \,,
\label{eq:breitwigner}
\end{equation}
where
\begin{equation}   
\Gamma(m_x)=\Gamma_0 \left(\frac{q}{q_0}\right)^{2\,L_{R}+1} \; \frac{m_0}{m_x} \; B^\prime_{L_{R}}(q,q_0,d)^2 \,.
\label{eq:mwidth}
\end{equation}
Here $m_0$ and $\Gamma_0$ denote the pole mass and width of the resonance, respectively.
In the case of the $\Lz(1405)$ resonance, which peaks below the $\proton\Km$ threshold, $m_0$ is replaced by an effective mass in the kinematically allowed region~\cite{LHCb-PAPER-2014-036}, 
\begin{equation}
m^{\rm eff}_0 = m^{\rm min} + (m^{\rm max} - m^{\rm min}) \left(1 + \tanh{\left(\frac{m_0 - \frac{m^{\rm max} + m^{\rm min}}{2}}{m^{\rm max} - m^{\rm min}}\right)}\right) \,,
\end{equation}
where $m^{\rm max}$ and $m^{\rm min}$  are the upper and lower limits of the kinematically allowed range, respectively.
In this case, the $q_0$ value in Eq.~\eqref{eq:mwidth} is the value of $q$ at $m = m^{\rm eff}_0$. 
This parameterisation ensures that only the tail of the RBW function enters the fit model as a virtual contribution.
Nonresonant components are modelled using an exponential lineshape,
\begin{equation}
  R_{\rm NR}(m^2_x) =
  \left(\frac{p}{m_{\Xib}}\right)^{L_{\Xib}^{R}} 
  \left(\frac{q}{m_{0}}\right)^{L_{R}}
  \exp(-\alpha m^2_x)\,, 
  \label{eq:NR}
\end{equation}
where $\alpha$ is a slope parameter that is determined from the fit and $m_0$ is fixed to be the midpoint of the \mpklow\ range, \ie\ $2.83 \gev$.

The primary outputs of the amplitude analysis are the \CP-conserving and \CP-violating components of the helicity couplings introduced in Eq.~\eqref{helcity_cpv}.
However, since these depend on the choice of phase convention, amplitude formalism and normalisation, they can be difficult to compare between analyses. 
It is therefore more useful to report the fit fractions $\mathcal{F}_{\rm i}$ for each intermediate component $i$ of the fit model, defined by
\begin{equation}
\label{eq:ff}
\mathcal{F}_{\rm i} = 
\frac{\int_{\Omega}(d\Gamma^+_{i}/d\Omega + d\Gamma^-_{i}/d\Omega) d\Omega}{\int_{\Omega}(d\Gamma^+/d\Omega + d\Gamma^-/d\Omega) d\Omega}\,,
\end{equation}
where 
\begin{equation}
\frac{d\Gamma^Q_i}{d\Omega} = 
\frac{1}{\left(8\pi m_\Xib\right)^3}
\sum_{M_\Xib,\lambda_{\proton}}
\left|
A^{Q}_{i,M_\Xib,\lambda_{\proton}}(\Omega)
\right|^2\,.
\end{equation}

It is also useful to report the interference fit fractions $\mathcal{I}_{ij}$ between the two intermediate components $i$ and $j$, defined by
\begin{equation}
\label{eq:iff}
\mathcal{I}_{ij} = 
\frac{ 
\int_{\Omega}(d\Gamma^+_{ij}/d\Omega + d\Gamma^-_{ij}/d\Omega) d\Omega
}
{\int_{\Omega}(d\Gamma^+/d\Omega + d\Gamma^-/d\Omega) d\Omega}\,,
\end{equation}
where 
\begin{equation}
\frac{d\Gamma^Q_{ij}}{d\Omega} = 
\frac{1}{\left(8\pi m_\Xib\right)^3}
\sum_{M_\Xib,\lambda_{\proton},i,j}
2\,{\rm Re}
\left(
A^{Q}_{i,M_\Xib,\lambda_{\proton}}
(A^{Q}_{j,M_\Xib,\lambda_{\proton}})^{*}
\right)
\,.
\end{equation}
The parameters of \CP\ violation, $A^{\CP}_{i}$, associated with each component $i$ of the model are also reported.
These are defined as
\begin{equation}
\label{eq:acpcomb}
	A^{\CP}_{i} = \frac{\int_{\Omega}(d\Gamma^+_{i}/d\Omega - d\Gamma^-_{i}/d\Omega) d\Omega}{\int_{\Omega}(d\Gamma^+_{i}/d\Omega + d\Gamma^-_{i}/d\Omega) d\Omega}\,.
\end{equation}

\subsection{Modelling of signal efficiency and background distributions}
\label{sec:bkg_shapes}

\begin{figure}[!tb]
\centering
\includegraphics[width=0.48\textwidth]{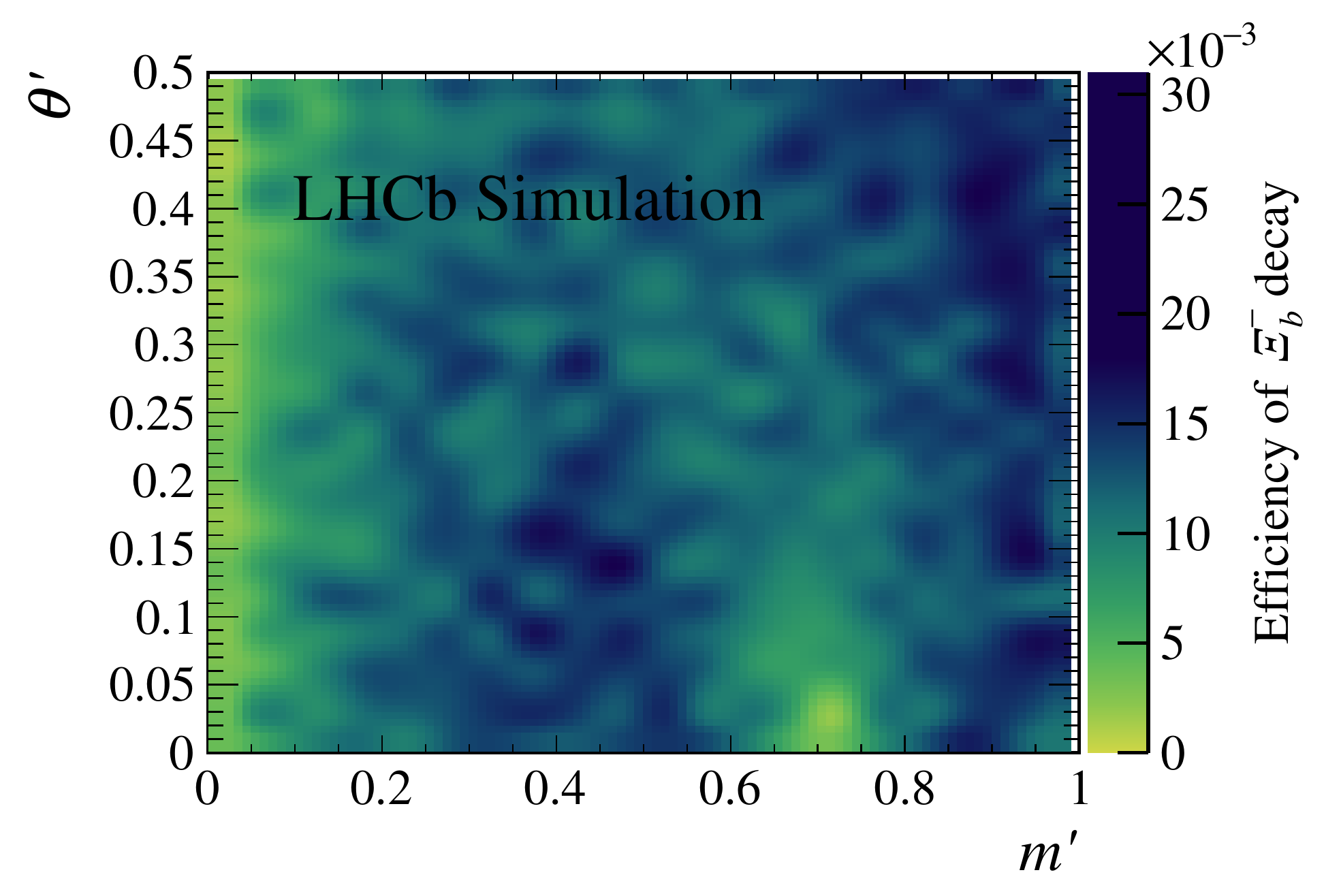}
\includegraphics[width=0.48\textwidth]{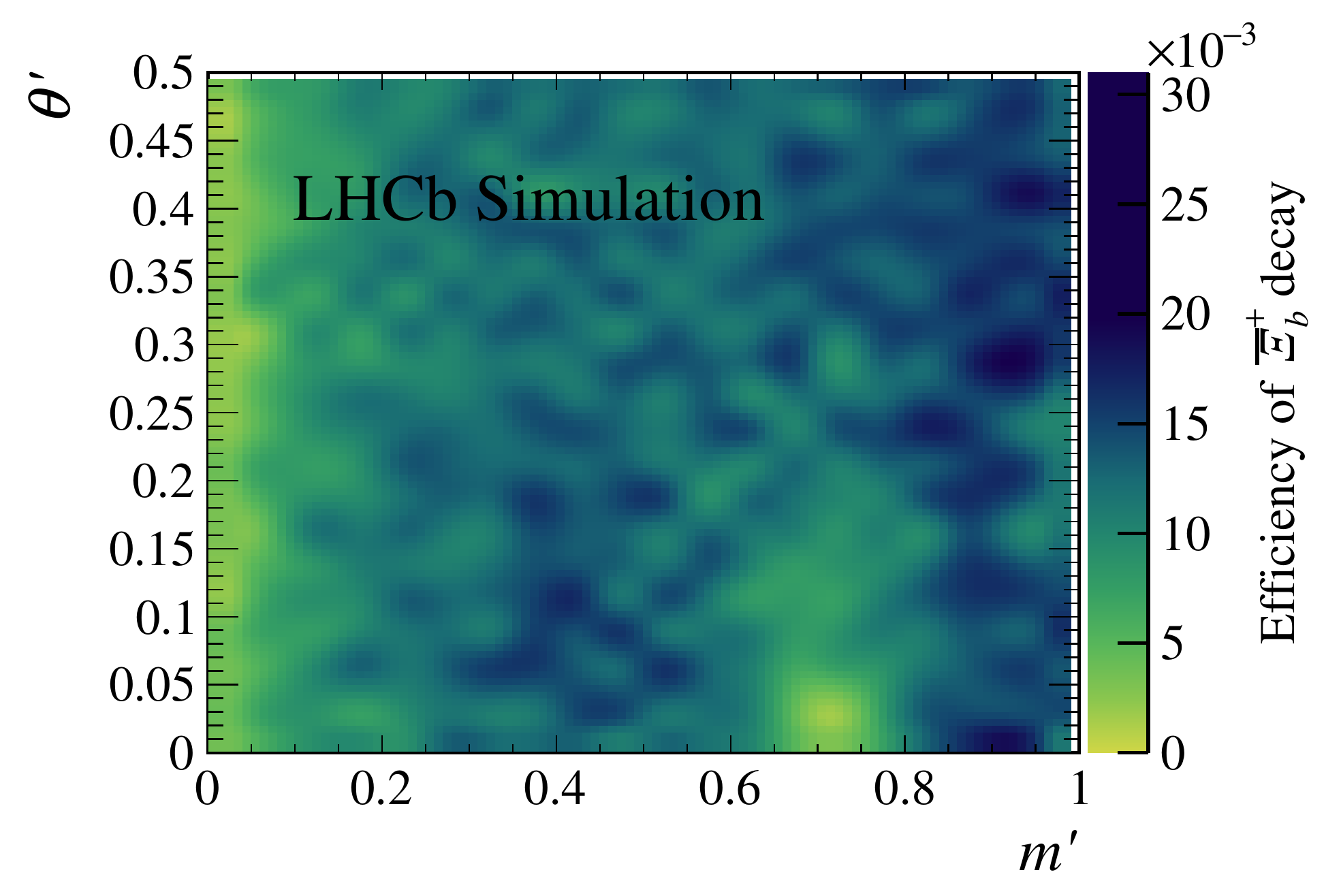} \\
\includegraphics[width=0.48\textwidth]{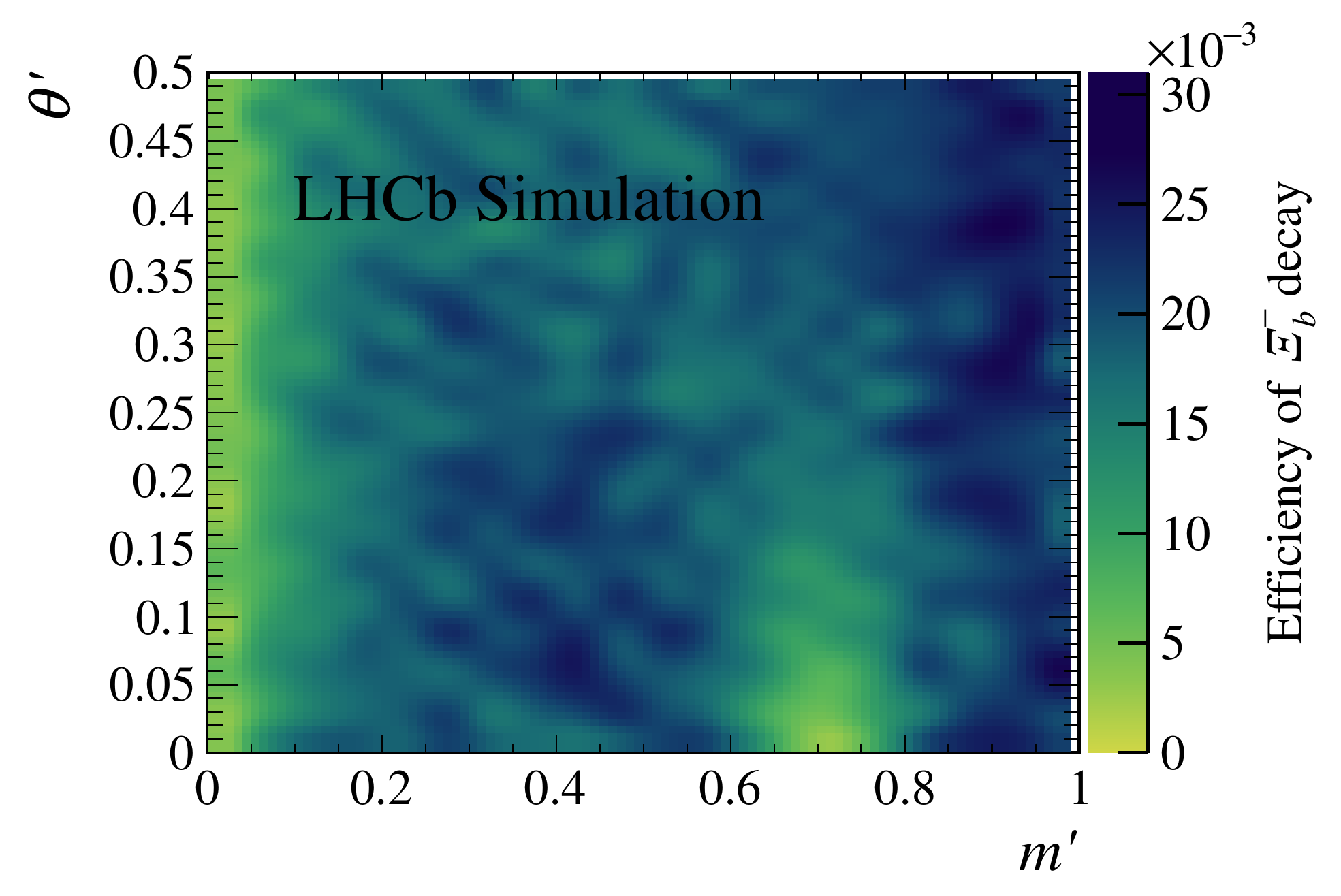} 
\includegraphics[width=0.48\textwidth]{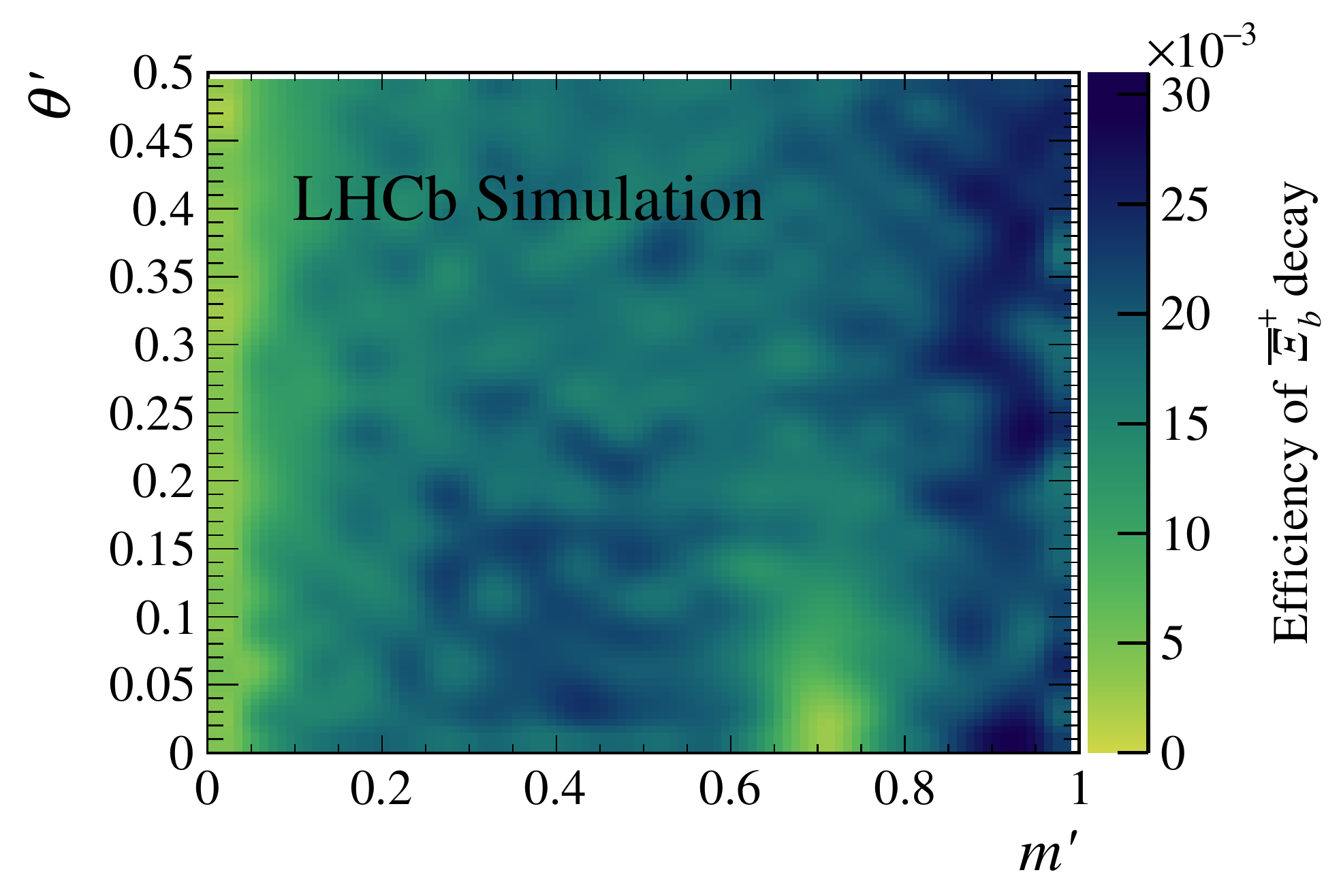} 
\caption{\small
Efficiency as a function of square Dalitz-plot position for (left)~\Xibm and (right)~\Xibp decays, for (top)~Run~1 and (bottom)~Run~2.
}
\label{fig:effmaps_splitcharge}
\end{figure}

The detector geometry and the online and offline selection procedure can induce variation in the signal efficiency across the phase space of the decay.
This is accounted for, as shown in Eq.~\eqref{eq:sig_cpv}, by determining the efficiency as a function of the SDP variables. 
The efficiency maps are obtained from simulation, but with effects related to PID calibrated using data as outlined in Sec.~\ref{sec:offselection}.
The efficiency maps for \Xibm\ and \Xibp\ decays can be seen in Fig.~\ref{fig:effmaps_splitcharge} separately for Run~1 and Run~2. 
These maps are obtained by employing a uniform $10\times10$ binning scheme and smoothing with a two-dimensional cubic spline to mitigate effects of discontinuity at the bin edges.
No significant detection asymmetry is observed.

\begin{figure}[!tb]
\centering
\includegraphics[width=0.48\textwidth]{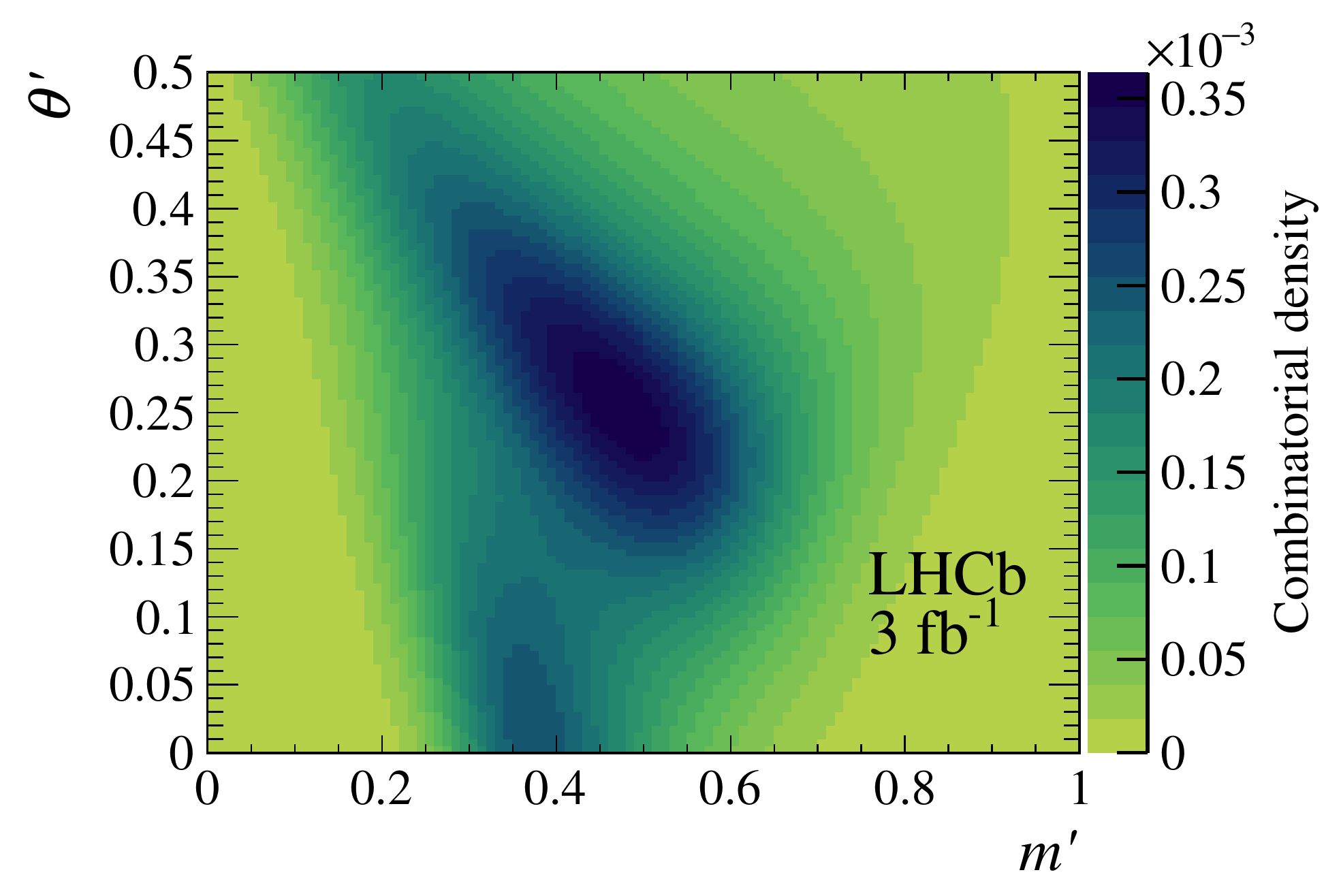} 
\includegraphics[width=0.48\textwidth]{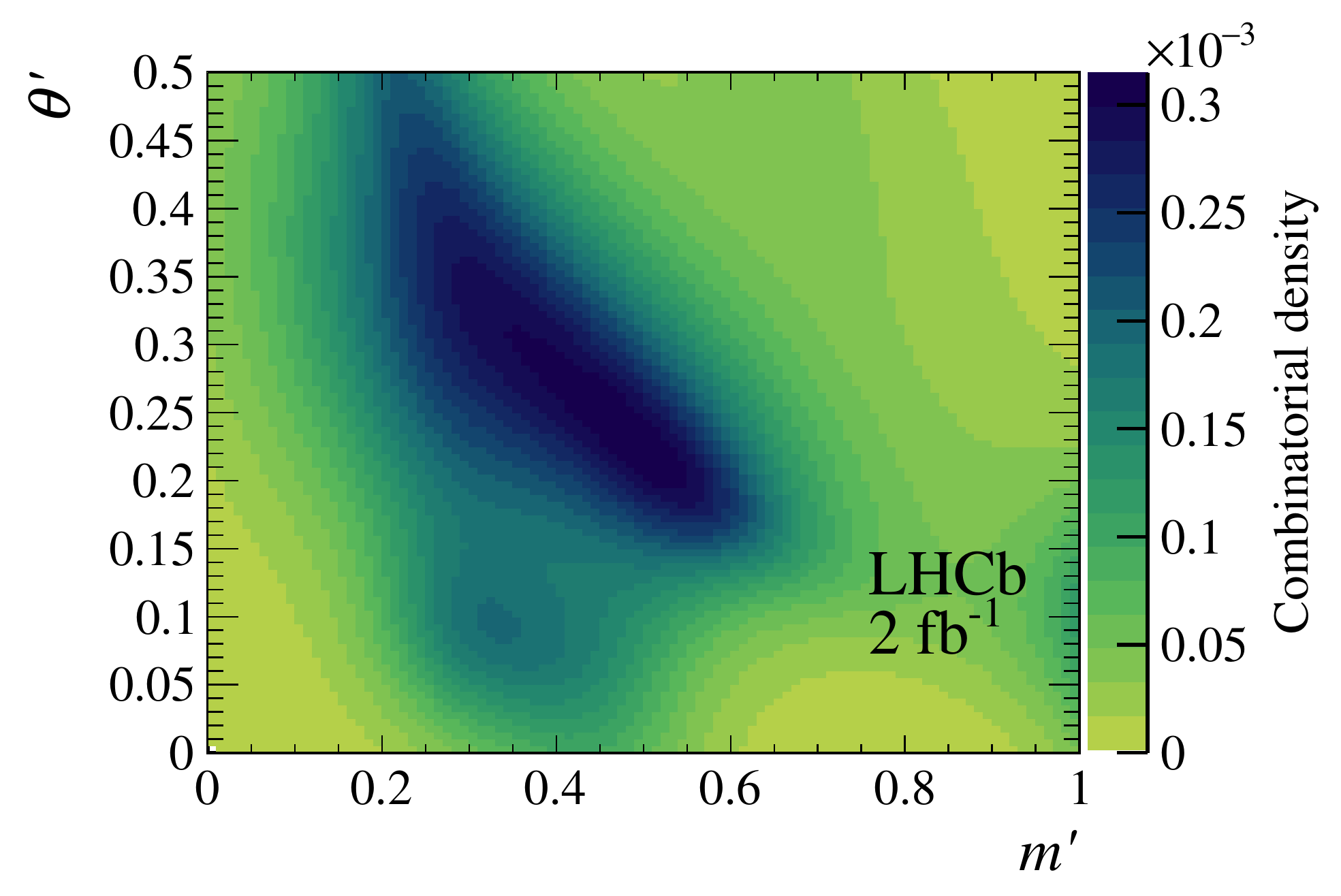} \\
\includegraphics[width=0.48\textwidth]{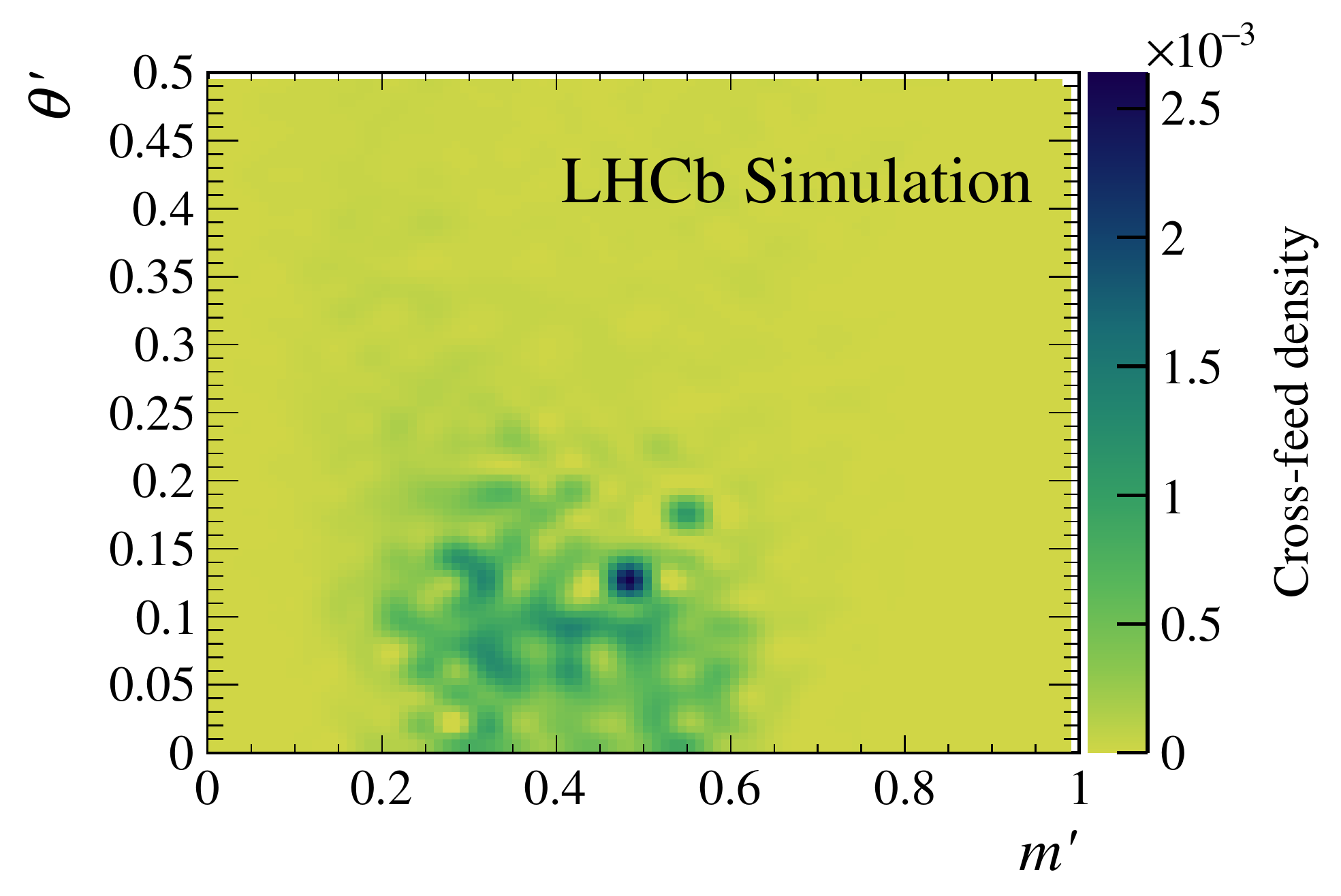}
\includegraphics[width=0.48\textwidth]{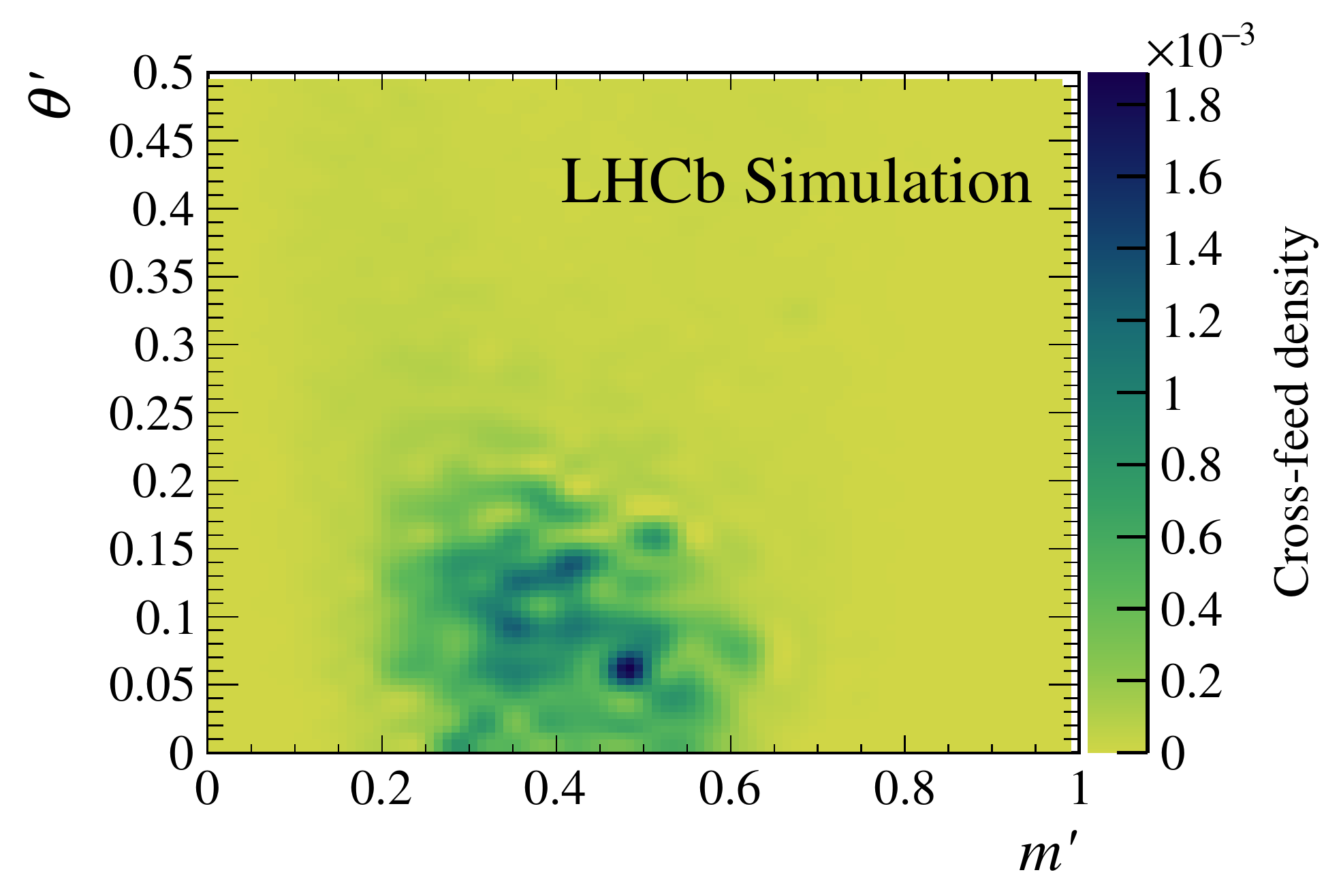}
\caption{\small
  SDP distributions of (top)~combinatorial and (bottom)~cross-feed background components for (left)~Run~1 and (right)~Run~2.
}
\label{fig:bkg_shapes}
\end{figure}

Candidates selected in data in the sideband $5890 < m(\KKp) < 6470 \mev$ are used to model the SDP distribution of the combinatorial background, which dominates this region as discussed in Sec.~\ref{sec:invmassfit}.
The effect of the \Xibm mass constraint used when calculating the SDP variables causes a distortion of the distribution from that of combinatorial background in the signal region. 
This is accounted for using a method~\cite{Mathad:2019rqj} in which the unnormalised function describing the $m(\KKp)$ and $\Omega^\prime = (m^\prime,\theta^\prime)$ space is expressed as
\begin{equation}
\label{eq:comb_extraNN}
	F(m(\KKp), \Omega^\prime) = \big|f_0(\Omega^\prime) + \exp\left(-\beta\,m(\KKp)\right) \times f_1(\Omega^\prime)\big|^2
\end{equation}
where $\beta$ is a free parameter determined from a fit to the sideband candidates. 
The functions $f_0$ and $f_1$ are modelled using neural networks that are trained using candidates from the data sideband region.
This model is then extrapolated to predict the PDF of the combinatorial background at the \Xibm mass, 
\ie $\mathcal{P}_{\rm comb}(\Omega^\prime) = F(m_{\Xib}, \Omega^\prime)/N$, where the normalisation factor $N=\int_{\Omega^\prime} F(m_{\Xib}, \Omega^\prime) d\Omega^\prime$.
The PDF in terms of DP variables is obtained using $\mathcal{P}_{\rm comb}(\Omega) = |J| \mathcal{P}_{\rm comb}(\Omega^\prime)$,
where $|J|$ is the Jacobian determinant of the transformation between variables, $d\Omega = |J| d\Omega^\prime$. 
The SDP distributions of the combinatorial background for Run~1 and Run~2 are shown in the top row of Fig.~\ref{fig:bkg_shapes}.

After imposing the selection criteria and splitting the data sample by the initial state charge, too few candidates are available in the sideband region to train the neural networks. 
As a result, the neural networks are trained using the combined sample of \Xibm\ and \Xibp\ candidates, and no asymmetry in the shape of the combinatorial background SDP distribution is assumed in the baseline model.

The SDP distribution of cross-feed background from misidentified \XibToKpip\ decays that enter the signal region is modelled using simulation, and is shown in the bottom row of Fig.~\ref{fig:bkg_shapes} separately for Run~1 and Run~2. 
These distributions are described in terms of a uniform $10\times10$ binned SDP histogram, smoothed with a two-dimensional cubic spline.
As described in Sec.~\ref{sec:invmassfit}, the simulation is weighted to reproduce resonance structures expected in the phase space of \XibToKpip\ decays.
Differences in selection requirements, together with the limited statistics of the \XibToKpip simulation samples, cause the PDFs to differ between Run~1 and Run~2.
In the baseline fit it is assumed that there is no asymmetry between \Xibm\ and \Xibp\ candidates in the cross-feed yields or SDP distributions.

\subsection{Fitting procedure}
\label{sec:fit_procedure}

The total PDF that is used to model the phase-space distributions of \XibToKKp and its conjugate decay is
\beq
\begin{aligned}
\label{eq:totpdf_cpv}
\mathcal{P}^{Q}_{\rm tot}(\Omega) = \frac{1}{N_{\rm tot}} 
\Big[ 
N_{\rm sig} \mathcal{P}^{Q}_{\rm sig}(\Omega) 
& + N_{\rm comb}\frac{(1 - Q A_{\rm comb})}{2} \mathcal{P}_{\rm comb}(\Omega) 
& + \frac{N_{\rm cf}}{2} \mathcal{P}_{\rm cf}(\Omega)
\Big]\,.
\end{aligned}
\eeq
The yields of signal, combinatorial background and cross-feed background components are denoted by $N_{\rm sig}$, $N_{\rm comb}$ and $N_{\rm cf}$, respectively, and are obtained as described in Sec.~\ref{sec:invmassfit}, separately for Run~1 and Run~2.
The quantity $N_{\rm tot} = N_{\rm sig} + N_{\rm comb} + N_{\rm cf}$ is the total yield in the signal region.
The PDFs for the signal, combinatorial background and cross-feed background components are denoted by $\mathcal{P}^{Q}_{\rm sig}(\Omega)$, $\mathcal{P}_{\rm comb}(\Omega)$ and $\mathcal{P}_{\rm cf}(\Omega)$, respectively, where the former is given in Eq.~\eqref{eq:sig_cpv} and the latter two are displayed in Fig.~\ref{fig:bkg_shapes} in terms of the SDP variables. 
In the baseline model, only the signal PDF can differ for \Xibm\ and \Xibp\ candidates, although a possible global combinatorial background asymmetry, $A_{\rm comb}$, is a free parameter of the model. 

An unbinned maximum-likelihood fit is performed to the combined sample of candidates for \XibToKKp\ and its conjugate decay to determine the parameters of the model, which are the \CP-conserving and \CP-violating coefficients of the helicity couplings of Eq.~\eqref{helcity_cpv}.
The fit is performed simultaneously to the Run~1 and Run~2 data samples, which have separate efficiency and background models as described above.
The fit model is implemented in a fitting package based on \textsc{TensorFlow}~\cite{tensorflow2015-whitepaper}, interfaced with the \textsc{Minuit} function minimisation algorithm~\cite{James:1975dr,Brun:1997pa}.
The function that is minimised, twice the negative log-likelihood, is
\begin{equation}
-2 \ln \mathcal{L} = 
-2 \sum_{r} \left( \sum_i^{N_r}  \ln\left(\mathcal{P}^{Q}_{\rm tot}(\Omega_{i})\right) \right)\,,
\label{eq:totpdf_lh}
\end{equation}
where the index $i$ runs over the $N_r$ candidates in the data sample from run period $r$ (Run~1 or Run~2), and $\Omega_i$ denotes the DP coordinates of candidate $i$.

\subsection{Model selection}
\label{sec:model_build}

The \XibToKKp\ decay can proceed via intermediate $\proton\Km$ resonances.
Various $\Lz^*$ and $\Sigmares^*$ resonances that are known to decay to $\proton\Km$ are considered as potential components of the signal model.\footnote{
    All $\Sigmares^*$ resonances considered are neutral; the conventional charge superscripts are omitted for brevity of notation.}
The Particle Data Group (PDG)~\cite{PDG2020} reports a large number of such states; those that are sufficiently well established 
are considered in this study and are shown in Table~\ref{tab:lambda-sigma}.
Masses and widths of all resonance components are fixed to either the central value or the midpoint in the range of values quoted in Table~\ref{tab:lambda-sigma}. 
Nonresonant components, labelled ${\rm NR}(J^P)$, with spin-parity $J^P=\frac{1}{2}^+,\frac{1}{2}^-,\frac{3}{2}^+,\frac{3}{2}^-$ are also considered.
Resonances with spin $J \geq 7/2$ are excluded from consideration as these would require $L_{\Xib} \ge 3$ and are therefore expected to be significantly suppressed.

\begin{table}[!tb]
  \centering
  \renewcommand{\arraystretch}{1.2}
  \caption{\small
    Summary of the considered $\Lz^*$ and $\Sigmares^*$ resonances, ranked either **** or *** by the PDG~\cite{PDG2020}.
    Note that the $\proton \Km$ threshold is at $1432 \mev$.
    Resonances marked $\dagger$ are included in the baseline model, as described in the text.
    The spin-parity of the $\Sigmares(2250)$ is not known and is assumed to be $\frac{3}{2}^+$.
    For many of these states, the PDG does not report masses and widths with central values and uncertainties, but rather gives real and imaginary parts of the pole position.
    This reflects the fact that a simple Breit--Wigner parameterisation of these resonances may not fully describe their lineshapes; however, more sophisticated parametrisations are beyond the scope of the current analysis.
  }
  \label{tab:lambda-sigma}
  \begin{tabular}{cccccc}
    \hline
    & Name & $J^P$ & Mass $(\mev)$ & Width $(\mev)$ & Main decay channels \\
    \hline
    & \multicolumn{5}{c}{****} \\
    \hline
    $\dagger$ & $\Lz(1405)$ & $\frac{1}{2}^-$ & $1405.1 \, ^{+1.3}_{-1.0}$ & $50.5 \pm 2.0$ & $\Sigmares \pi$ \\
    $\dagger$ & $\Lz(1520)$ & $\frac{3}{2}^-$ & 1518 to 1520 & 15 to 17 & $N\Kbar$, $\Sigmares \pi$ \\
    $\dagger$ & $\Lz(1670)$ & $\frac{1}{2}^-$ & 1660 to 1680 & 25 to 50 & $N\Kbar$, $\Sigmares \pi$, $\Lz\eta$ \\
    & $\Lz(1690)$ & $\frac{3}{2}^-$ & 1685 to 1695 & 50 to 70 & $N\Kbar$, $\Sigmares \pi$, $\Lz\pi\pi$, $\Sigmares \pi\pi$ \\
    & $\Lz(1820)$ & $\frac{5}{2}^+$ & 1815 to 1825 & 70 to 90 & $N\Kbar$ \\
    & $\Lz(1830)$ & $\frac{5}{2}^-$ & 1810 to 1830 & \phantom{1}60 to 110 & $\Sigmares \pi$ \\ 
    & $\Lz(1890)$ & $\frac{3}{2}^+$ & 1850 to 1910 & \phantom{1}60 to 200 & $N\Kbar$ \\
    \hline
    $\dagger$ & $\Sigmares(1385)$ & $\frac{3}{2}^+$ & $1383.7 \pm 1$ & $36 \pm 5$ & $\Lz \pi,\Sigmares \pi$ \\ 
    & $\Sigmares(1670)$ & $\frac{3}{2}^-$ & 1665 to 1685 & 40 to 80 & $\Sigmares \pi$ \\ 
    $\dagger$ & $\Sigmares(1775)$ & $\frac{5}{2}^-$ & 1770 to 1780 & 105 to 135 & $N\Kbar$, $\Lz^{(*)}\pi$ \\
    $\dagger$ & $\Sigmares(1915)$ & $\frac{5}{2}^+$ & 1900 to 1935 & \phantom{1}80 to 160 & not clear \\
    \hline
    & \multicolumn{5}{c}{***} \\
    \hline
    & $\Lz(1600)$ & $\frac{1}{2}^+$ & 1560 to 1700 & \phantom{1}50 to 250 & $N\Kbar$, $\Sigmares \pi$ \\
    & $\Lz(1800)$ & $\frac{1}{2}^-$ & 1720 to 1850 & 200 to 400 & $N\Kbar^{(*)}$, $\Sigmares \pi$, $\Lz\eta$ \\
    & $\Lz(1810)$ & $\frac{1}{2}^+$ & 1750 to 1850 & \phantom{1}50 to 250 & $N\Kbar^{(*)}$, $\Sigmares \pi$, $\Lz\eta$, $\Xi\kaon$ \\
    & $\Lz(2110)$ & $\frac{5}{2}^+$ & 2090 to 2140 & 150 to 250 & $N\Kbar^{(*)}$, $\Sigmares \pi$, $\Lz\Omega$\\
    \hline
    & $\Sigmares(1660)$ & $\frac{1}{2}^-$ & 1630 to 1690 & \phantom{1}40 to 200 & $N\Kbar$, $\Sigmares \pi$, $\Lz\pi$ \\
    & $\Sigmares(1750)$ & $\frac{1}{2}^-$ & 1730 to 1800 & \phantom{1}60 to 160 & $N\Kbar$, $\Sigmares \pi$, $\Lz\pi$, $\Sigmares \eta$ \\
    & $\Sigmares(1940)$ & $\frac{3}{2}^-$ & 1900 to 1950 & 150 to 300 & $N\Kbar$, $\Sigmares \pi$, $\Lz\pi$\\
    & $\Sigmares(2250)$ & $?^?$ & 2210 to 2280 & \phantom{1}60 to 150 & $N\Kbar$, $\Sigmares \pi$, $\Lz\pi$\\
    \hline
  \end{tabular}
\end{table}

In this analysis, as the normalisation of the decay density is arbitrary, the $\Lz(1520)$ resonance is chosen as the reference component.
This implies that the coupling with the positive helicity of the $\Lz(1520)$ resonance $a^{Q}_{R,\lambda_{R}=+1/2}$ is real. 
Explicitly, for the reference $\Lz(1520)$ resonance, $y_{R,\lambda_{R}=+1/2} = \delta y_{R,\lambda_{R}=+1/2} = 0$ and $x_{R,\lambda_{R}=+1/2} = 1$, while $\delta x_{R,\lambda_{R}=+1/2}$ is free to vary in the fit to allow for \CP\ violation in the $\Lz(1520)$ amplitude.
The analysis is found to be insensitive to the coupling of the $\Lz(1520)$ component with negative helicity, and therefore $x_{R,\lambda_{R}=-1/2} = \delta x_{R,\lambda_{R}=-1/2} = y_{R,\lambda_{R}=-1/2} = \delta y_{R,\lambda_{R}=-1/2} = 0$ for the reference resonance.  
The helicity couplings of all other resonant and nonresonant components are left free to vary in the fit.

To establish a baseline fit model, the $\Lz(1520)$ component alone is initially included in the model, with additional components added iteratively in the order that maximises the change in $-2\ln \mathcal{L}$ obtained from fits to the data with prospective models.
Components with different spin and parity should have zero interference fit fractions due to the orthogonality relation satisfied by the small Wigner d-matrix elements.
However, the symmetrisation of the Dalitz plot can lead to non-zero values for such interference fit fractions in this analysis.
As a result, in establishing the baseline model, it is possible to encounter ``unphysical'' interference fit fractions ($>40\%$) between two components.
When such a case occurs, the component that gives the minimal change in $-2\ln \mathcal{L}$ when removed from the fit model is discarded.
The procedure is terminated when the change in $-2\ln \mathcal{L}$ from including any further contribution is less than 9 units, limiting the potential for the model to be influenced by statistical fluctuations.
This approach leads to a model that contains $\Sigmares(1385)$, $\Lz(1405)$, $\Lz(1520)$, $\Lz(1670)$, $\Sigmares(1775)$ and $\Sigmares(1915)$ components.
The potential for additional components to be present in the true underlying model is considered as a source of systematic uncertainty.

\subsection{Fit to data}
\label{sec:amplitude_fit}

In an attempt to find the global minimum, a large number of fits to data are performed, where the initial values of the helicity couplings are randomised.
The baseline results are obtained from the fit that returns the smallest $-2\ln \mathcal{L}$ value out of this ensemble.
This procedure is found to converge successfully to the global minimum without any secondary minima.

\begin{figure}[!tb]
\begin{center}
\includegraphics[width=0.49\textwidth]{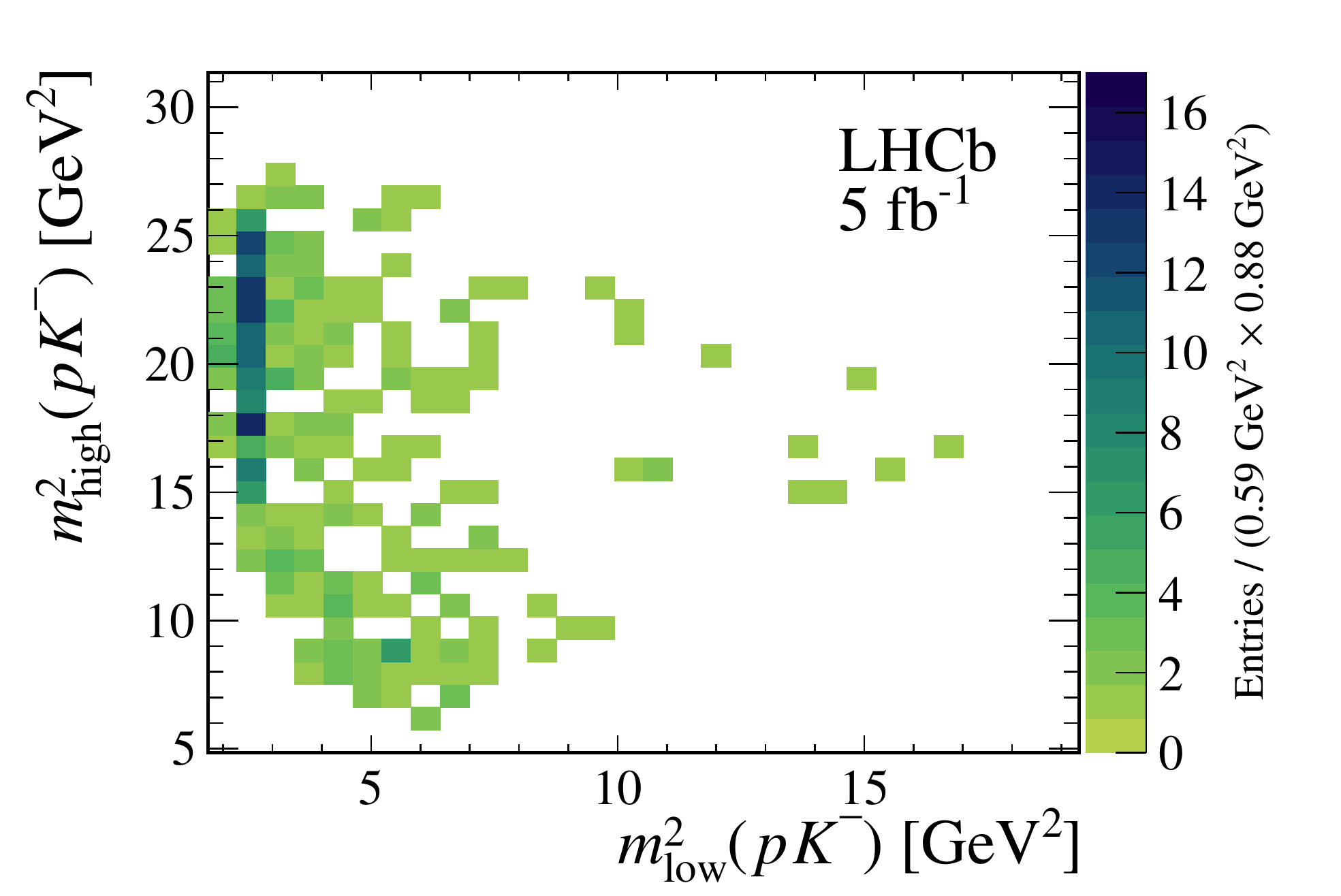}
\includegraphics[width=0.49\textwidth]{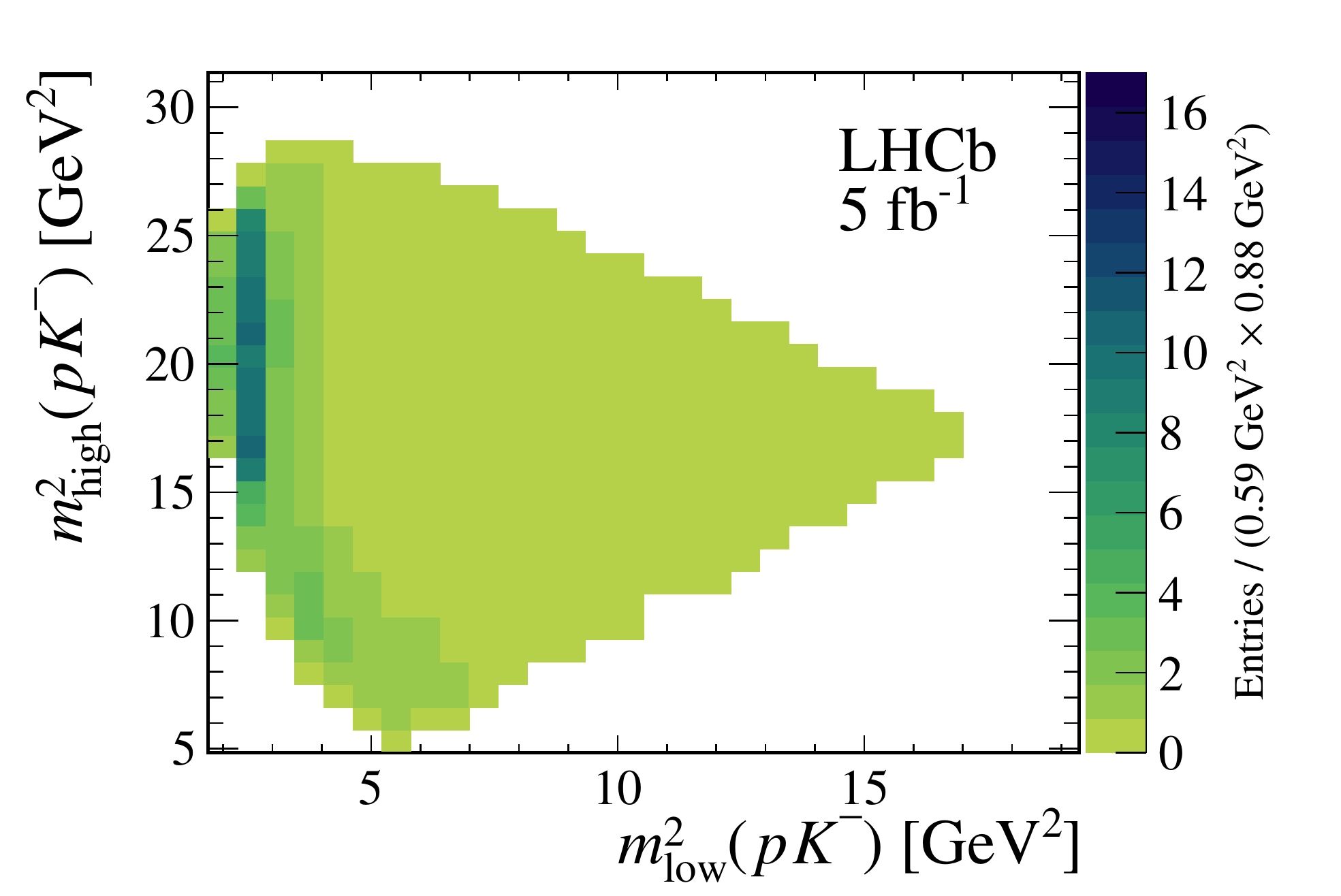} \\
\includegraphics[width=0.49\textwidth]{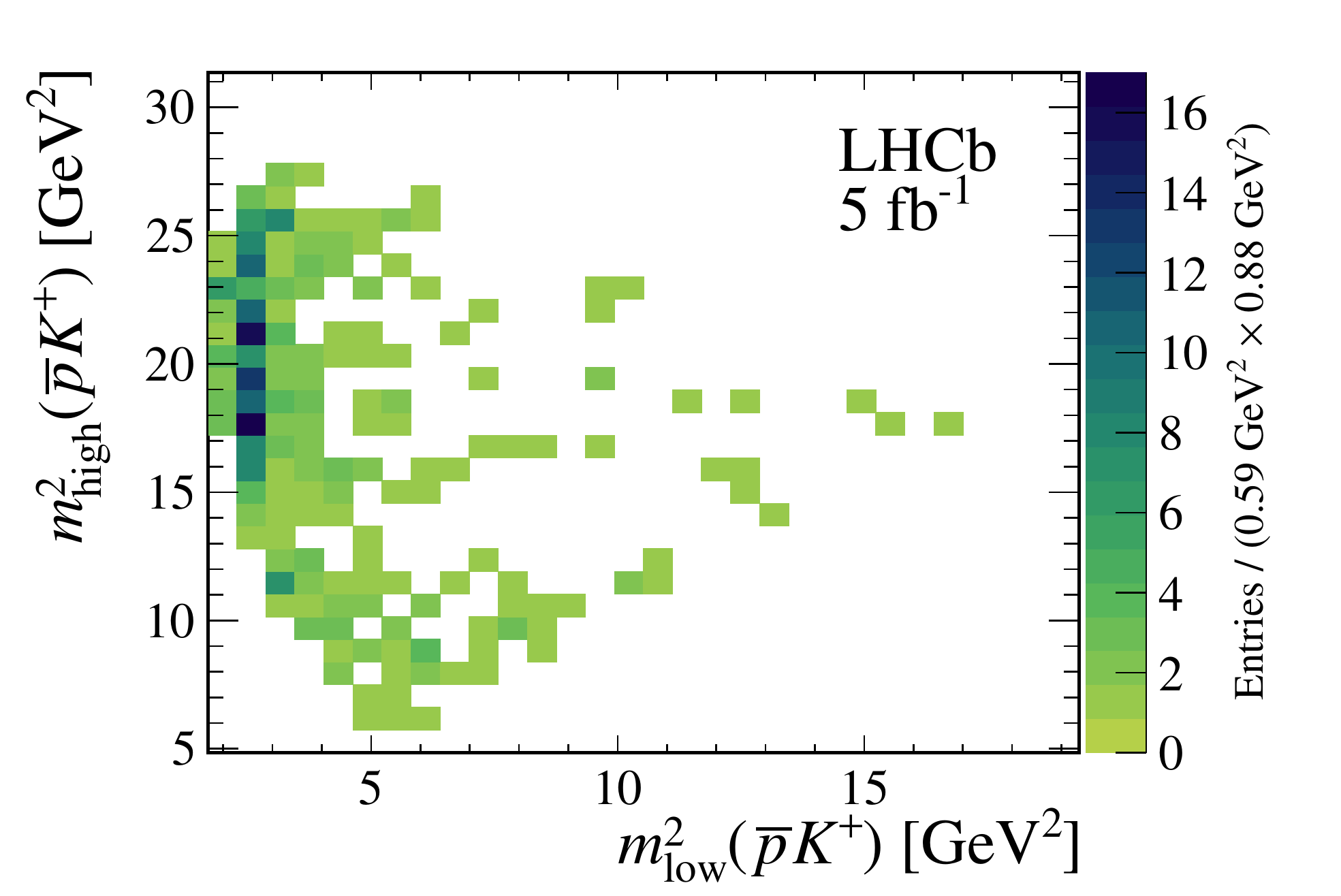}
\includegraphics[width=0.49\textwidth]{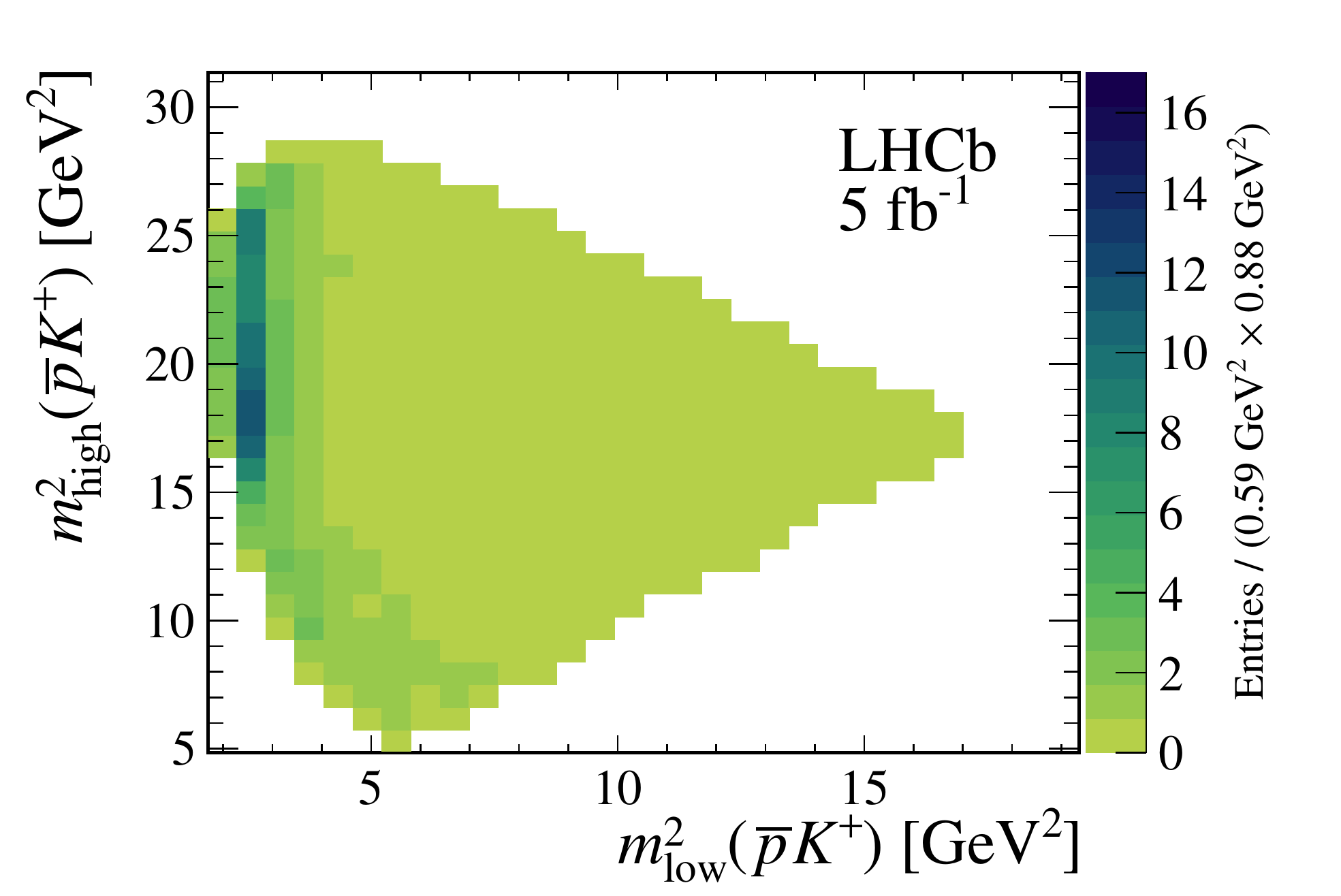}
\caption{\small
  Dalitz-plot distributions from (left) data and (right) the fit model for (top) \Xibm\ and (bottom) \Xibp\ candidates.
}
\label{fig:dp_proj_cpv1}
\end{center}
\end{figure}

\begin{figure}[!tb]
\begin{center}
\includegraphics[width=0.48\textwidth]{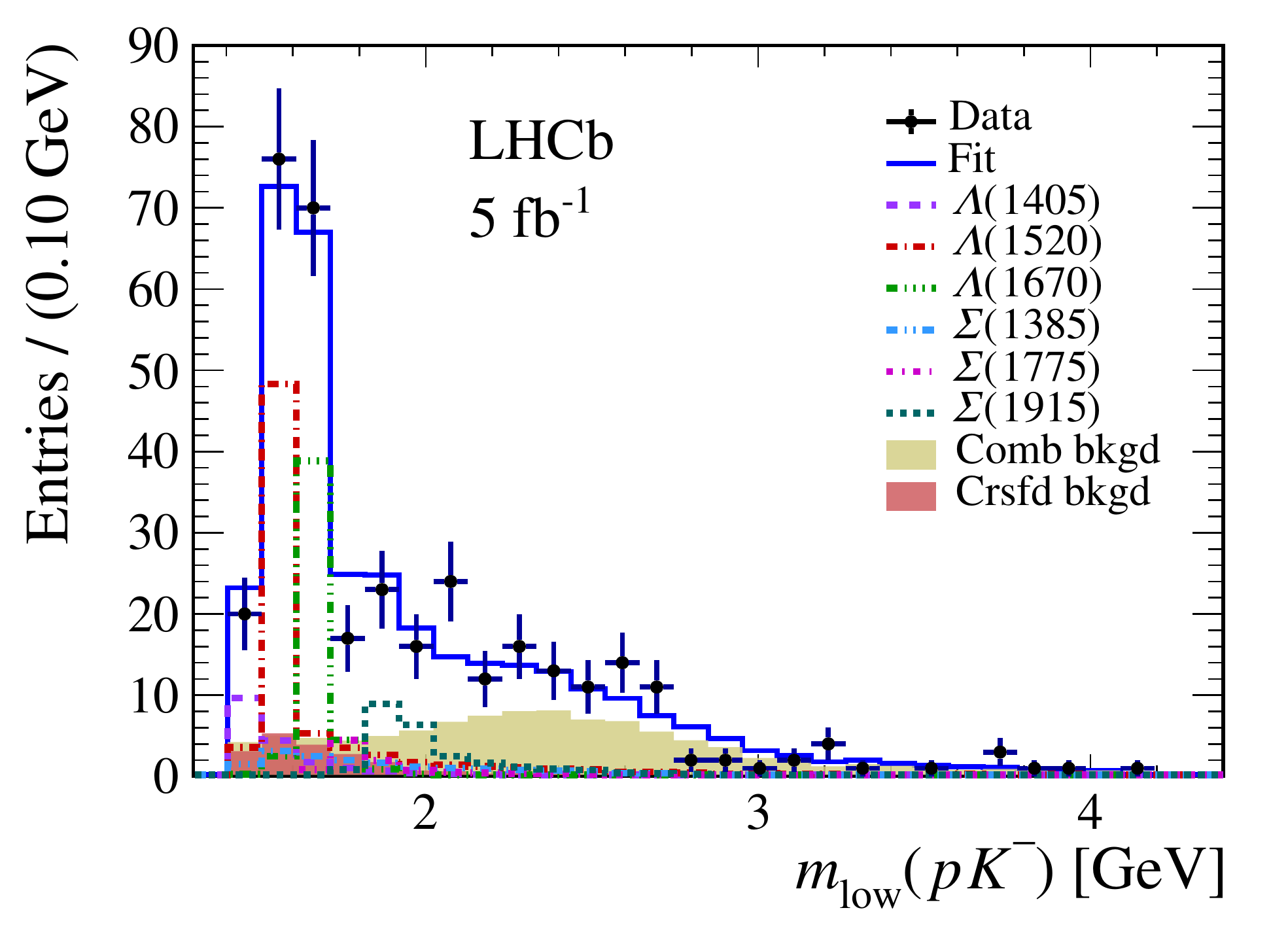} 
\includegraphics[width=0.48\textwidth]{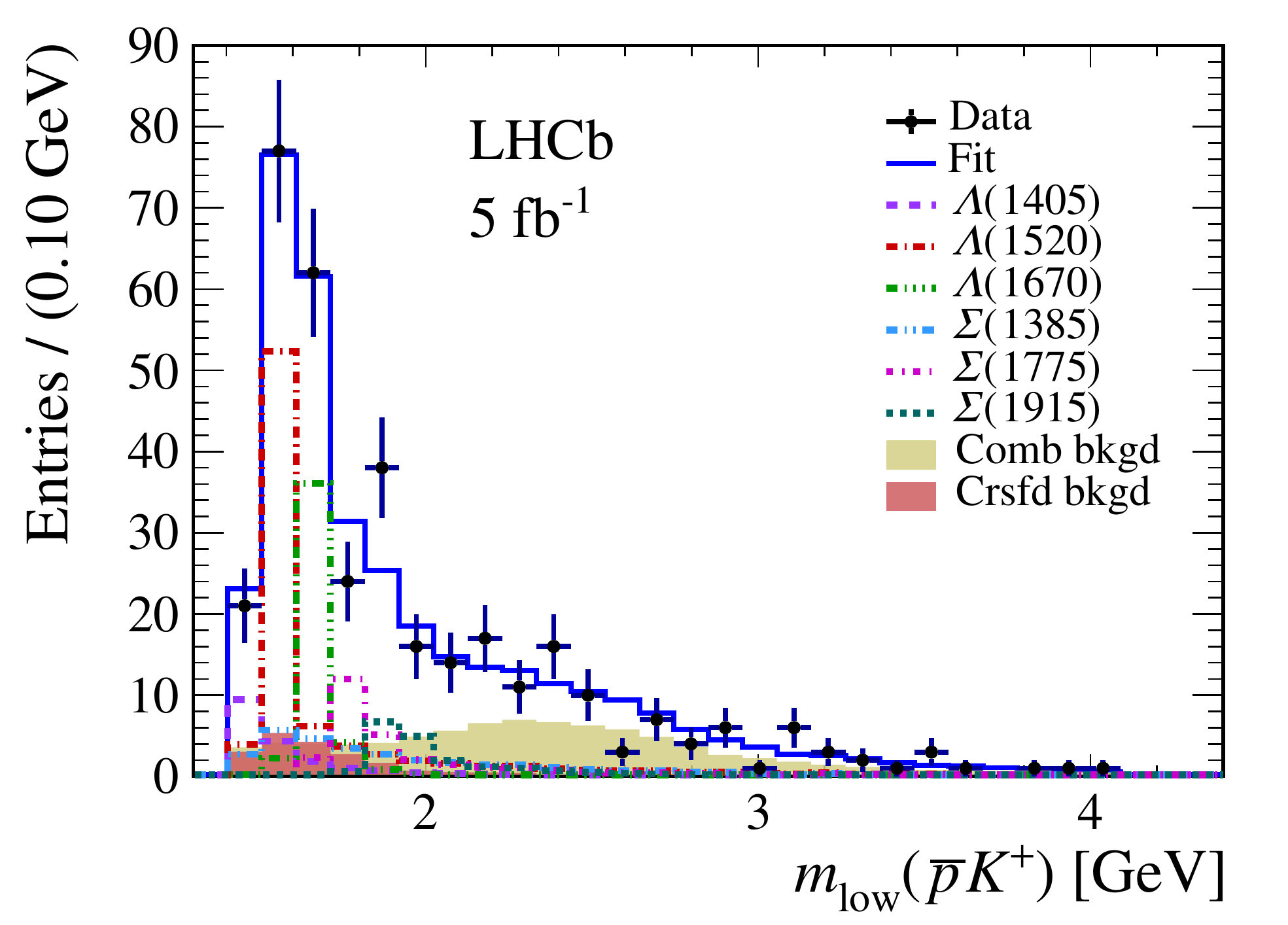} \\
\includegraphics[width=0.48\textwidth]{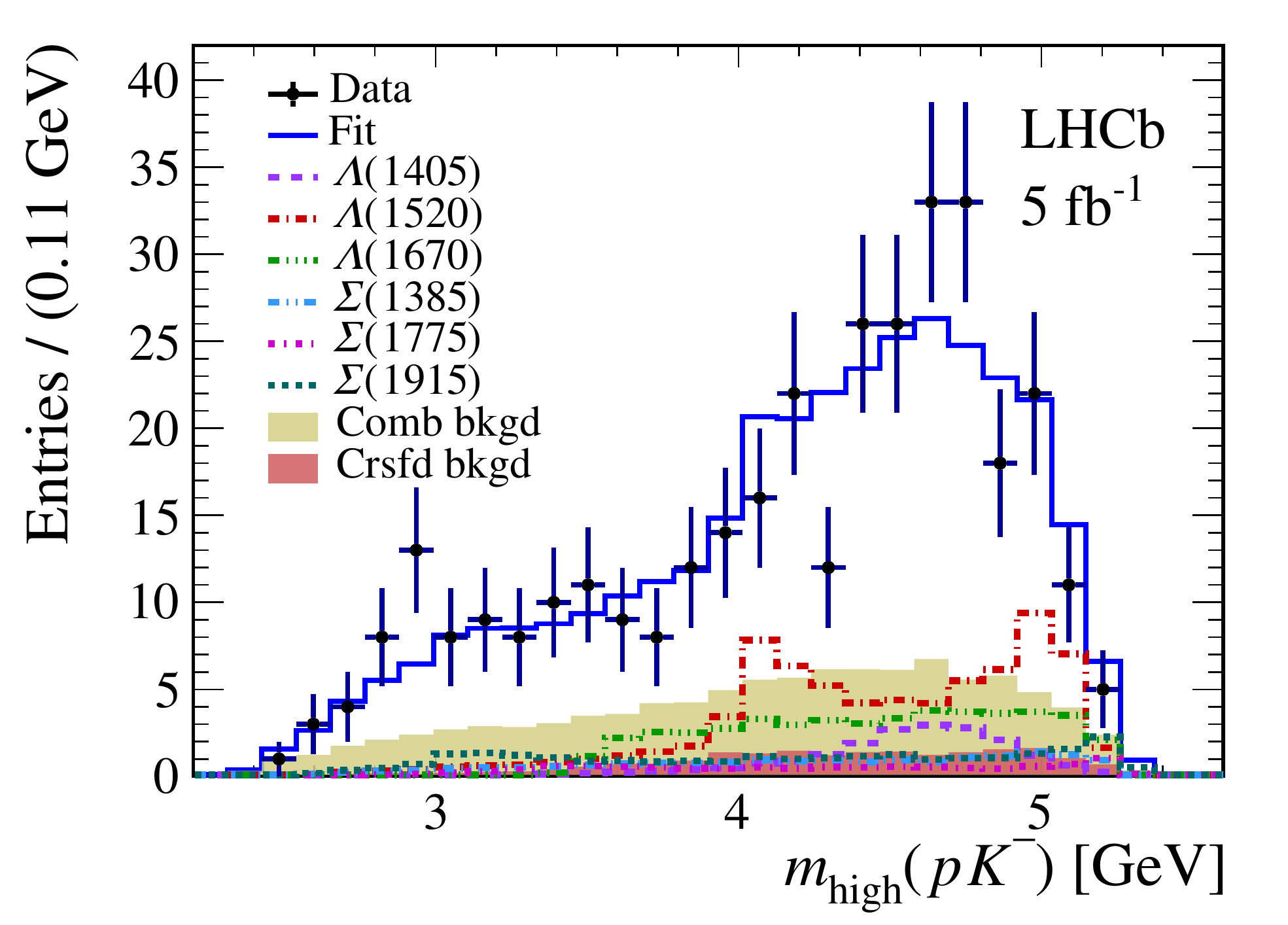}
\includegraphics[width=0.48\textwidth]{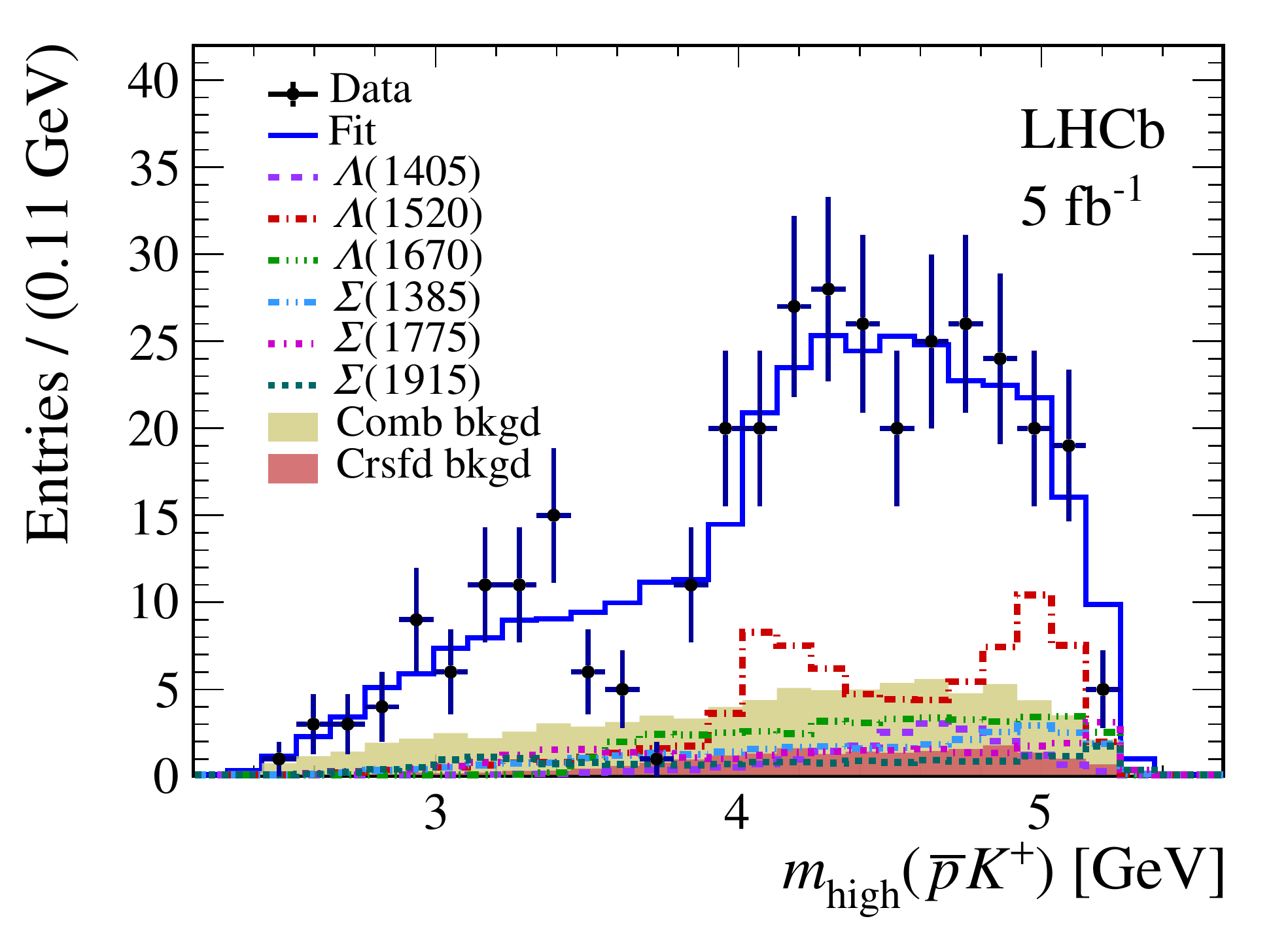}
\caption{\small
  Distributions of (top) $\mpklow$ and (bottom) $\mpkhigh$ for (left) \Xibm\ and (right) \Xibp\ candidates, with results of the fits superimposed.
  The total fit result is shown as the blue solid curve, with contributions from individual signal components and from combinatorial (Comb) and cross-feed (Crsfd) background shown as indicated in the legend.
}
\label{fig:dp_proj_cpv2}
\end{center}
\end{figure}

The Dalitz-plot distribution of the combined Run~1 and Run~2 data sample is compared to the model obtained from the fit to data, separately for \Xibm\ and \Xibp\ candidates, in Fig.~\ref{fig:dp_proj_cpv1}.
Projections of the fit results onto \mpklow\ and \mpkhigh\ are compared to the data in Fig.~\ref{fig:dp_proj_cpv2}.
Further comparisons of the fit result and the data in regions of the phase space are presented in Appendix~\ref{appd:fitprojects}.
There is no indication of \CP\ violation, \ie\ no significant difference between \Xibm\ and \Xibbarp\ decays, in the distributions.

The overall agreement between the data and the model is good, with unbinned goodness-of-fit tests using the mixed-sample and point-to-point dissimilarity approaches~\cite{Williams:2010vh} giving $p$-values of 0.20 and 0.25, respectively.
In Fig.~\ref{fig:dp_proj_cpv2}, there is an apparent discrepancy between the model and the data at $\mpkhigh$ between $3.4\gev$ to $3.7\gev$, predominantly in the \Xibbarp\ sample.
However, due to the symmetry of the final state, any structure that appears in \mpkhigh should also appear in \mpklow, where no such structure is observed.
Addition of extra components to the fit model does not significantly improve the data description.
Moreover, the apparent discrepancy in Fig.~\ref{fig:dp_proj_cpv2} does not take into account the systematic uncertainty in the mismodelling of the combinatorial background, which is the largest component at $\mpkhigh \sim 3.7 \gev$.
Therefore, this feature is not considered to be significant and is not investigated further.

\section{Systematic uncertainties}
\label{sec:systematics}

The outcomes of the analysis are the ratio $\mathcal{R}$ of \Omegab and \Xibm branching fractions and fragmentation fractions (see Eq.~\eqref{eq:bfr}), as well as the fit fractions, interference fit fractions and \CP-asymmetry parameters obtained from the amplitude analysis.
Various sources of systematic uncertainty can affect these measurements.
These are discussed one by one in this section, concluding with a summary. 
Systematic uncertainties are evaluated separately for Run~1 and Run~2, where appropriate.


\subsection{Invariant mass fits}

The fits to the $m(\KKp)$ invariant mass distributions determine the signal and background yields, which are used in both the calculation of $\mathcal{R}$ and the amplitude analysis.
Four sources of systematic uncertainty arising from these fits are considered.
The first is due to the limited size of the data sample.
This enters the calculation of $\mathcal{R}$ as statistical uncertainty, but is a source of systematic uncertainty in the amplitude analysis where the signal and background yields are fixed parameters.
To evaluate the associated systematic uncertainty, these yields are varied according to the covariance matrix obtained from the $m(\KKp)$ fit and for each variation the fit to the phase-space distribution is repeated.
The root mean square (RMS) of the distribution of the change in each result of the amplitude analysis is assigned as the corresponding systematic uncertainty.  

The second source relates to the $m(\KKp)$ fit model.
Models for each of the components are varied to evaluate associated systematic uncertainties: the signal model is replaced with a Hypatia function~\cite{Santos:2013gra}; the combinatorial background model is replaced with a second-order Chebyshev polynomial; the cross-feed and partially reconstructed background models are replaced with kernel density estimates; additional partially reconstructed background components are included; the model used to describe the phase-space distribution of the \XibToKpip\ cross-feed background is varied. 
In each case, the change in each result from its baseline value is taken as the systematic uncertainty. 

The third source concerns fixed parameters in the $m(\KKp)$ fit that are taken from simulation.
An ensemble of pseudoexperiments is generated using the nominal values of these parameters, and each is then fitted many times with parameters fixed to alternative values obtained using the covariance matrices of the fits to the simulation samples.
The standard deviation of the change in each result is evaluated for every pseudoexperiment, and its average value over the ensemble is assigned as the systematic uncertainty.

Finally, potential fit bias is investigated by generating multiple pseudoexperiments with yield and fit parameters obtained from the nominal $m(\KKp)$ fit.
The difference between the mean fit result of the ensemble and the nominal value is assigned as the associated systematic uncertainty.
  
\subsection{Selection efficiency maps}

The efficiency maps are altered to evaluate systematic uncertainties from six sources.  
In each case the difference between the results obtained using the alternative efficiency maps and that with the baseline efficiency maps is assigned as the systematic uncertainty.
    
The first source reflects uncertainties in the \pt\ distribution of \Xibm\ baryons produced in the LHCb acceptance. 
Alternative efficiency maps are obtained where the simulation samples are weighted so that the \pt\ distribution matches that of the background-subtracted data.  
Since there is no significant signal of \Omegab\ baryons, they are assumed to have the same \pt\ distribution as \Xibm\ baryons and the \OmegabToKKp\ efficiency map is altered in the same way.

The second source is a possible mismatch of the hardware trigger efficiency between simulation and data, which could arise due to miscalibration of transverse energy measurements from the calorimeter.
Alternative  efficiency maps are obtained by applying corrections that are calculated, as a function of track \pt, using control samples of kaons from \decay{\Dstarp}{\Dz(\Km\pip)\pip} decays and protons from \decay{\Lb}{\Lc(\proton\Km\pip)\pim} decays.

The third source is due to uncertainty arising from binning the phase space when evaluating the efficiency maps.
Alternative efficiency maps are obtained employing a different SDP binning scheme.  
An additional systematic uncertainty is associated to the efficiency of \OmegabToKKp decays, and arises from their unknown phase-space distribution.
The standard deviation of the variation of the efficiency across the binned SDP histogram is assigned as the corresponding uncertainty.
    
The remaining sources relate to particle identification.
The PID variables used in the MVA are drawn from data calibration samples accounting for dependence on the signal kinematics.
Systematic uncertainties in this procedure arise from the limited statistics of both the simulation and calibration samples, and the modelling of the PID variables in the calibration samples.  
The limitations due to both simulation and calibration sample size are evaluated by bootstrapping to create multiple samples, and repeating the procedure for each sample. 
The impact of potential mismodelling of the PID variables in the calibration samples is evaluated by describing the corresponding distributions using density estimates with different kernel widths.
For each of these cases, alternative efficiency maps are produced to determine the associated uncertainties on the results of the analysis.

In principle, mismodelling of the proton and kaon reconstruction efficiencies, and associated asymmetries, could be a source of systematic uncertainty.
However, such effects are known to be negligible at the level of precision achieved in this analysis~\cite{LHCb-PAPER-2014-013,LHCb-PAPER-2021-016}, and therefore are not accounted for explicitly.  

\subsection{Background shapes}

The combinatorial background SDP distribution is obtained by extrapolating from an $m(\KKp)$ sideband region, and has uncertainties related to the available yield in the sideband and the extrapolation procedure itself.  
The former is evaluated by bootstrapping to create multiple combinatorial background samples, and repeating the amplitude fit with each.
The RMS of the distribution of the change in each result is taken as the systematic uncertainty.  
The latter is evaluated by changing the architecture of the neural network, with the change in each result with respect to its baseline value assigned as the associated systematic uncertainty.

In the baseline fit, the \XibToKpip\ cross-feed background is described with a model consisting of $\Lz(1405)$, $\Lz(1520)$, $\Lz(1690)$, $N(1440)$, $N(1520)$, $N(1535)$ and $N(1650)$ resonances.
To evaluate the systematic uncertainty arising from this assumption, the model is modified by adding $\Lz(1600)$, $\Lz(1670)$, $\Lz(1800)$ and $N(1720)$ components, and removing the $N(1520)$ component.
The change in each result with respect to its baseline value is assigned as the associated systematic uncertainty.

\subsection{Background asymmetry}

In the baseline model, it is assumed that there is no local asymmetry in the combinatorial background as described in Sec.~\ref{sec:bkg_shapes}.  
The associated systematic uncertainty is evaluated by considering separate background distributions for \Xibm\ and \Xibp\ candidates.
In order to obtain sufficiently large background samples to determine these separate distributions, the MVA output requirement for candidates in the sideband region is relaxed.

A possible global combinatorial background asymmetry is accounted for in the baseline fit, while cross-feed background is assumed to have no asymmetry.
A fit allowing for a global cross-feed background asymmetry is performed, and the differences between the results in this fit and their nominal values is assigned as the systematic uncertainty arising from this assumption.

\subsection{Production asymmetry}

The baseline fit model assumes no asymmetry in the \Xibm\ production rates in the LHCb acceptance, consistent with measurements~\cite{LHCb-PAPER-2018-047}.
To evaluate the associated systematic uncertainty, the model is adjusted to include production asymmetries within the experimentally allowed range by introducing a global asymmetry in the efficiency maps.
An ensemble of fits with varied  \Xibm\ production asymmetries is performed, and the RMS of the distribution of the change in each result with respect to its baseline value is assigned as the systematic uncertainty.  

\subsection{Polarisation}

The transverse polarisation of \Xibm\ baryons produced in $pp$ collisions is assumed to be consistent with zero, as observed for \Lb baryons~\cite{LHCb-PAPER-2012-057,LHCb-PAPER-2020-005}.  
The distributions of $\mpklowsq$ and $\mpkhighsq$ are independent of the \Xibm\ polarisation.
However, if the \Xibm baryons produced in LHC collisions are polarised, efficiency variation across additional phase-space variables should be considered in the analysis. 
To evaluate the systematic uncertainty due to potential \Xibm\ polarisation, two sets of pseudoexperiments are generated, with the \Xibm polarisation set to $20\%$ in one case and $-8\%$ in the other.
This corresponds to the $\pm 2\sigma$ range measured for the \Lb\ baryon in Ref.~\cite{LHCb-PAPER-2012-057,LHCb-PAPER-2020-005}, where $\sigma$ indicates the Gaussian standard deviation.
A conservatively broad range is taken to allow for differences between \Xibm\ and \Lb\ polarisation.
The pseudoexperiments are generated using a signal model whose helicity couplings are set to the values from the nominal fit and where the measured efficiency variation over the additional phase-space observables is introduced. 
A fit to the Dalitz-plot variables is then performed using the baseline model. 
The largest deviation of the parameters from the nominal case is assigned as the systematic uncertainty.

\subsection{Modelling of the lineshapes}

Each resonant contribution has fixed parameters in the amplitude fit.
These include masses, widths and Blatt--Weisskopf radius parameters.  
An ensemble of fits is obtained varying the masses and widths of all resonances within the range of values quoted by the PDG and given in Table~\ref{tab:lambda-sigma}.  
The Blatt--Weisskopf radius parameter associated with the \Xibm\ baryon is varied in the range $3$--$7 \gev^{-1}$ and that associated with the resonances is varied in the range $0$--$3.5 \gev^{-1}$.  
The RMS of the distribution of the change in each fitted parameter is taken as the systematic uncertainty.
  
For the $\Lz(1405)$ and $\Sigmares(1385)$ resonances that peak below the $\mpklow$ threshold, the effective RBW used in the baseline fit is replaced with a lineshape equivalent to the Flatt\'{e} parameterisation as done in Ref.~\cite{LHCb-PAPER-2015-029}. 
The total width is modified to account for the $\Sigmapi$ channel, \ie $\Gamma(m) = \Gamma_{\pK}(m) + \Gamma_{\Sigmapi}(m)$, assuming equal couplings to both channels. 
For each of these lineshape variations, the differences in the results between fits with the alternative and baseline models are assigned as the associated systematic uncertainties.

\subsection{Alternative fit model}

The effect of including additional signal components in the fit model is examined to assign systematic uncertainty due to the composition of the baseline model.
The $\Lz(1690)$, $\Lz(1820)$, $\Sigmares(1670)$ and $\NRthreehalfpos$ components are added to the nominal model individually.
These modifications of the model are chosen since they improve $-2\ln\mathcal{L}$, although not by a significant amount.
The largest deviation, among the four cases, from the nominal value of each measured quantity is taken as the associated systematic uncertainty.
  
\subsection{Summary of systematic uncertainties}
\label{sec:syst:summary}

\paragraph*{Ratio of fragmentation and branching fractions:}

Since separate fits are performed to the $m(\KKp)$ distributions from the Run~1 and Run~2 samples, and the signal efficiencies are also determined separately, results for $\mathcal{R}$ in each of the two samples are obtained.
Systematic uncertainties on $\mathcal{R}$ are considered as being either completely uncorrelated or 100\% correlated between the two results.
The systematic uncertainties that are uncorrelated between Run~1 and Run~2 are folded into their respective likelihood functions, by convolution with a Gaussian of appropriate width.
The correlated systematic uncertainties are later folded into the combined likelihood that is obtained by multiplying the likelihood functions of the two samples. 

The uncorrelated systematic uncertainties are those that are related to the fixed parameters in the fit model, the fit bias and the impact on the efficiency of the \Xibm\ production kinematics, and the descriptions of the hardware trigger and particle identification response.
The correlated systematic uncertainties are those related to knowledge of the phase-space distributions of the decays and the fit model choice.
Slightly different procedures are used to obtain the total uncertainty for the two sources of correlated systematic uncertainty.
That related to knowledge of the phase-space distribution affects the efficiency, and hence is a constant relative uncertainty.
The method to evaluate the uncertainty due to fit model choice gives different relative uncertainties between Run~1 and Run~2.
Since the two samples have approximately equal statistical weight in the combination, the average of the relative uncertainties is taken and assigned to the combined result.
Table~\ref{table:syst} summarises the systematic uncertainties on $\mathcal{R}$.

\begin{table}[!tb]
  \caption{\small
    Absolute systematic uncertainties on $\mathcal{R}$, in units of $10^{-3}$, from (top) uncorrelated and (bottom) correlated sources.
    The total is the sum in quadrature of all contributions.
  }
  \label{table:syst}
  \centering
  \renewcommand{\arraystretch}{1.1}
  \begin{tabular}{lcc}
    \hline
    Uncorrelated sources 	  & Run~1  & Run~2  \\
    \hline
    \Xibm \pt distribution	  & ${<\,}0.1\phantom{\,<}$  & 0.7   \\
    Hardware trigger efficiency	  & 0.1  & 1.6   \\
    PID efficiency        & 0.1  & 0.6   \\
    Fixed parameters 	  & 0.8  & 0.5   \\
    Fit bias 		  & 0.5  & ${<\,}0.1\phantom{\,<}$   \\
    \hline
    Total                 & 1.0  & 1.9   \\
    \hline
  \end{tabular}

  \vspace{1ex}
  
  \begin{tabular}{lccc}
    \hline
    Correlated sources 	    & Run~1  & Run~2  & Combined  \\
    \hline
    Phase-space distribution    & 8.9       & 22.5   & 	10.6	\\
    Fit model choice 	        & 9.1       & 13.1   & 	\phantom{1}8.6	\\
    \hline
    Total   &  -- 	& -- 	 &  	13.6		\\
    \hline
  \end{tabular}
\end{table}

\paragraph*{Amplitude analysis:}

The results of the amplitude analysis are the \CP\ asymmetry parameters $A^{\CP}$, defined in Eq.~\eqref{eq:acpcomb}, the fit fractions $\mathcal{F}$ defined in Eq.~\eqref{eq:ff} and the interference fit fractions $\mathcal{I}$ defined in Eq.~\eqref{eq:iff}.
A summary of the systematic uncertainties 
on these quantities is shown in Table~\ref{table:summarysyst}.

The most precise results are those related to the $\Lz(1520)$ and $\Lz(1670)$ resonances.  
For $\Lz(1520)$, the dominant systematic uncertainty on $A^{\CP}$ is due to ignoring the efficiency variation over angular variables if \Xibm\ baryons are produced polarised, whereas for $\mathcal{F}$ it is due to the limited size of the sample used for combinatorial background modelling and the variation of the Blatt--Weisskopf radius parameters.
For $\Lz(1670)$, the largest systematic uncertainty on $A^{\CP}$ is due to variation of the $\Lz(1405)$ lineshape which has the same spin and parity as this component, whereas for $\mathcal{F}$ it is due to use of an alternate fit model.
For $\Sigmares(1385)$, the dominant systematic uncertainty on $A^{\CP}$ is due to use of an alternate fit model, whereas for $\mathcal{F}$ it is 
due to variation of the Blatt--Weisskopf radius parameters.
For $\Lz(1405)$, the largest systematic uncertainty on $A^{\CP}$ is due to use of an alternate fit model, whereas for $\mathcal{F}$ it is 
due to variation in its lineshape.
For $\Sigmares(1775)$, the dominant systematic uncertainty on $A^{\CP}$ is due to the limited size of the sample used for modelling of the combinatorial background, whereas for $\mathcal{F}$ it is due to use of an alternate fit model.
For $\Sigmares(1915)$, the dominant systematic uncertainty on both $A^{\CP}$ and $\mathcal{F}$ is due to use of an alternate fit model.

For interference fit fractions, the largest systematic uncertainties are mainly due to 
the use of an alternate fit model, the limited size of the sample used for modelling of the combinatorial background, variation of the resonance lineshapes
and variation of Blatt--Weisskopf radius parameters. 

\begin{sidewaystable}[!tb] 
  \caption{\small
    	   Summary of absolute systematic uncertainties, in units of $10^{-2}$, on the results of the amplitude analysis: 
	    (top) \CP\ asymmetry parameters $A^{\CP}$ and fit fractions $\mathcal{F}$, (bottom) interference fit fractions $\mathcal{I}$.
           The total is the sum in quadrature of all individual sources.}
  \label{table:summarysyst}
  \centering
  \resizebox{\textwidth}{!}{
    \renewcommand{\arraystretch}{1.1}
    \begin{tabular}{l|c|ccccccccc|c}
      \hline
\multicolumn{2}{c|}{Component \& Parameter}  	& Mass fits  & Bkg shapes  & Bkg asym  & Eff  & Prod asym  & Polarisation  & RBW params  & Lineshapes  & Alt fit model    & Total   \\
\hline

\multirow{2}{*}{$\Sigmares(1385)$}      & $A^{\CP}$        		&3.3     &20.6\phantom{2}    &4.4     &8.2     &6.9     &15.0\phantom{1}    &7.0     &12.6\phantom{1}    &65.5\phantom{6}    &72.7\phantom{7} \\    \cdashline{2-12}
                                        & ${\cal F}$           		&1.4     &3.1     &0.5     &0.7     &0.2     &1.0     &5.0     &0.6     &4.4     &7.6  \\    \hline
\multirow{2}{*}{$\Lz(1405)$}            & $A^{\CP}$        		&2.4     &9.6     &2.7     &5.1     &5.1     &5.5     &5.1     &19.5\phantom{1}    &20.6\phantom{2}    &31.9\phantom{3} \\    \cdashline{2-12}
                                        & ${\cal F}$           		&0.3     &1.4     &0.3     &0.8     &${<\,}0.1\phantom{\,<}$ &0.3     &0.3     &2.3     &0.9     &3.0  \\    \hline
\multirow{2}{*}{$\Lz(1520)$}            & $A^{\CP}$        		&0.3     &0.9     &0.6     &2.9     &4.3     &5.0     &0.7     &1.2     &1.3     &7.6  \\    \cdashline{2-12}
                                        & ${\cal F}$           		&1.1     &1.8     &${<\,}0.1\phantom{\,<}$ &1.3     &${<\,}0.1\phantom{\,<}$ &0.6     &1.8     &0.6     &1.7     &3.6  \\    \hline
\multirow{2}{*}{$\Lz(1670)$}            & $A^{\CP}$        		&1.8     &4.2     &1.4     &2.9     &4.4     &3.3     &3.7     &4.9     &1.6     &10.1\phantom{1} \\    \cdashline{2-12}
                                        & ${\cal F}$           		&0.8     &2.3     &0.1     &0.7     &0.1     &0.6     &1.4     &1.8     &4.4     &5.6  \\    \hline
\multirow{2}{*}{$\Sigmares(1775)$}      & $A^{\CP}$        		&2.5     &7.8     &1.7     &3.1     &3.4     &7.0     &3.8     &4.7     &3.7     &13.8\phantom{1} \\    \cdashline{2-12}
                                        & ${\cal F}$           		&0.5     &1.5     &0.1     &0.4     &0.2     &0.2     &1.0     &0.9     &3.5     &4.1  \\    \hline
\multirow{2}{*}{$\Sigmares(1915)$}      & $A^{\CP}$        		&2.5     &6.7     &5.0     &6.4     &4.8     &5.2     &10.5\phantom{1}    &2.1     &13.9\phantom{1}    &21.8\phantom{2} \\    \cdashline{2-12}
                                        & ${\cal F}$           		&0.2     &2.3     &0.1     &1.3     &0.2     &0.2     &2.2     &1.5     &8.4     &9.2  \\    \hline
\multicolumn{2}{c|}{$\Lz(1405), \Lz(1520)$\ ${\cal I}$}       		&0.2     &0.6     &0.1     &0.1     &${<\,}0.1\phantom{\,<}$ &0.1     &0.2     &0.2     &0.4     &0.8  \\ \hline
\multicolumn{2}{c|}{$\Lz(1405), \Lz(1670)$\ ${\cal I}$}       		&0.3     &0.9     &0.1     &0.5     &${<\,}0.1\phantom{\,<}$ &${<\,}0.1\phantom{\,<}$ &1.2     &1.6     &0.9     &2.4  \\ \hline
\multicolumn{2}{c|}{$\Lz(1405), \Sigmares(1385)$\ ${\cal I}$} 		&0.2     &0.4     &0.1     &0.2     &${<\,}0.1\phantom{\,<}$ &0.1     &0.5     &0.7     &1.8     &2.0  \\ \hline
\multicolumn{2}{c|}{$\Lz(1405), \Sigmares(1775)$\ ${\cal I}$} 		&${<\,}0.1\phantom{\,<}$ &0.3     &${<\,}0.1\phantom{\,<}$ &0.1     &${<\,}0.1\phantom{\,<}$ &0.1     &0.4     &0.3     &1.1     &1.2  \\ \hline
\multicolumn{2}{c|}{$\Lz(1405), \Sigmares(1915)$\ ${\cal I}$} 		&0.1     &0.4     &${<\,}0.1\phantom{\,<}$ &0.1     &${<\,}0.1\phantom{\,<}$ &${<\,}0.1\phantom{\,<}$ &0.1     &0.3     &0.6     &0.8  \\ \hline
\multicolumn{2}{c|}{$\Lz(1520), \Lz(1670)$\ ${\cal I}$}       		&0.1     &0.2     &${<\,}0.1\phantom{\,<}$ &${<\,}0.1\phantom{\,<}$ &${<\,}0.1\phantom{\,<}$ &${<\,}0.1\phantom{\,<}$ &0.2     &0.1     &0.8     &0.8  \\ \hline
\multicolumn{2}{c|}{$\Lz(1520), \Sigmares(1385)$\ ${\cal I}$} 		&0.7     &0.9     &0.2     &0.2     &0.2     &0.1     &1.6     &0.9     &3.6     &4.2  \\ \hline
\multicolumn{2}{c|}{$\Lz(1520), \Sigmares(1775)$\ ${\cal I}$} 		&0.3     &0.6     &0.1     &0.1     &0.1     &0.2     &0.5     &0.5     &1.6     &1.9  \\ \hline
\multicolumn{2}{c|}{$\Lz(1520), \Sigmares(1915)$\ ${\cal I}$} 		&${<\,}0.1\phantom{\,<}$ &0.1     &${<\,}0.1\phantom{\,<}$ &${<\,}0.1\phantom{\,<}$ &${<\,}0.1\phantom{\,<}$ &${<\,}0.1\phantom{\,<}$ &0.1     &${<\,}0.1\phantom{\,<}$ &0.1     &0.2  \\ \hline
\multicolumn{2}{c|}{$\Lz(1670), \Sigmares(1385)$\ ${\cal I}$} 		&0.2     &0.3     &0.1     &0.1     &${<\,}0.1\phantom{\,<}$ &${<\,}0.1\phantom{\,<}$ &0.4     &0.6     &0.6     &1.0  \\ \hline
\multicolumn{2}{c|}{$\Lz(1670), \Sigmares(1775)$\ ${\cal I}$} 		&0.1     &0.1     &${<\,}0.1\phantom{\,<}$ &${<\,}0.1\phantom{\,<}$ &${<\,}0.1\phantom{\,<}$ &${<\,}0.1\phantom{\,<}$ &0.1     &0.1     &0.2     &0.3  \\ \hline
\multicolumn{2}{c|}{$\Lz(1670), \Sigmares(1915)$\ ${\cal I}$} 		&${<\,}0.1\phantom{\,<}$ &0.1     &${<\,}0.1\phantom{\,<}$ &${<\,}0.1\phantom{\,<}$ &${<\,}0.1\phantom{\,<}$ &${<\,}0.1\phantom{\,<}$ &0.3     &0.1     &0.4     &0.5  \\ \hline
\multicolumn{2}{c|}{$\Sigmares(1385), \Sigmares(1775)$\ ${\cal I}$} 	&0.1     &0.3     &${<\,}0.1\phantom{\,<}$ &${<\,}0.1\phantom{\,<}$ &${<\,}0.1\phantom{\,<}$ &${<\,}0.1\phantom{\,<}$ &0.2     &0.1     &0.2     &0.4  \\ \hline
\multicolumn{2}{c|}{$\Sigmares(1385), \Sigmares(1915)$\ ${\cal I}$} 	&0.2     &0.6     &${<\,}0.1\phantom{\,<}$ &0.2     &${<\,}0.1\phantom{\,<}$ &0.1     &0.6     &0.2     &1.0     &1.3  \\ \hline
\multicolumn{2}{c|}{$\Sigmares(1775), \Sigmares(1915)$\ ${\cal I}$} 	&0.1     &0.4     &${<\,}0.1\phantom{\,<}$ &0.1     &${<\,}0.1\phantom{\,<}$ &0.1     &0.7     &0.2     &1.0     &1.3  \\ 
\hline
\end{tabular} 
}
\end{sidewaystable}
\afterpage{\clearpage}

\section{Results}
\label{sec:results}

\paragraph*{Ratio of fragmentation and branching fractions:}

The results for the ratio $\mathcal{R}$ of the relative fragmentation and branching fractions for \OmegabToKKp\ and \XibToKKp\ decays are
\begin{eqnarray*}
  \mathcal{R} &=& ({-20} \pm 30 \stat \pm 1 \,(\text{uncorr syst})) \times 10^{-3}~~~\text{for Run~1 and} \\
  \mathcal{R} &=& ({\phantom{-}51} \pm 28 \stat \pm 2 \,(\text{uncorr syst})) \times 10^{-3}~~~\text{for Run~2}\,,
\end{eqnarray*}
where the first uncertainty is statistical and the second includes only the uncorrelated systematic effects presented in Table~\ref{table:syst}.
The negative log-likelihood functions for these two results are added to obtain a combined result, 
\begin{equation*}
    \mathcal{R} \equiv \frac{f_{\Omegab}}{f_{\Xibm}} \times \frac{{\cal B}(\OmegabToKKp)}{{\cal B}(\XibToKKp)} = (24 \pm 21 \stat \pm 14 \syst) \times 10^{-3}\,,
\end{equation*}
where both uncorrelated and correlated systematic uncertainties are included.
In the combined result, it is implied that $f_{\Omegab}/f_{\Xibm}$, which may vary with centre-of-mass energy of the LHC $pp$ collisions, is an effective value averaged over the Run~1 and Run~2 data samples.
This result is found to be consistent with, and more precise than, the previous measurement~\cite{LHCb-PAPER-2016-050}.
No significant evidence of the \OmegabToKKp\ decay is found, and therefore an upper limit on $\mathcal{R}$ is calculated at 90~(95)\,\% confidence level by integrating the likelihood in the physical region of non-negative branching fraction,
\begin{equation*}
  \mathcal{R} \equiv \frac{f_{\Omegab}}{f_{\Xibm}} \times \frac{{\cal B}(\OmegabToKKp)}{{\cal B}(\XibToKKp)} < 62~(71) \times 10^{-3}\,.
\end{equation*}

\paragraph*{Amplitude analysis:}

The results for the \CP-asymmetry parameters for each component of the signal model are shown in Table~\ref{table:cpasymmetry_results_syst}.
No significant \CP\ asymmetry is observed.
The fit fraction matrix is reported in Table~\ref{table:ff_results_syst}.
The diagonal elements correspond to the fit fractions of the respective components, and the off-diagonal elements are the interference fit fractions.
These results are derived from the helicity couplings that are the free parameters of the amplitude fit.
Their statistical uncertainties are evaluated from an ensemble of pseudoexperiments, while systematic uncertainties are obtained as described in Sec.~\ref{sec:systematics}.

\begin{table}[!tb] 
\caption{\small
  Results for the \CP-asymmetry parameters.
  The statistical uncertainties are obtained from pseudoexperiments while the systematic uncertainties are obtained following the procedure described in Sec.~\ref{sec:systematics}.
}
\label{table:cpasymmetry_results_syst}
\centering
\renewcommand{\arraystretch}{1.1}
\begin{tabular}{lr@{$\,\pm\,$}c@{$\,\pm\,$}l}
\hline 
Component & \multicolumn{3}{c}{$A^{\CP}$ ($10^{-2}$)}\\
\hline
$\Sigmares(1385)$ & $-27$ & 34 (stat) 		& 73 (syst)\\
$\Lz(1405)$       & $ -1$ & 24 (stat) 		& 32 (syst)\\
$\Lz(1520)$       & $ -5$ & \phantom{1}9 (stat) & \phantom{1}8  (syst)\\
$\Lz(1670)$       & $  3$ & 14 (stat)           & 10 (syst)\\
$\Sigmares(1775)$ & $-47$ & 26 (stat)        	& 14 (syst)\\
$\Sigmares(1915)$ & $ 11$ & 26 (stat)        	& 22 (syst)\\ [0.3ex]
\hline
\end{tabular} 
\end{table}

\begin{sidewaystable}[!tb] 
\caption{\small
  Results for the fit fractions (diagonal elements) and interference fit fractions (off-diagonal elements) obtained from the amplitude analysis.
  Identical values for the interference fit fractions in the upper triangle are omitted.
  All values are in units of $10^{-2}$, with the first uncertainty being statistical and the second systematic.
}
\label{table:ff_results_syst}
\centering
\begin{tabular}{lr@{$\,\pm\,$}c@{$\,\pm\,$}lr@{$\,\pm\,$}c@{$\,\pm\,$}lr@{$\,\pm\,$}c@{$\,\pm\,$}lr@{$\,\pm\,$}c@{$\,\pm\,$}lr@{$\,\pm\,$}c@{$\,\pm\,$}lr@{$\,\pm\,$}c@{$\,\pm\,$}l}
\hline \\ [-2.5ex]
Component 
		& \multicolumn{3}{c}{$\Sigmares(1385)$ } 
		& \multicolumn{3}{c}{$\Lz(1405)$    } 
		& \multicolumn{3}{c}{$\Lz(1520)$    } 
		& \multicolumn{3}{c}{$\Lz(1670)$    } 
		& \multicolumn{3}{c}{$\Sigmares(1775)$ } 
		& \multicolumn{3}{c}{$\Sigmares(1915)$ } \\ [0.3ex]
\hline
$\Sigmares(1385)$& $ 11.4$ &  4.9 &  7.6 \\
$\Lz(1405)$      & $ -1.3$ &  0.8 &  2.0 & $ 8.1$ & 2.7 & 3.0 \\
$\Lz(1520)$      & $  3.4$ &  1.6 &  4.2 & $ 0.1$ & 0.5 & 0.8 & $33.0$ & 4.1 & 3.6 \\
$\Lz(1670)$      & $ -0.1$ &  0.6 &  1.0 & $ 3.0$ & 1.8 & 2.4 & $-0.1$ & 0.4 & 0.8 & $19.5$ &  3.2 &  5.6 \\
$\Sigmares(1775)$& $  0.1$ &  0.3 &  0.4 & $-0.7$ & 0.5 & 1.2 & $ 1.1$ & 1.0 & 1.9 & $-0.3$ &  0.2 &  0.3 & $ 9.7$ & 3.5 & 4.1 \\
$\Sigmares(1915)$& $  0.6$ &  0.6 &  1.3 & $ 0.3$ & 0.3 & 0.8 & $ 0.1$ & 0.1 & 0.2 & $-0.1$ &  0.2 &  0.5 & $ 1.0$ & 0.5 & 1.3 & $11.3$ & 3.7 & 9.2 \\ [0.3ex]
\hline
\end{tabular} 
\end{sidewaystable}

The significance of each component in the baseline model is evaluated using pseudoexperiments.
These are generated, each with a sample size corresponding to the data, according to the best fit model with the component of interest removed from the model.
They are then fitted both with the model used to generate and with the model including the component of interest.
Twice the difference between the negative log-likelihood values obtained in these two fits ($-2 \Delta \ln \mathcal{L}$) is used as a test statistic. 
A $p$-value, corresponding to the probability of observing $-2 \Delta \ln \mathcal{L}$ values as large or larger than that found in the fit to data, is found by extrapolating the tail of the distribution obtained from the ensemble of pseudoexperiments.
In order to account for dominant systematic uncertainties, this procedure is performed for the alternative model that gives the smallest value of $-2 \ln \mathcal{L}$ in fits to data.
The outcome is that the $\Lz(1520)$ and $\Lz(1670)$ components have $p$-values corresponding to $12.0\,\sigma$ and $6.1\,\sigma$, respectively.
All other components have significance below $3.5\,\sigma$.

The branching fraction of each quasi-two-body contribution to the \XibToKKp decay, corresponding to an intermediate resonance $R$, can be obtained from its fit fraction~$\mathcal{F}_i$,
\begin{equation}
\label{eq:bf_twobody}
{\cal B}(\decay{\Xibm}{R\Km}) = {\cal B}(\XibToKKp) \times \mathcal{F}_i\,.
\end{equation}
The branching fraction of \XibToKKp\ has not been measured directly, but the ratio of fragmentation and branching fractions relative to the $\Bm \to \Kp\Km\Km$ decay is known~\cite{LHCb-PAPER-2016-050}.
This can be combined with the known values of ${\cal B}\left( \Bm \to \Kp\Km\Km \right)$~\cite{Garmash:2004wa,Lees:2012kxa,PDG2020}, $f_{\Xibm}/f_{\Lb}$~\cite{LHCb-PAPER-2018-047} and $f_{\Lb}/\left(f_u + f_d\right)$~\cite{LHCb-PAPER-2018-050}, assuming that $f_u = f_d$, to obtain 
\begin{equation*}
{\cal B}\left( \Xibm\to \proton \Km\Km \right) = \left( 2.3 \pm 0.9 \right) \times 10^{-6} \, ,
\end{equation*}
where the dominant uncertainty is that due to possible SU(3)-breaking effects which affect $f_{\Xibm}/f_{\Lb}$~\cite{LHCb-PAPER-2018-047}.
Consequently, the values of the quasi-two-body branching fractions are found to be  
\begin{eqnarray*}
{\cal B}\left(\Xibm \to \Sigmares(1385)\Km\right) & = & \left(0.26 \pm 0.11 \pm 0.17 \pm 0.10\right) \times 10^{-6} \,, \\
{\cal B}\left(\Xibm \to \Lz(1405)\Km\right)       & = & \left(0.19 \pm 0.06 \pm 0.07 \pm 0.07\right) \times 10^{-6} \,, \\
{\cal B}\left(\Xibm \to \Lz(1520)\Km\right)       & = & \left(0.76 \pm 0.09 \pm 0.08 \pm 0.30\right) \times 10^{-6} \,, \\
{\cal B}\left(\Xibm \to \Lz(1670)\Km\right)       & = & \left(0.45 \pm 0.07 \pm 0.13 \pm 0.18\right) \times 10^{-6} \,, \\
{\cal B}\left(\Xibm \to \Sigmares(1775)\Km\right) & = & \left(0.22 \pm 0.08 \pm 0.09 \pm 0.09\right) \times 10^{-6} \,, \\
{\cal B}\left(\Xibm \to \Sigmares(1915)\Km\right) & = & \left(0.26 \pm 0.09 \pm 0.21 \pm 0.10\right) \times 10^{-6} \,,
\end{eqnarray*}
where the uncertainties are statistical, systematic and due to the knowledge of ${\cal B}\left( \Xibm\to \proton \Km\Km \right)$, respectively.

\section{Summary}
\label{sec:summary}

The structure of \XibToKKp\ decays has been studied through an amplitude analysis.
This is the first amplitude analysis of any \bquark-baryon decay mode allowing for \CP-violation effects.
The analysis uses $pp$ collision data recorded with the \lhcb detector, corresponding to integrated luminosities 
of $1 \invfb$ at $\sqrt{s} = 7 \tev$, $2 \invfb$ at $\sqrt{s} = 8 \tev$ and $2 \invfb$ at $\sqrt{s} = 13 \tev$.
Due to the inclusion of more data and significantly improving the selection procedure compared to the previous study of this channel~\cite{LHCb-PAPER-2016-050}, a yield of about 460 signal decays within the $m(\KKp)$ signal region is obtained, with a signal to background ratio of about $2:1$.
A good description of the data is obtained with an amplitude model containing contributions from $\Sigmares(1385)$, $\Lz(1405)$, $\Lz(1520)$, $\Lz(1670)$, $\Sigmares(1775)$ and $\Sigmares(1915)$ resonances.
The \CP\ asymmetry for each contributing component is evaluated and no significant \CP-violation effect is observed.
The $\decay{\Xibm}{\Lz(1520)\Km}$ and $\decay{\Xibm}{\Lz(1670)\Km}$ decays are observed with significance greater than $5\,\sigma$, and their branching fractions measured, together with those of $\decay{\Xibm}{\Sigmares(1385)\Km}$, $\decay{\Xibm}{\Lz(1405)\Km}$, $\decay{\Xibm}{\Sigmares(1775)\Km}$ and $\decay{\Xibm}{\Sigmares(1915)\Km}$ decays.
No significant signal for \OmegabToKKp\ decays is found and an upper limit on the ratio of fragmentation and branching fractions of \OmegabToKKp\ and \XibToKKp\ decays is set.
With the substantially larger samples that are anticipated following the upgrade of \lhcb~\cite{LHCb-TDR-012,LHCb-PII-EoI}, it will be possible to reduce both statistical and systematic uncertainties on \CP-violation observables in three-body \bquark-baryon decays, and thereby test the Standard Model using the methods pioneered in this study.

\clearpage
\appendix
\section{Fit projections} 
\label{appd:fitprojects}

\begin{figure}[!hb]
\centering
\includegraphics[width=0.48\textwidth]{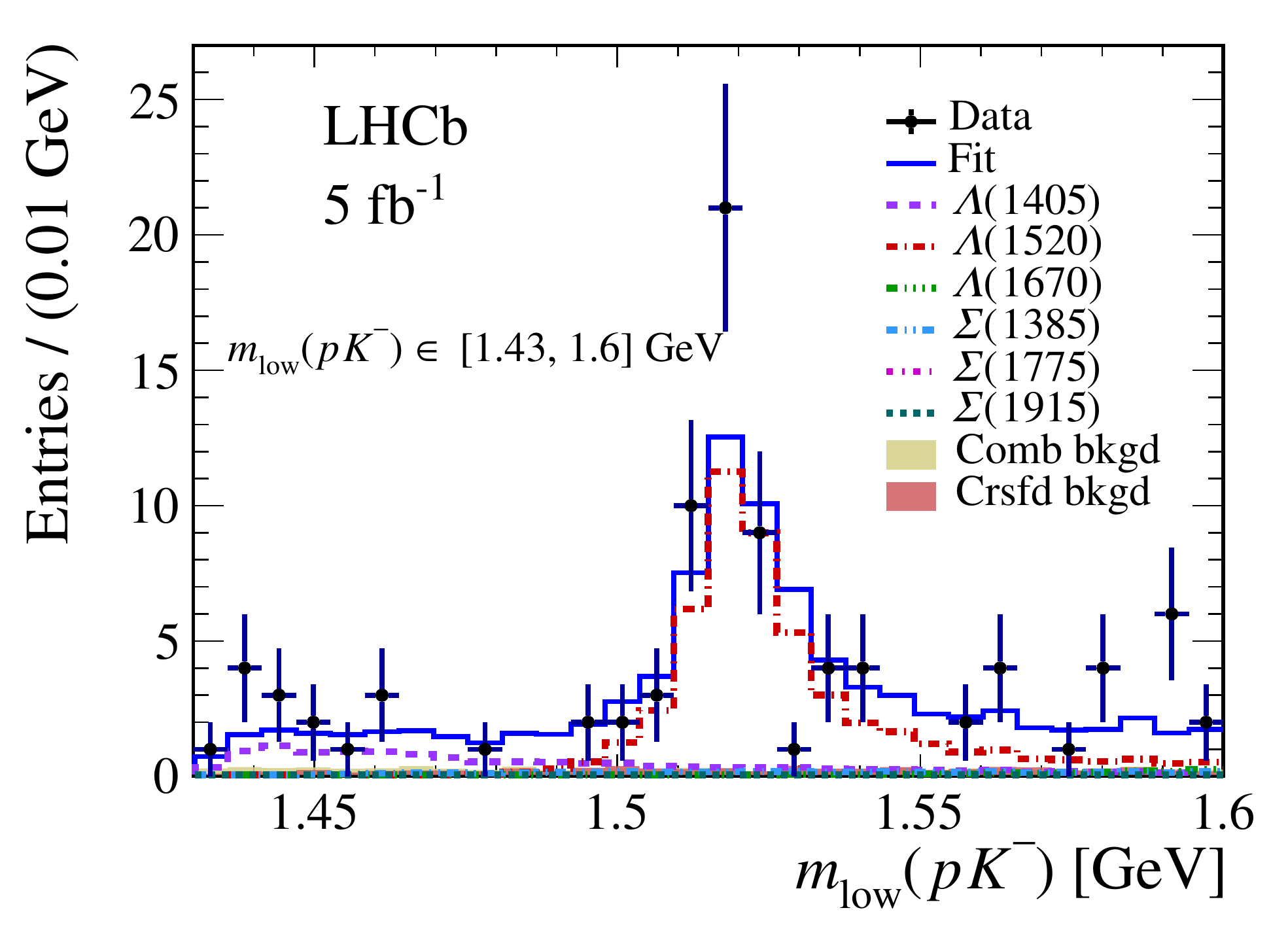} 
\includegraphics[width=0.48\textwidth]{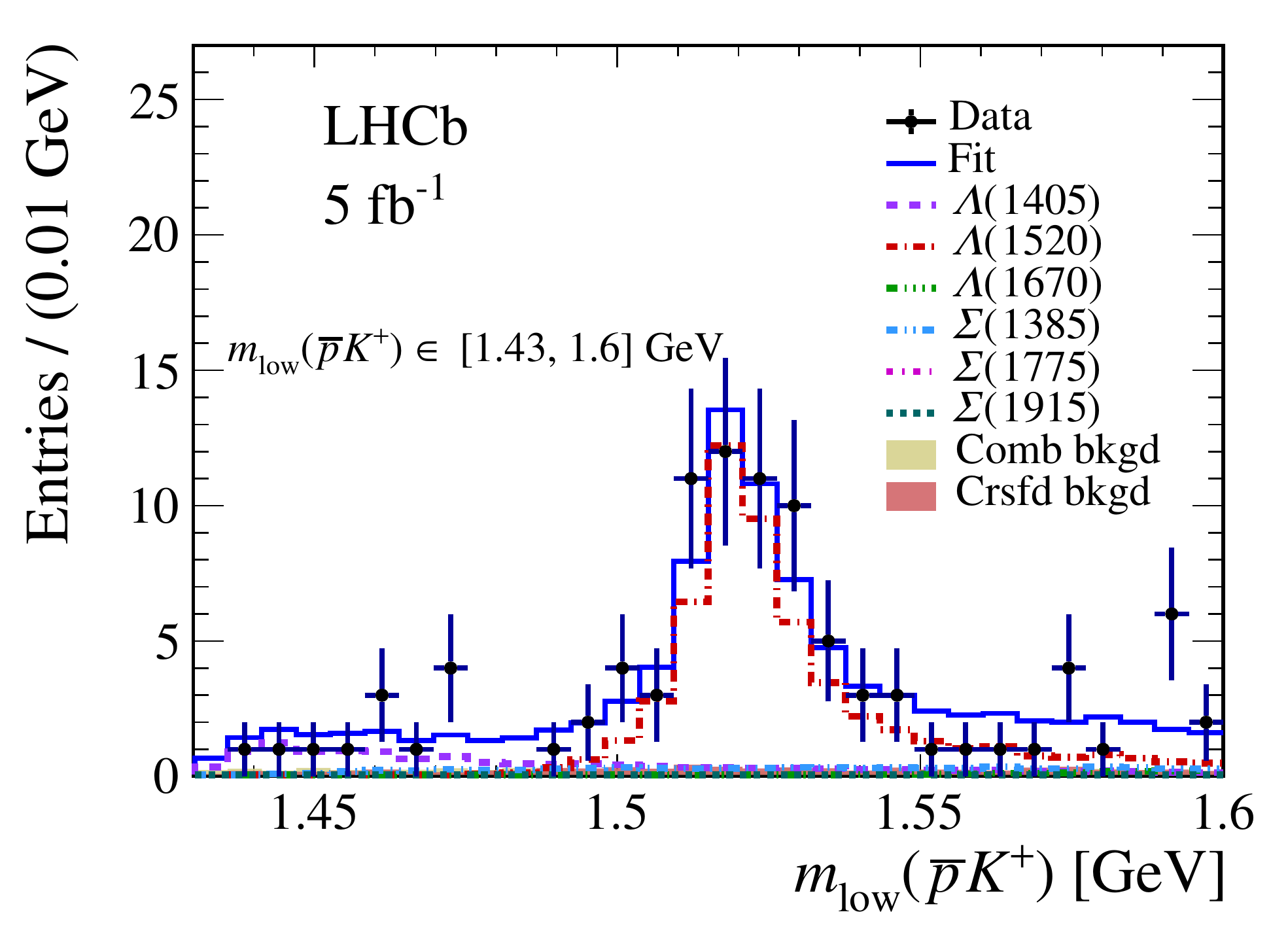} \\
\includegraphics[width=0.48\textwidth]{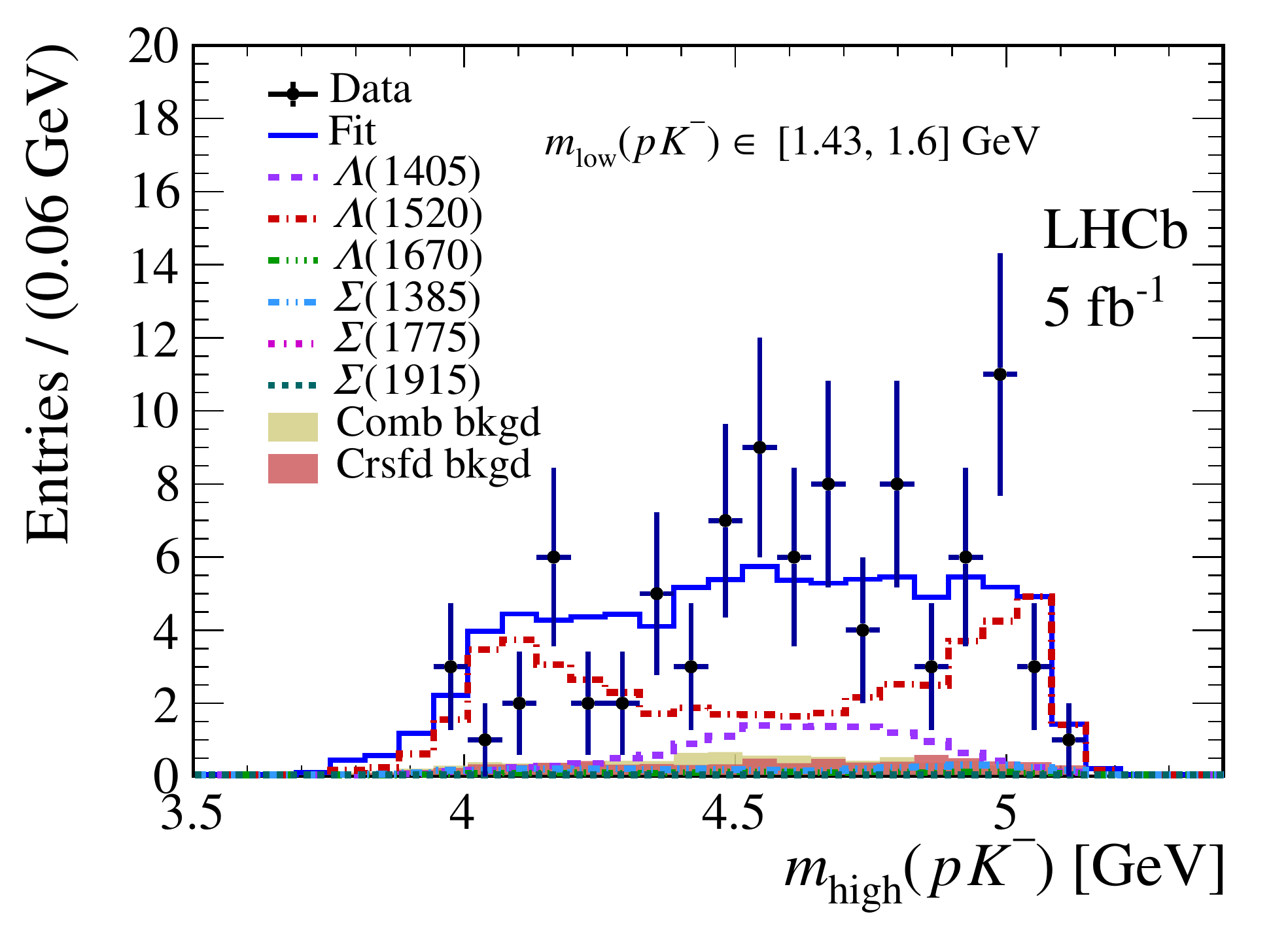} 
\includegraphics[width=0.48\textwidth]{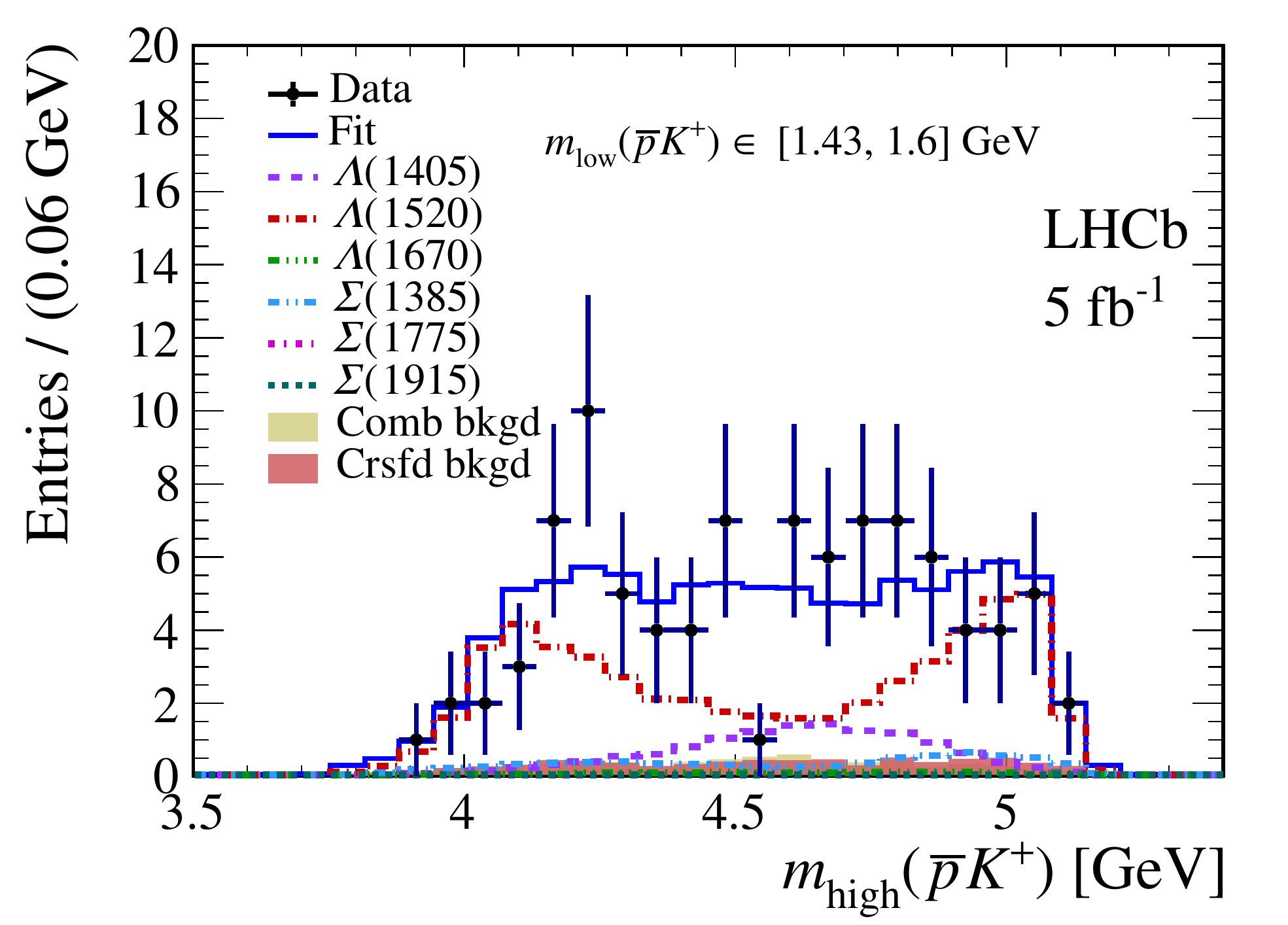} 
\caption{\small
   Distributions of (top)~$\mpklow$ and (bottom)~$\mpkhigh$, with $1.43 < \mpklow < 1.60 \gev$, for (left)~\Xibm\ and (right)~\Xibp\ candidates, with results of the fits superimposed.
  The total fit result is shown as the blue solid curve, with contributions from individual signal components and from combinatorial (Comb) and cross-feed (Crsfd) background shown as indicated in the legend.
}
\label{fig:mpklow_diffregions_low}
\end{figure}

\begin{figure}[!tb]
\centering
\includegraphics[width=0.48\textwidth]{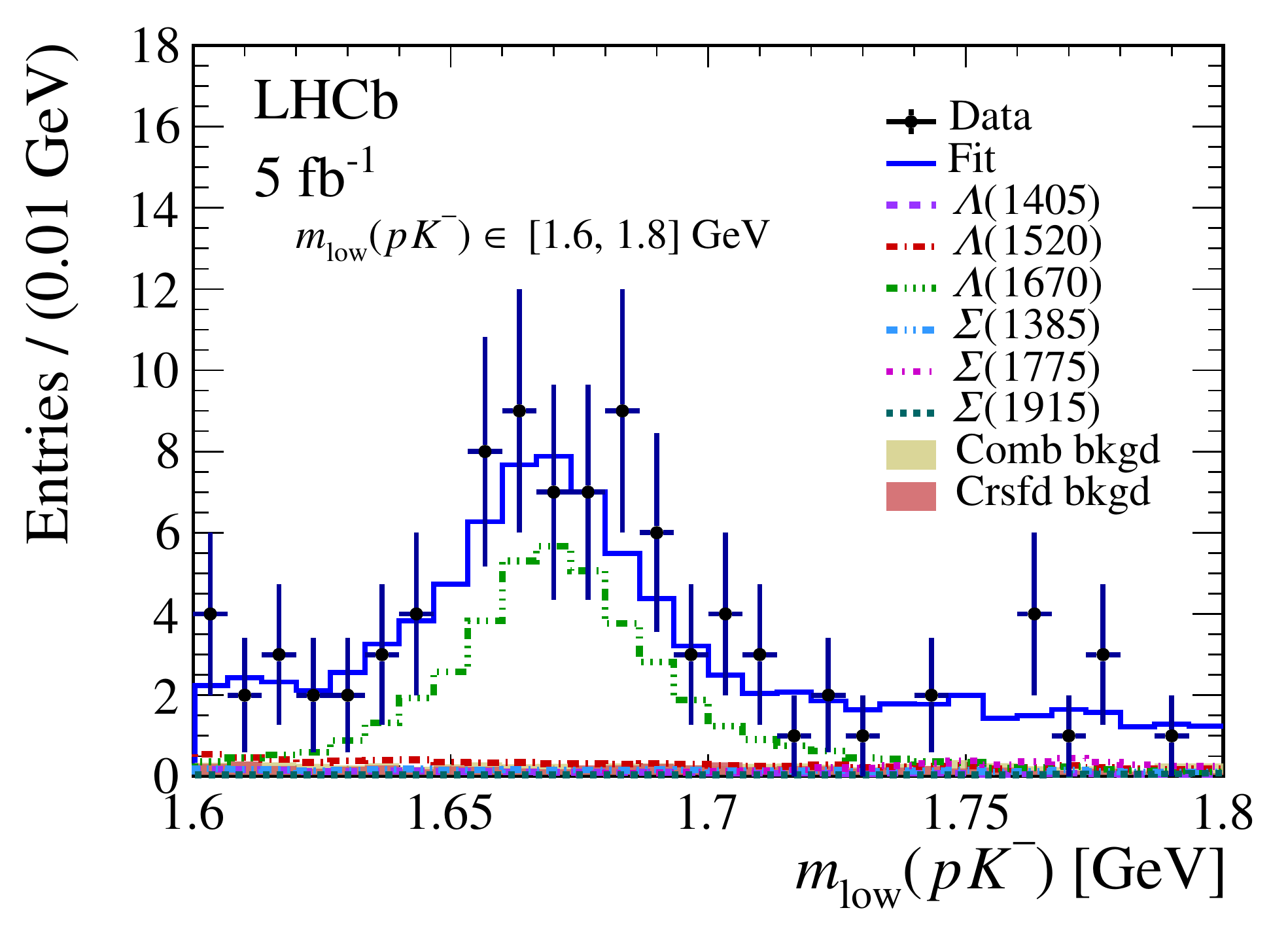} 
\includegraphics[width=0.48\textwidth]{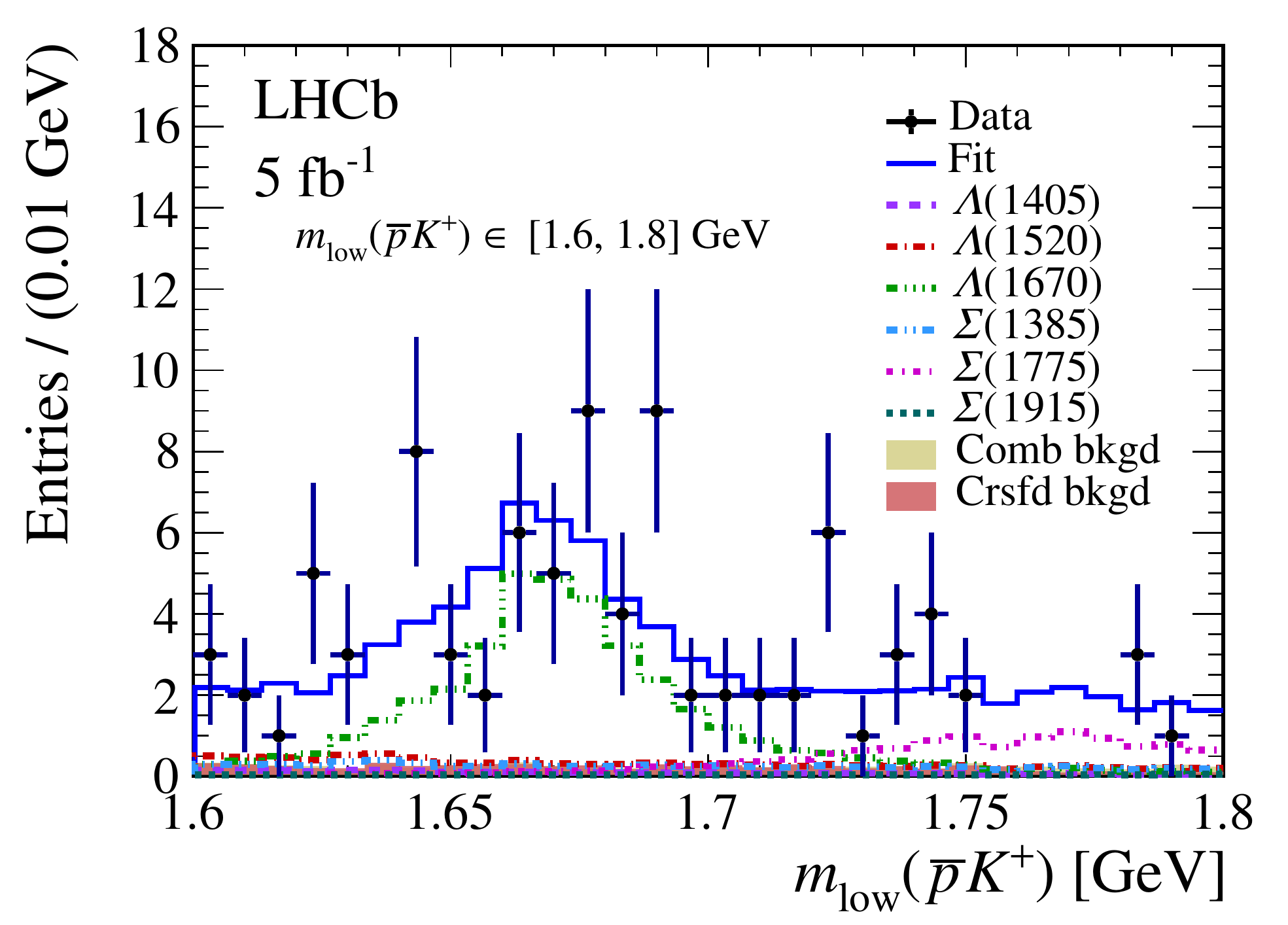} \\
\includegraphics[width=0.48\textwidth]{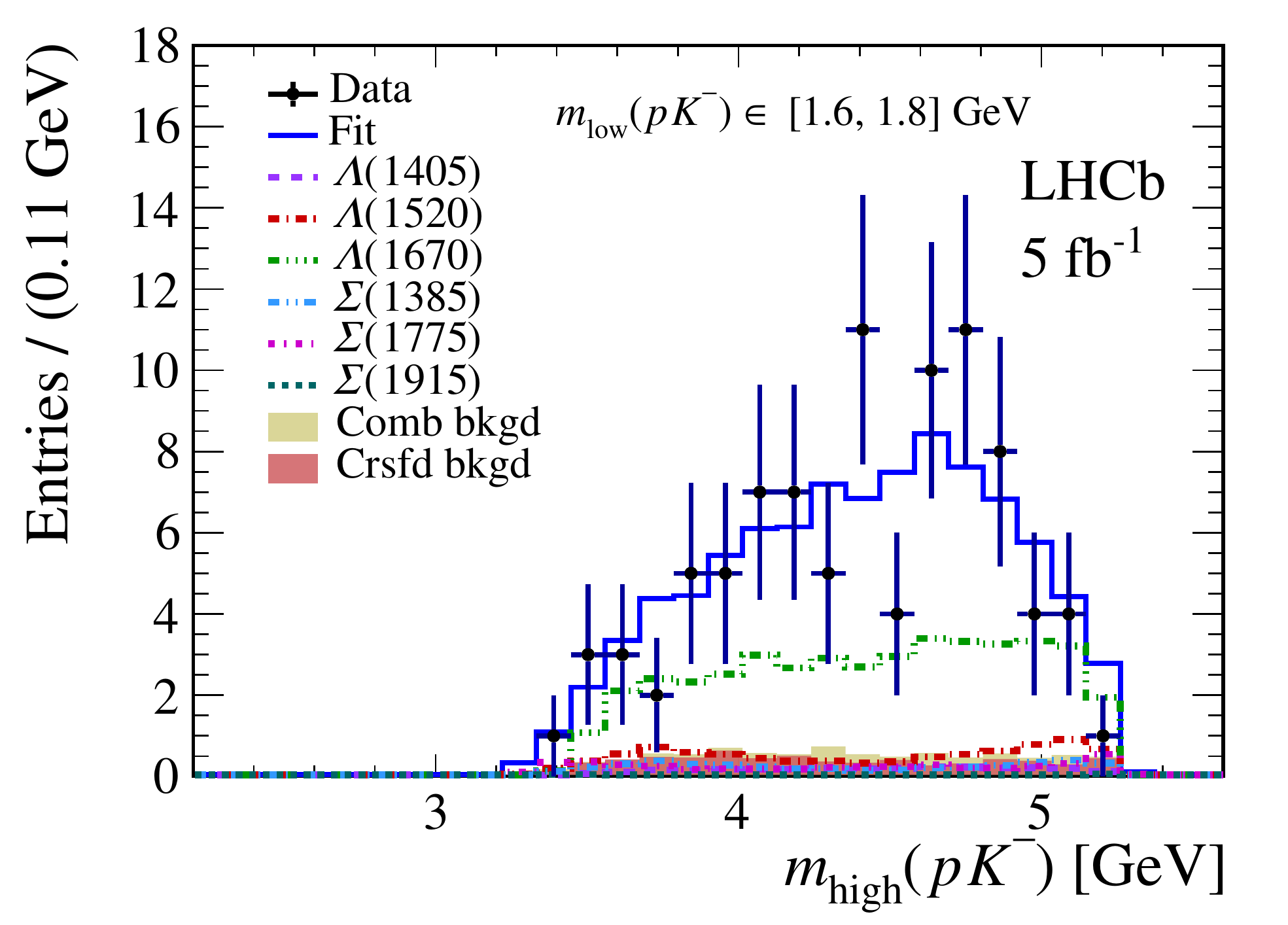} 
\includegraphics[width=0.48\textwidth]{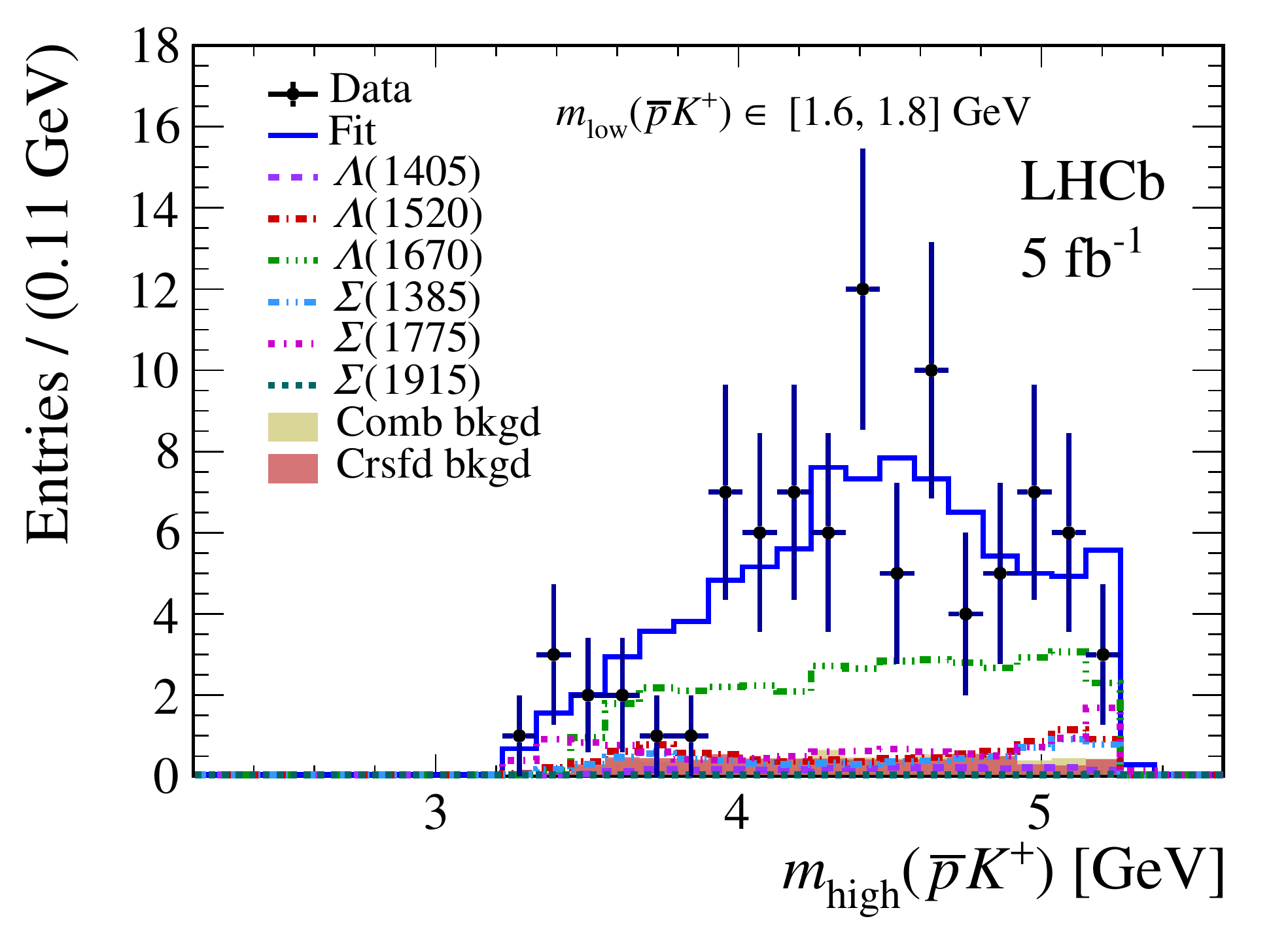} 
\caption{\small
  Distributions of (top)~$\mpklow$ and (bottom)~$\mpkhigh$, with $1.6 < \mpklow < 1.8 \gev$, for (left)~\Xibm\ and (right)~\Xibp\ candidates, with results of the fits superimposed.
  The total fit result is shown as the blue solid curve, with contributions from individual signal components and from combinatorial (Comb) and cross-feed (Crsfd) background shown as indicated in the legend.
}
\label{fig:mpklow_diffregions_mid1}
\end{figure}

\begin{figure}[!tb]
\centering
\includegraphics[width=0.48\textwidth]{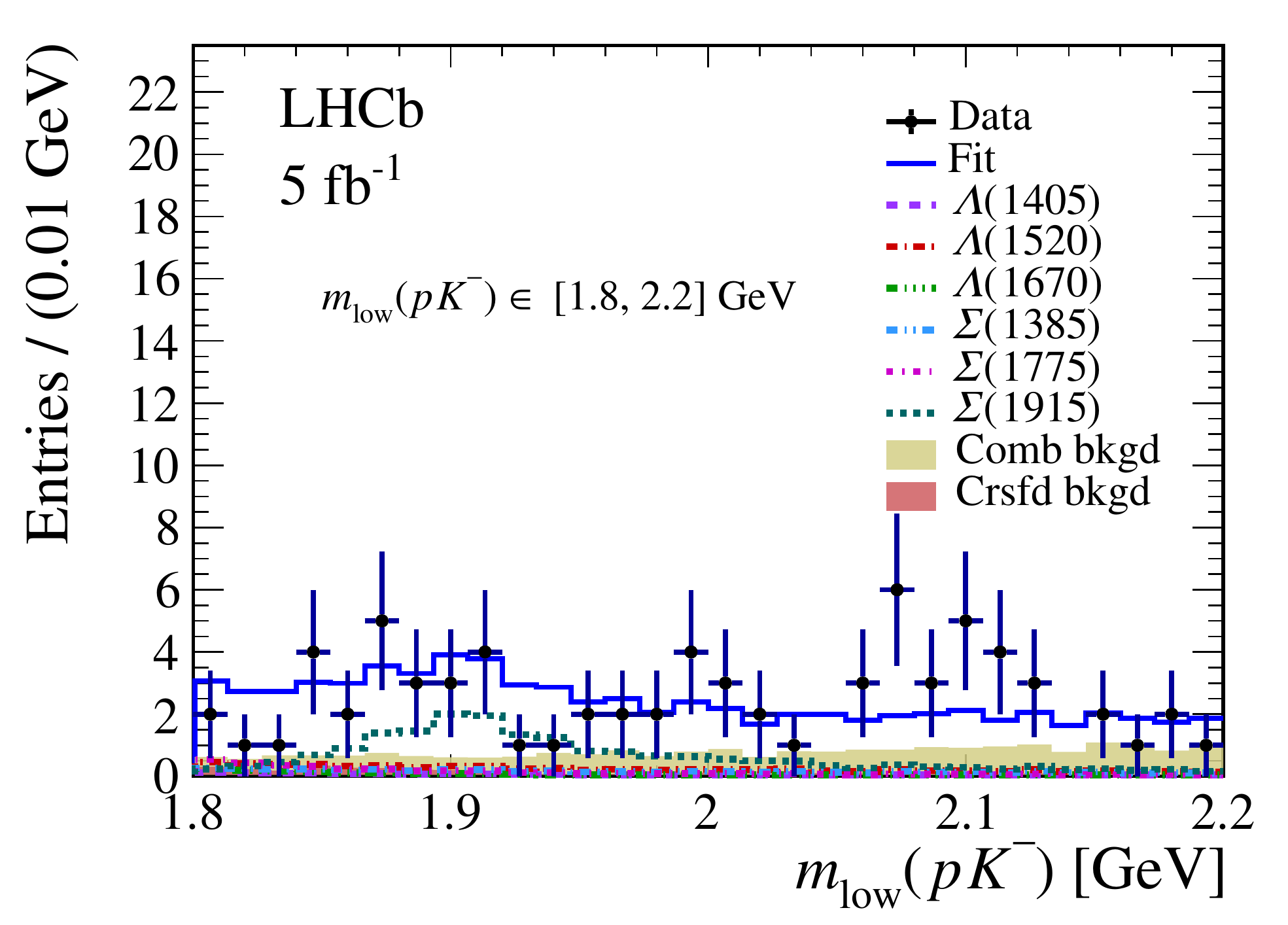} 
\includegraphics[width=0.48\textwidth]{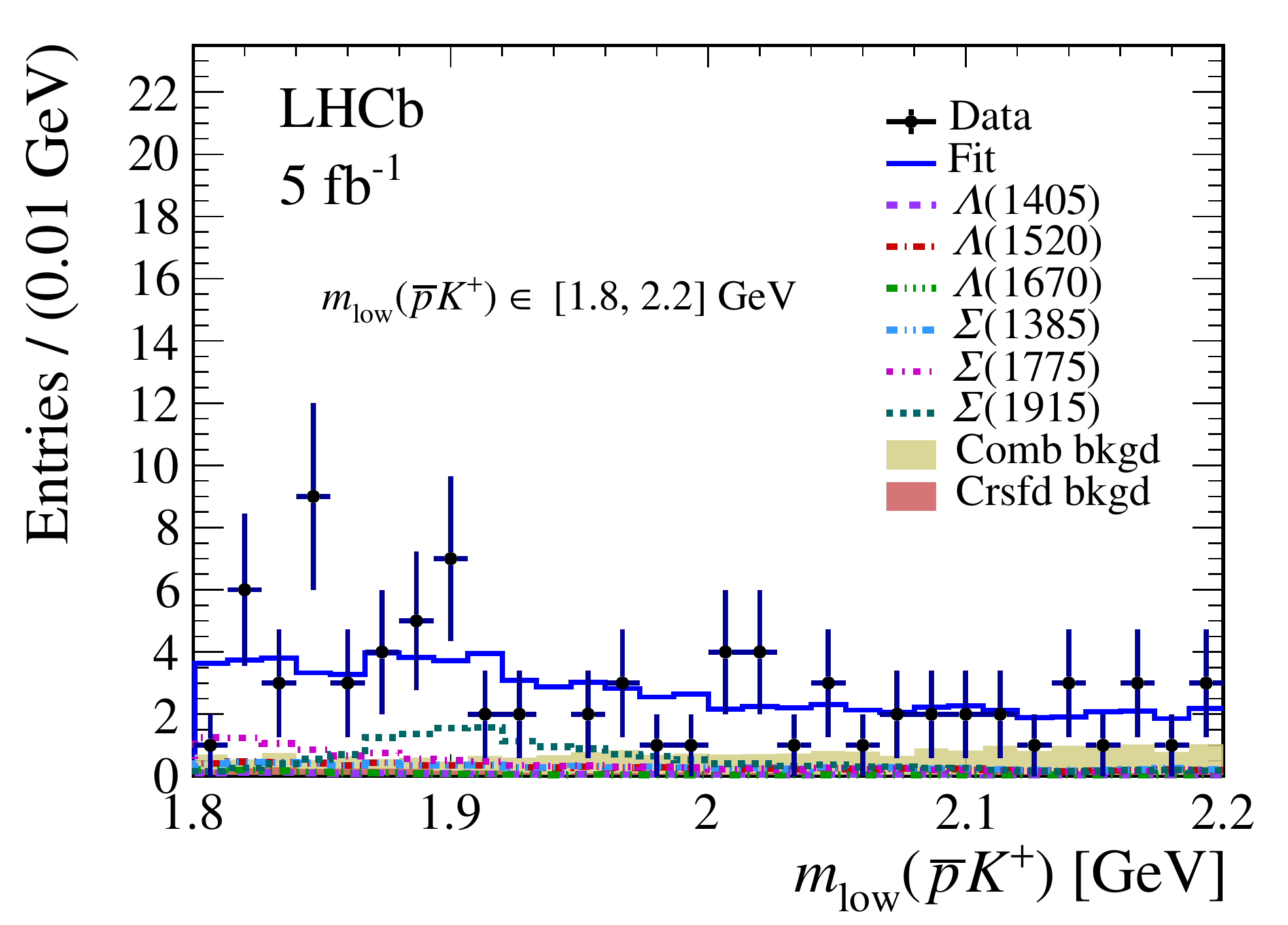} \\
\includegraphics[width=0.48\textwidth]{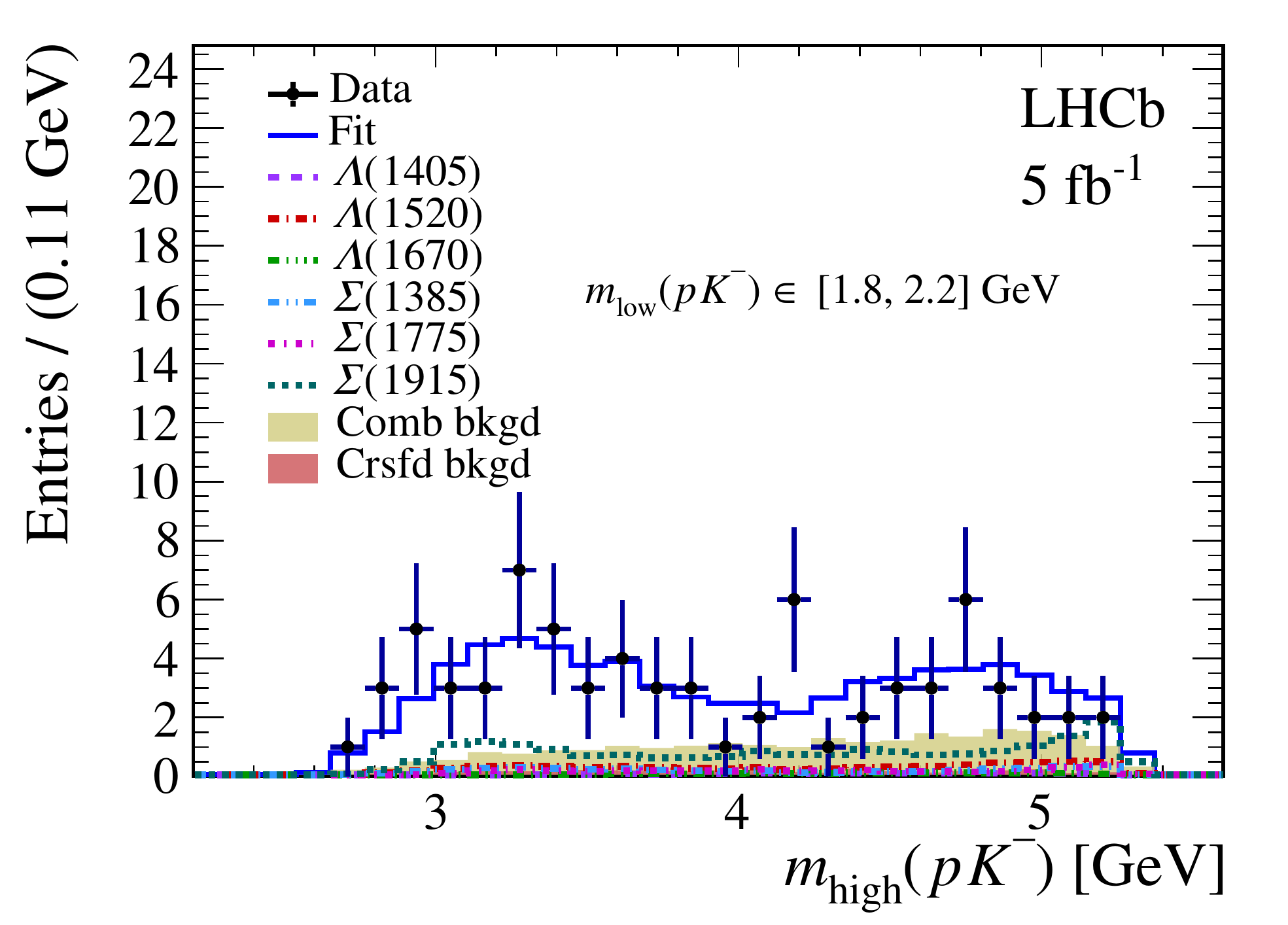} 
\includegraphics[width=0.48\textwidth]{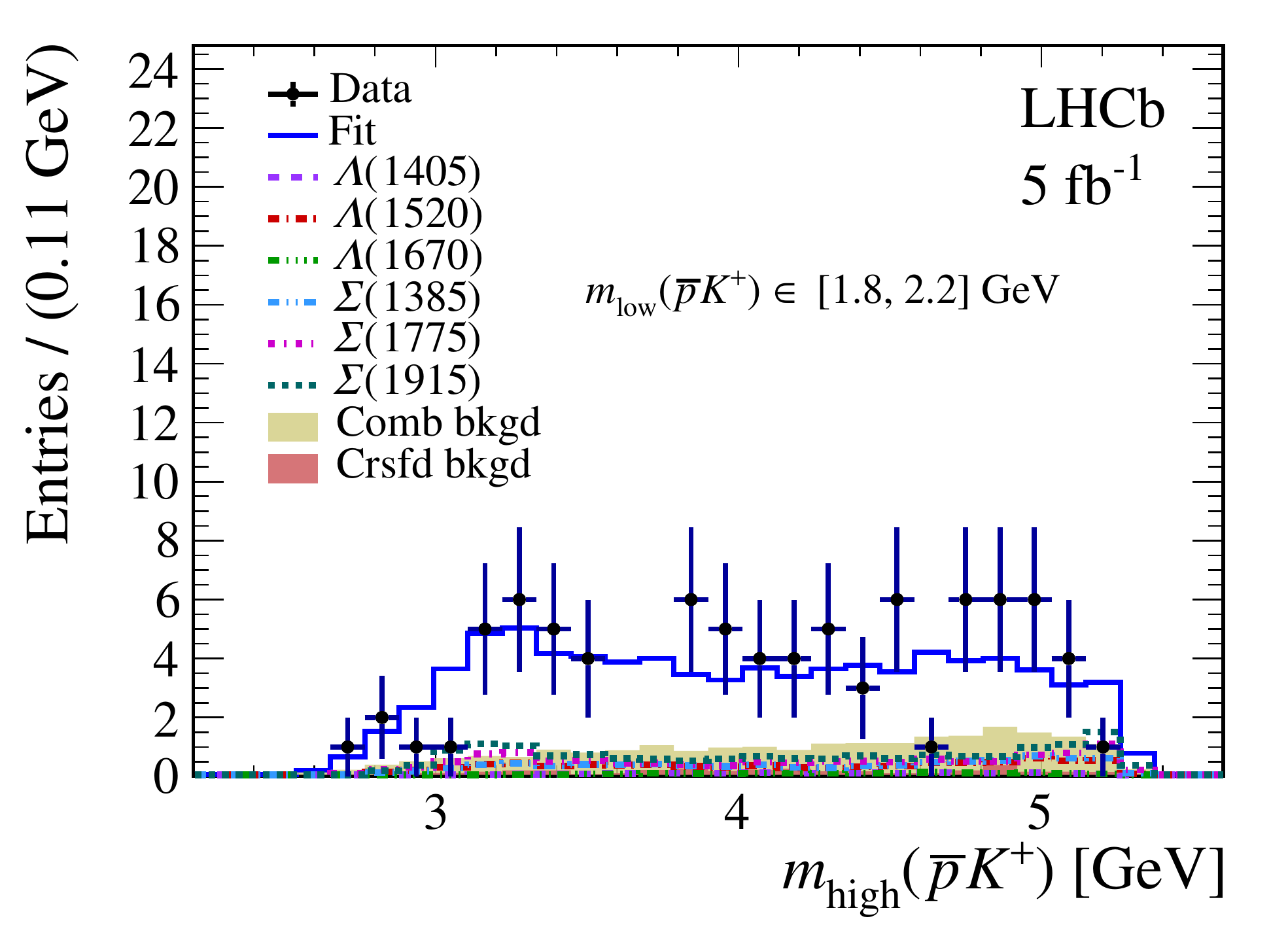} 
\caption{\small
  Distributions of (top)~$\mpklow$ and (bottom)~$\mpkhigh$, with $1.8 < \mpklow < 2.2 \gev$, for (left)~\Xibm\ and (right)~\Xibp\ candidates, with results of the fits superimposed.
  The total fit result is shown as the blue solid curve, with contributions from individual signal components and from combinatorial (Comb) and cross-feed (Crsfd) background shown as indicated in the legend.
}
\label{fig:mpklow_diffregions_mid2}
\end{figure}

\begin{figure}[!tb]
\centering
\includegraphics[width=0.48\textwidth]{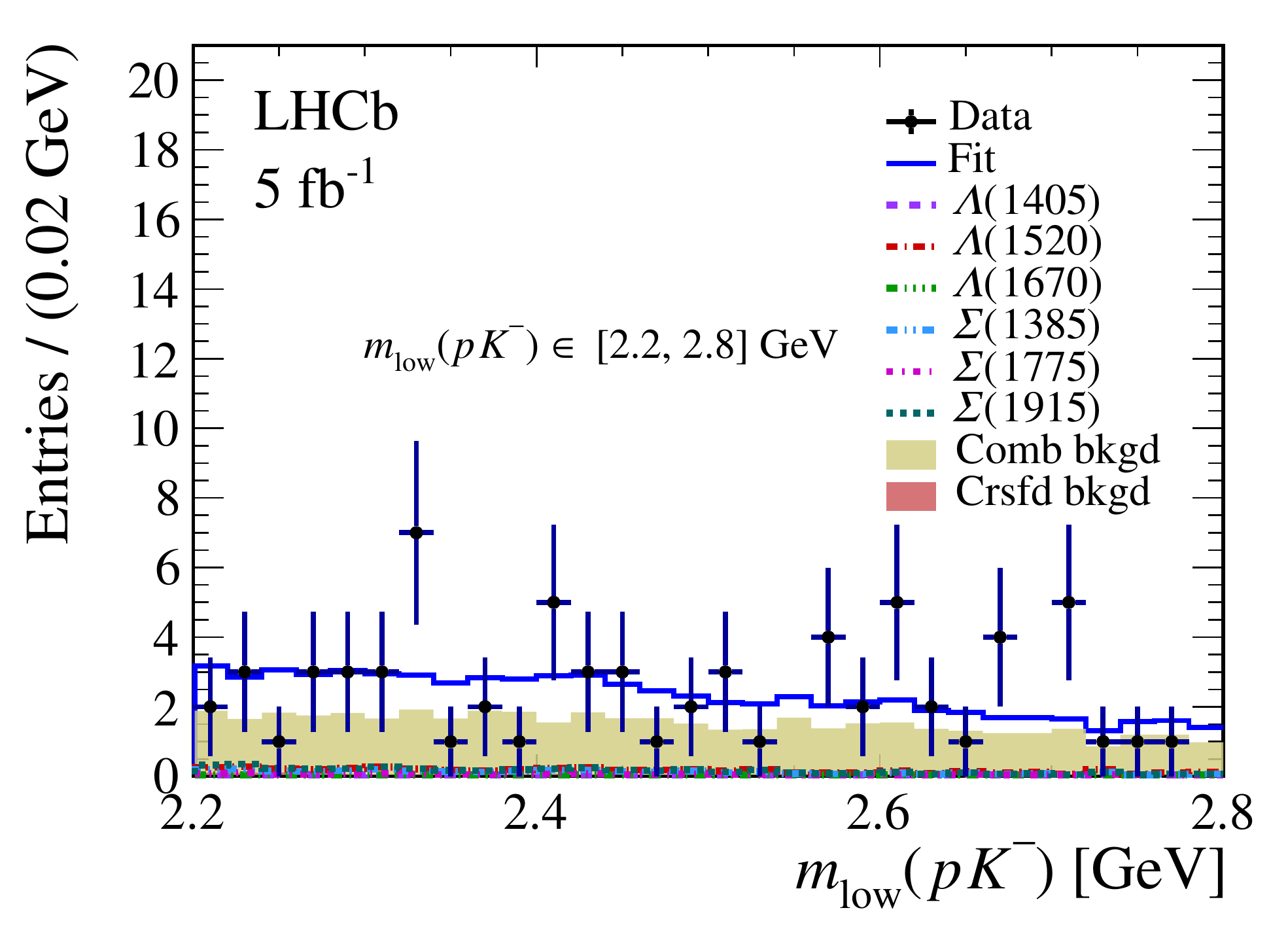} 
\includegraphics[width=0.48\textwidth]{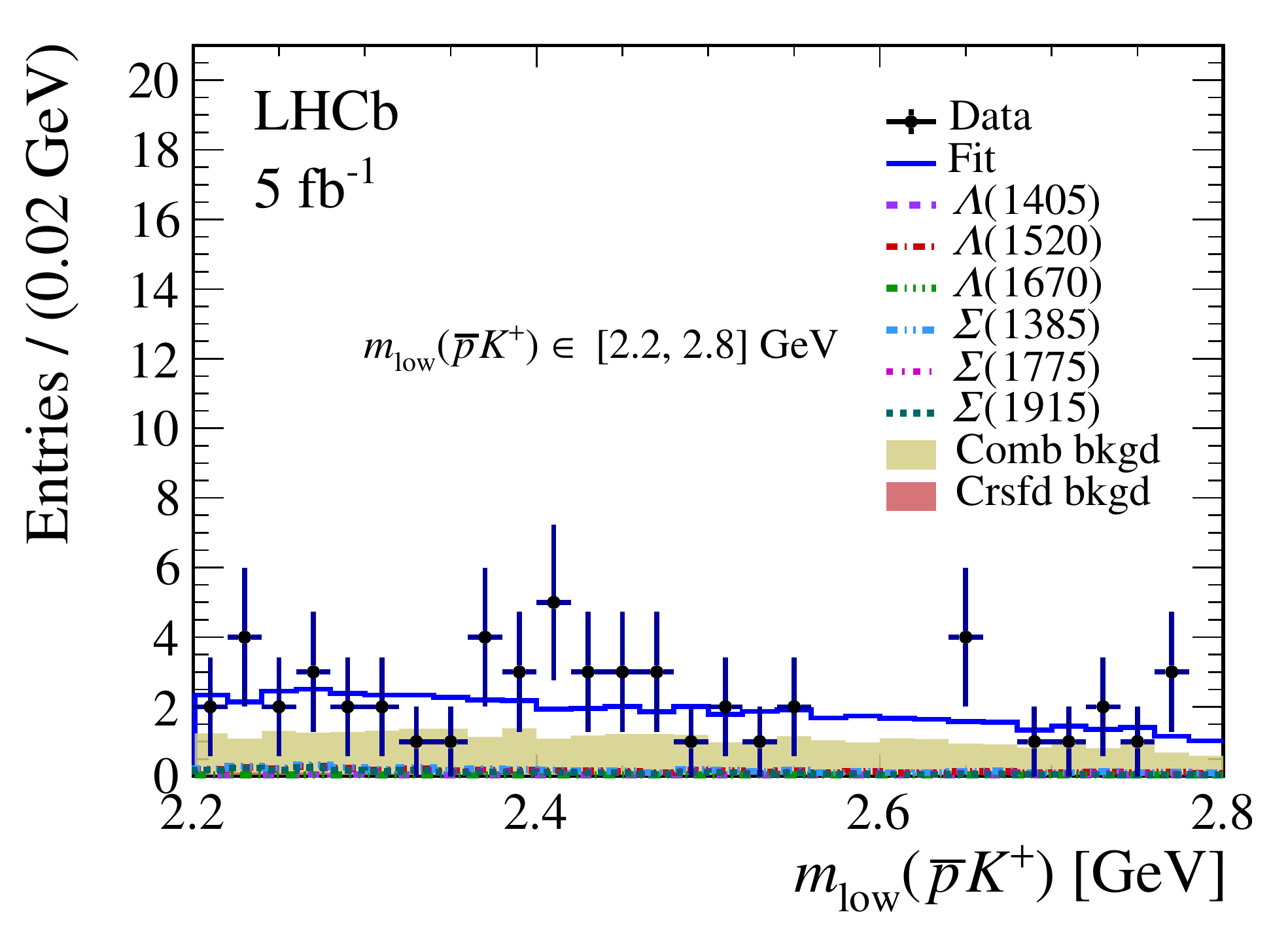} \\
\includegraphics[width=0.48\textwidth]{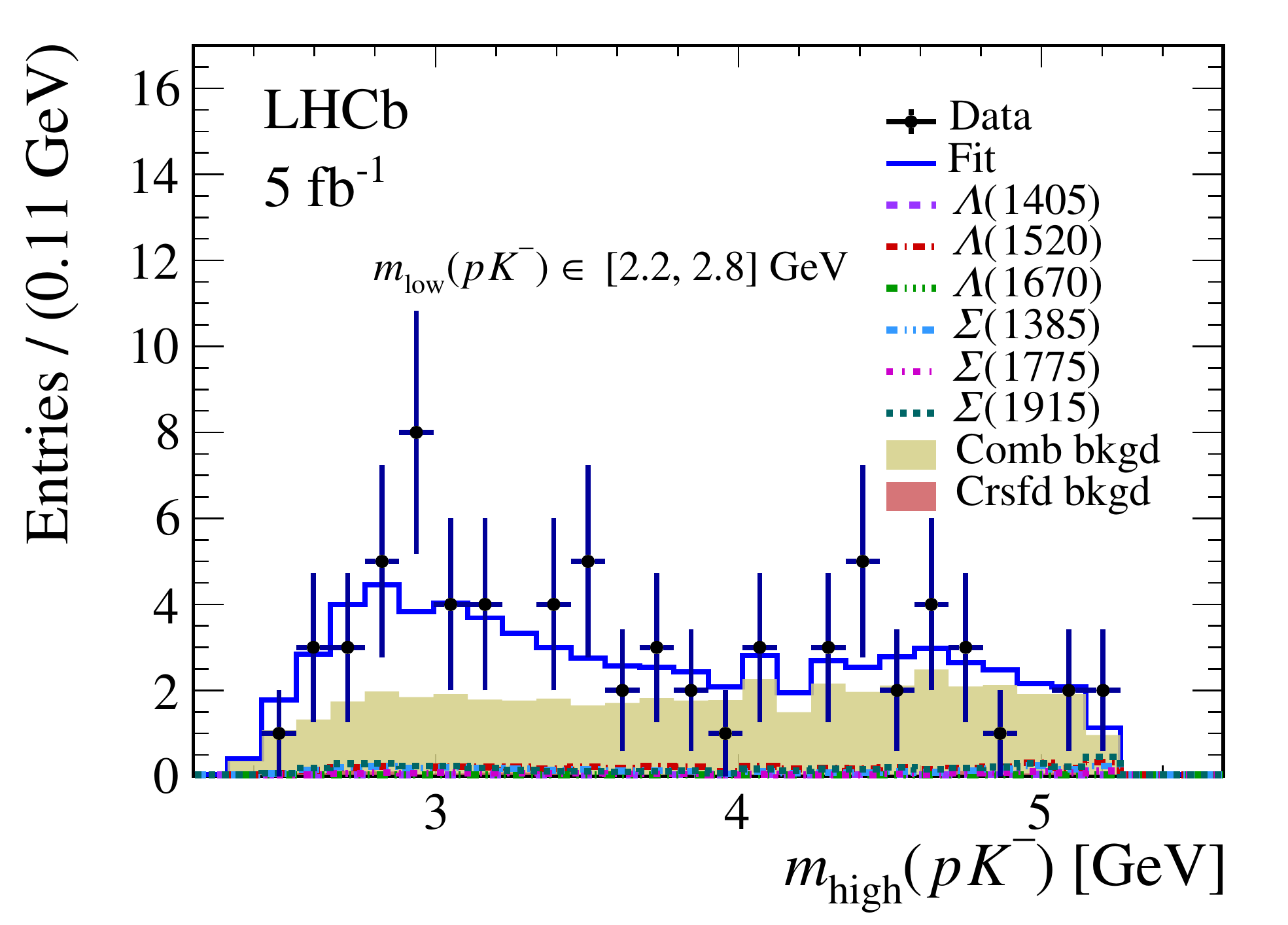} 
\includegraphics[width=0.48\textwidth]{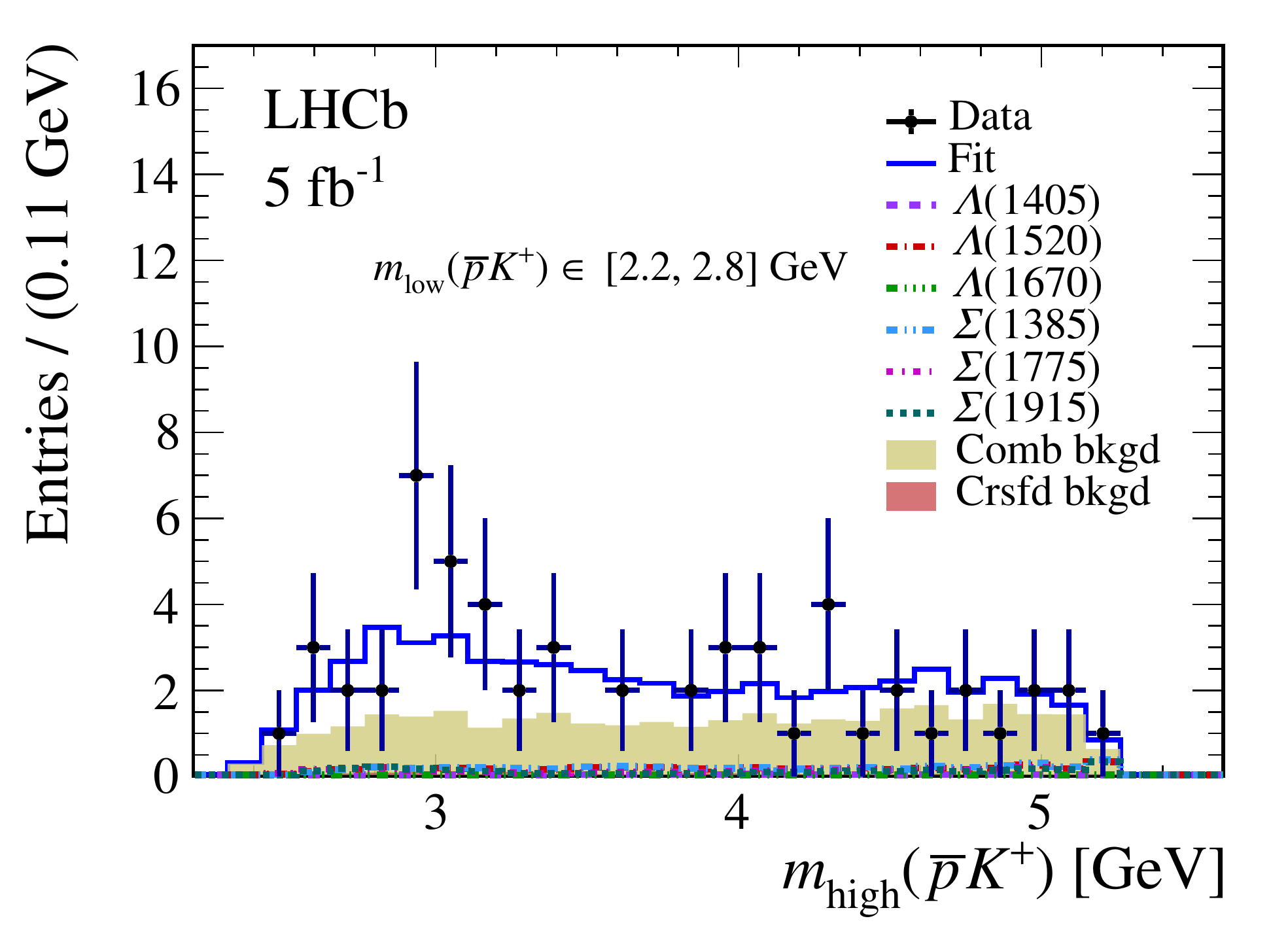} 
\caption{\small
  Distributions of (top)~$\mpklow$ and (bottom)~$\mpkhigh$, with $2.2 < \mpklow < 2.8 \gev$, for (left)~\Xibm\ and (right)~\Xibp\ candidates, with results of the fits superimposed.
  The total fit result is shown as the blue solid curve, with contributions from individual signal components and from combinatorial (Comb) and cross-feed (Crsfd) background shown as indicated in the legend.
}
\label{fig:mpklow_diffregions_mid3}
\end{figure}

\begin{figure}[!tb]
\centering
\includegraphics[width=0.48\textwidth]{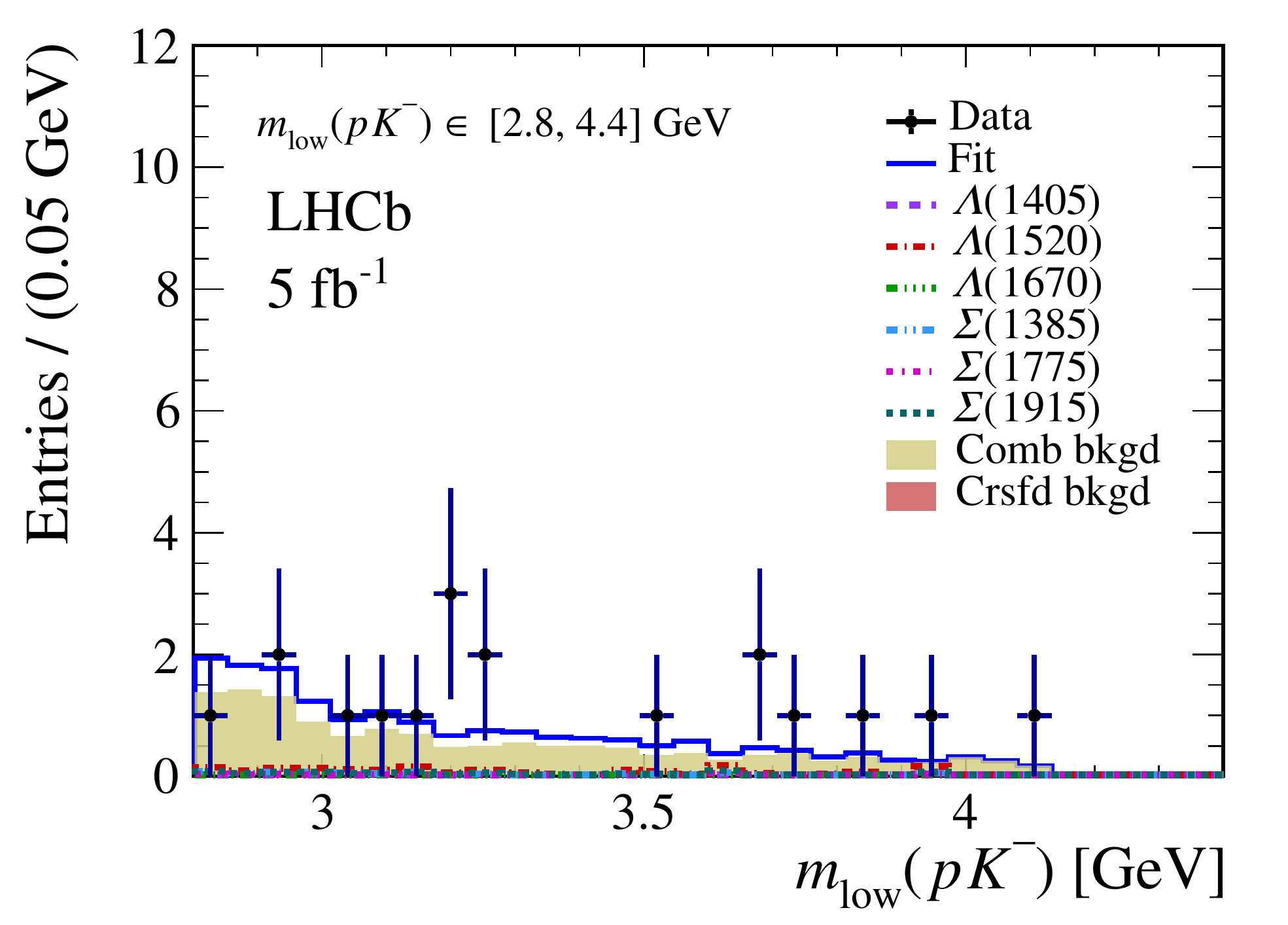} 
\includegraphics[width=0.48\textwidth]{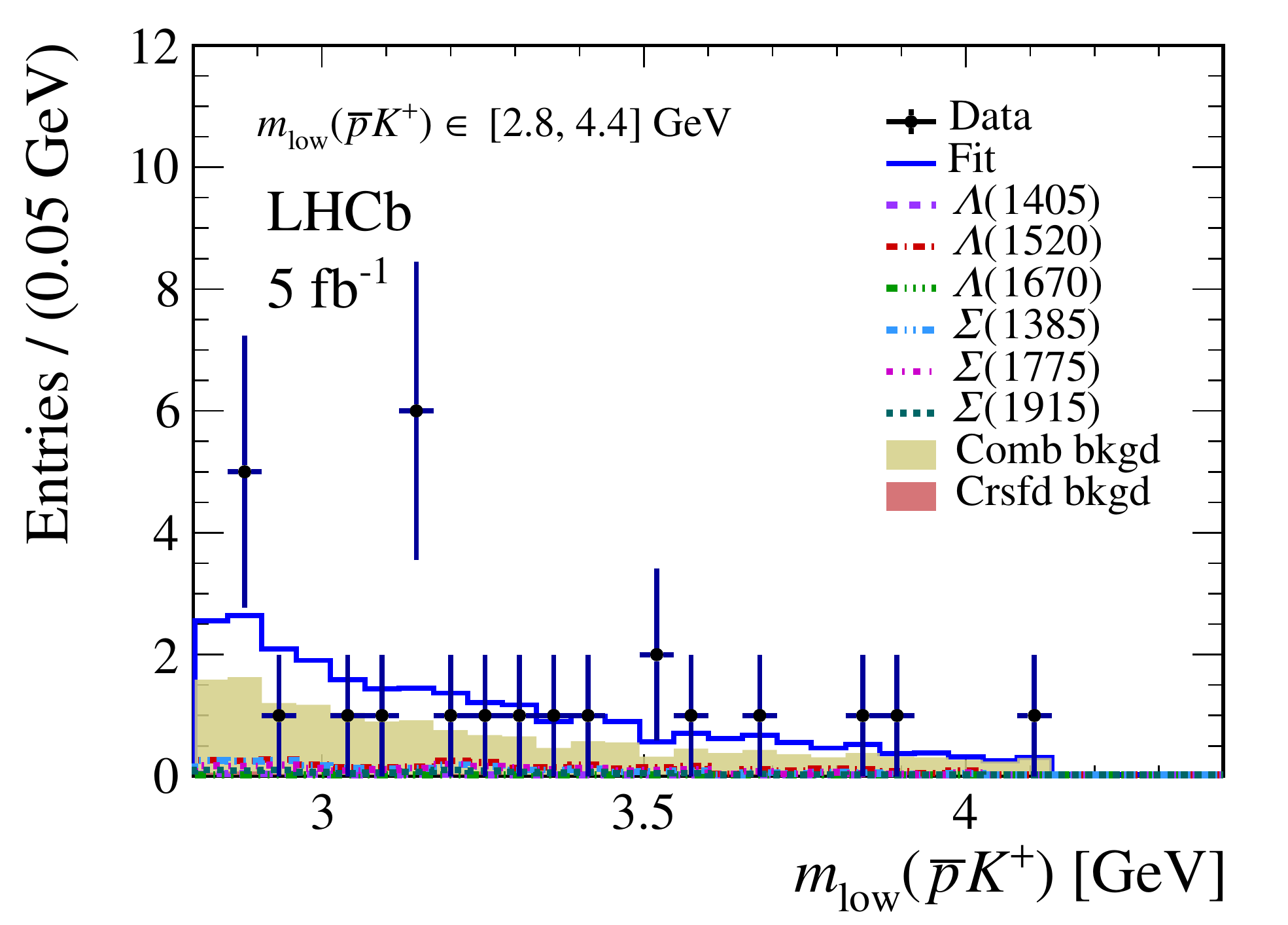} \\
\includegraphics[width=0.48\textwidth]{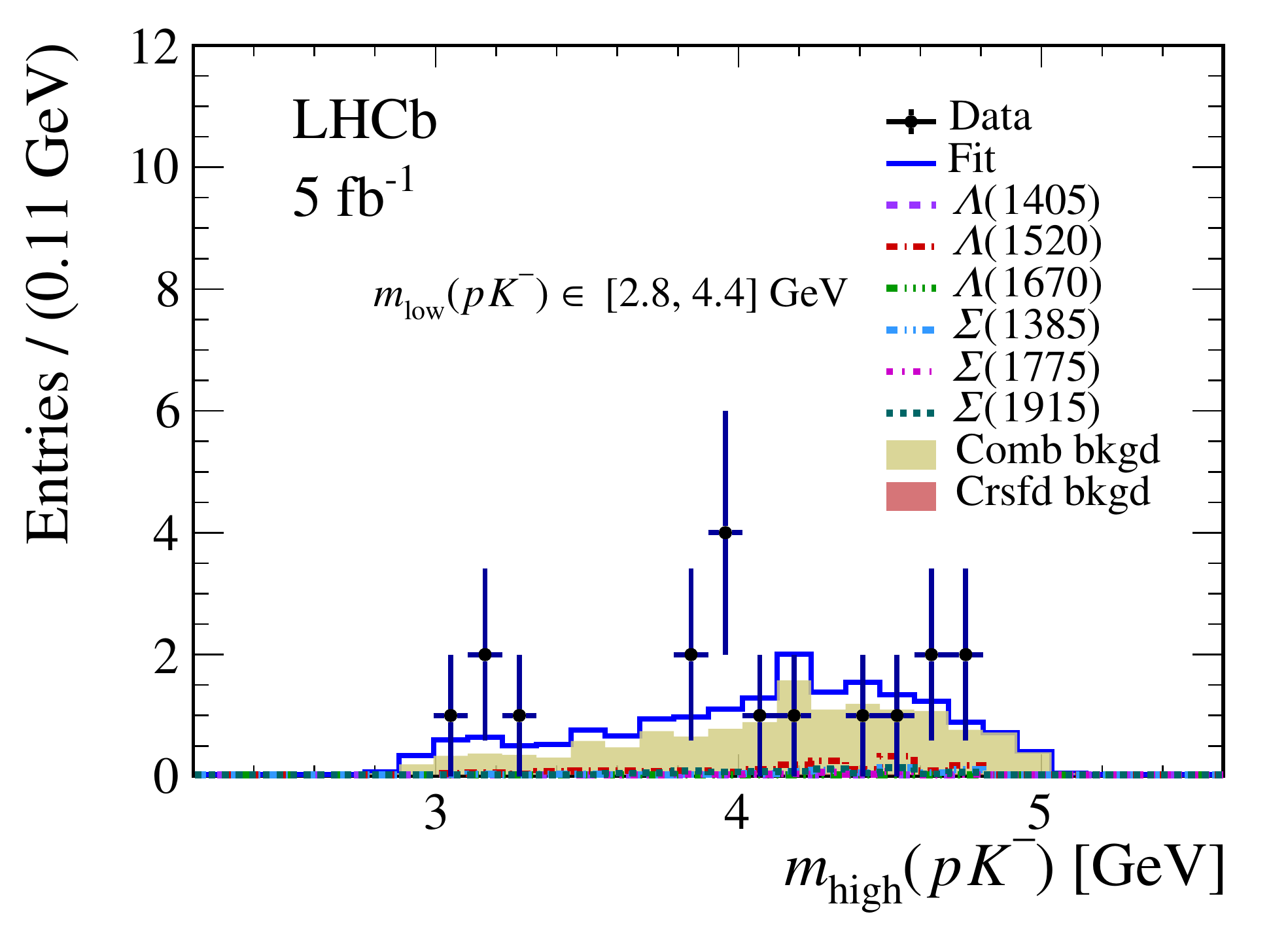}
\includegraphics[width=0.48\textwidth]{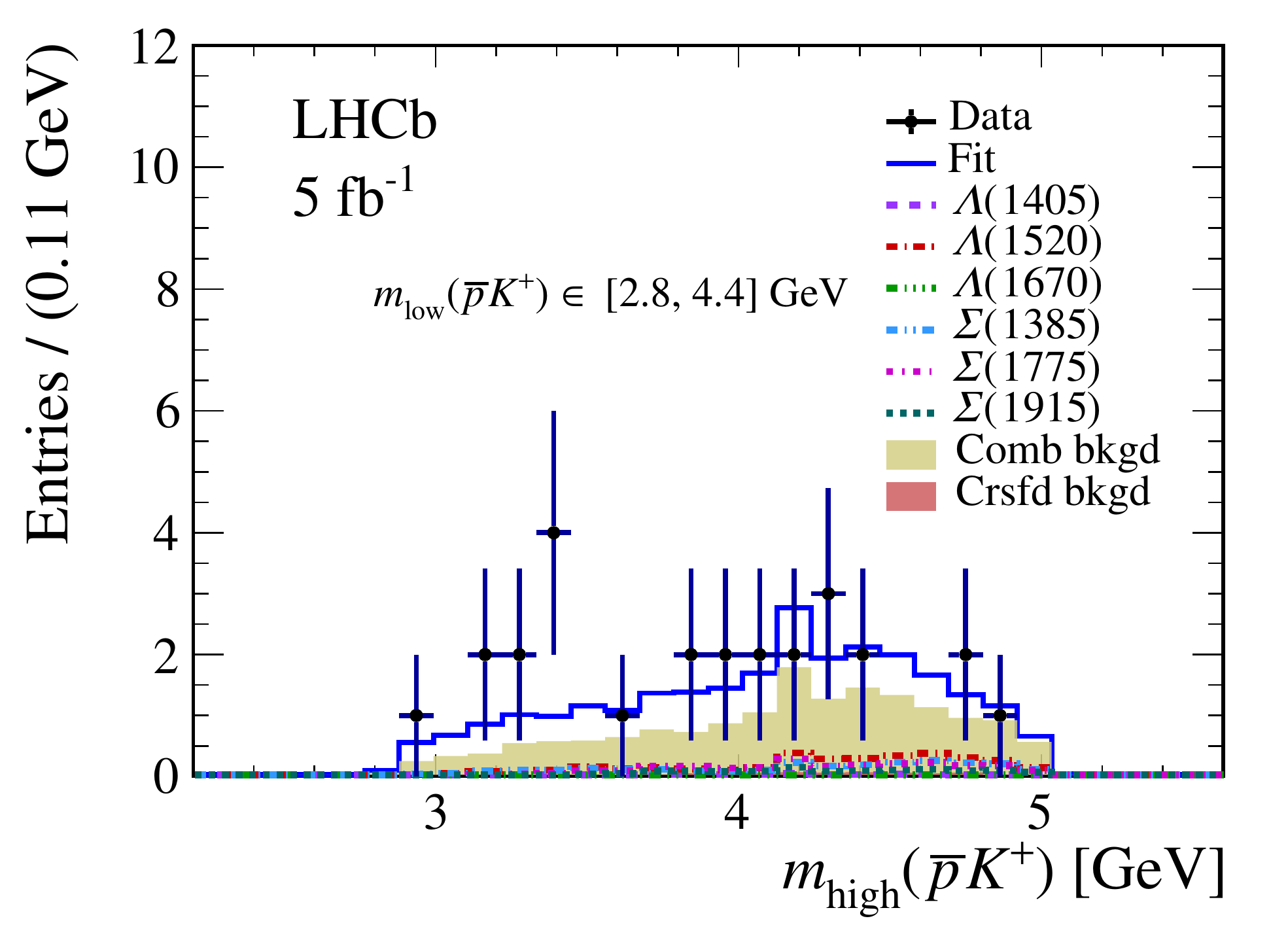}
\caption{\small
  Distributions of (top)~$\mpklow$ and (bottom)~$\mpkhigh$, with $\mpklow > 2.8 \gev$, for (left)~\Xibm\ and (right)~\Xibp\ candidates, with results of the fits superimposed.
  The total fit result is shown as the blue solid curve, with contributions from individual signal components and from combinatorial (Comb) and cross-feed (Crsfd) background shown as indicated in the legend.
}
\label{fig:mpklow_diffregions_high}
\end{figure}

\begin{figure}[!tb]
\centering
\includegraphics[width=0.48\textwidth]{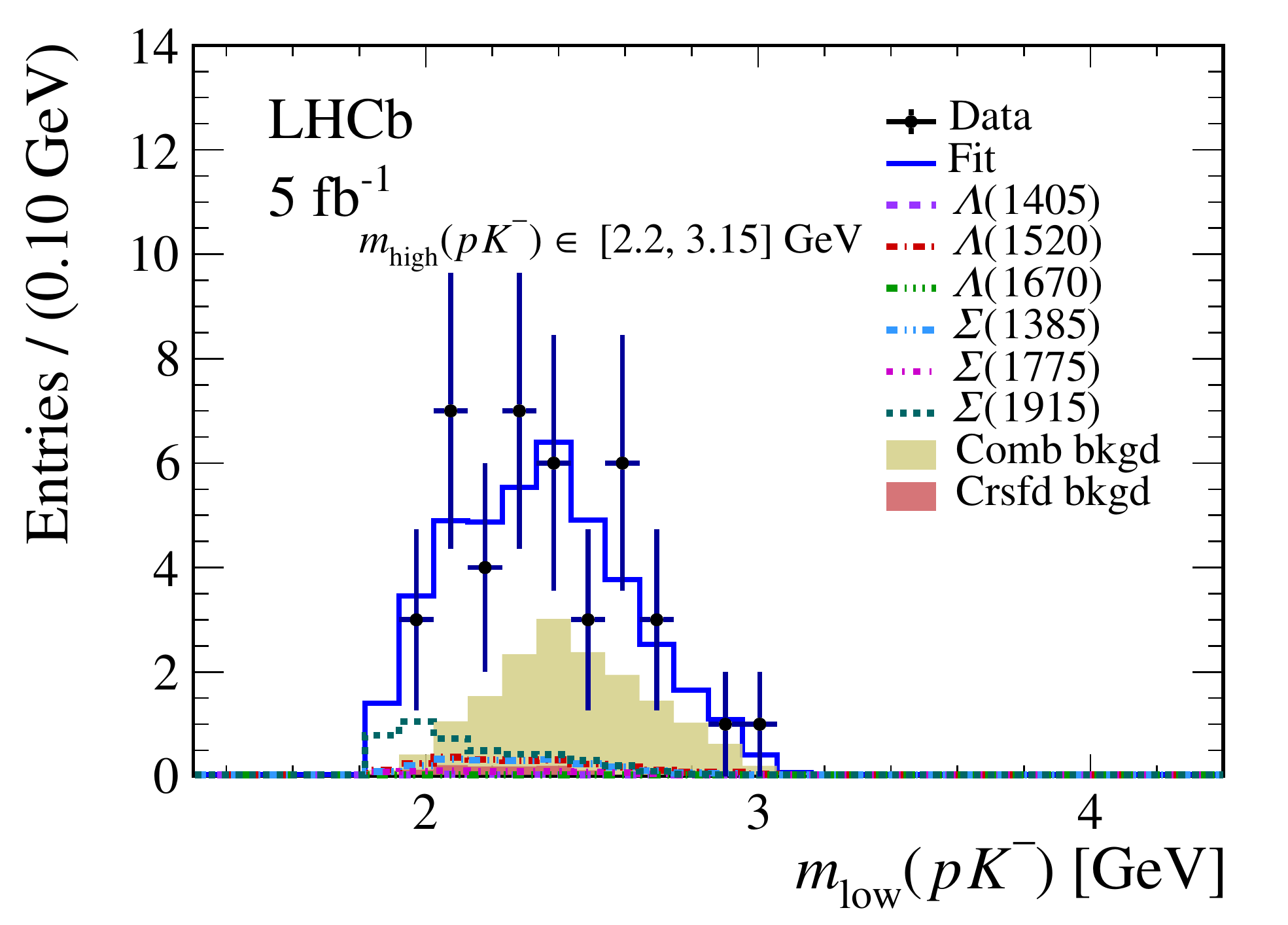} 
\includegraphics[width=0.48\textwidth]{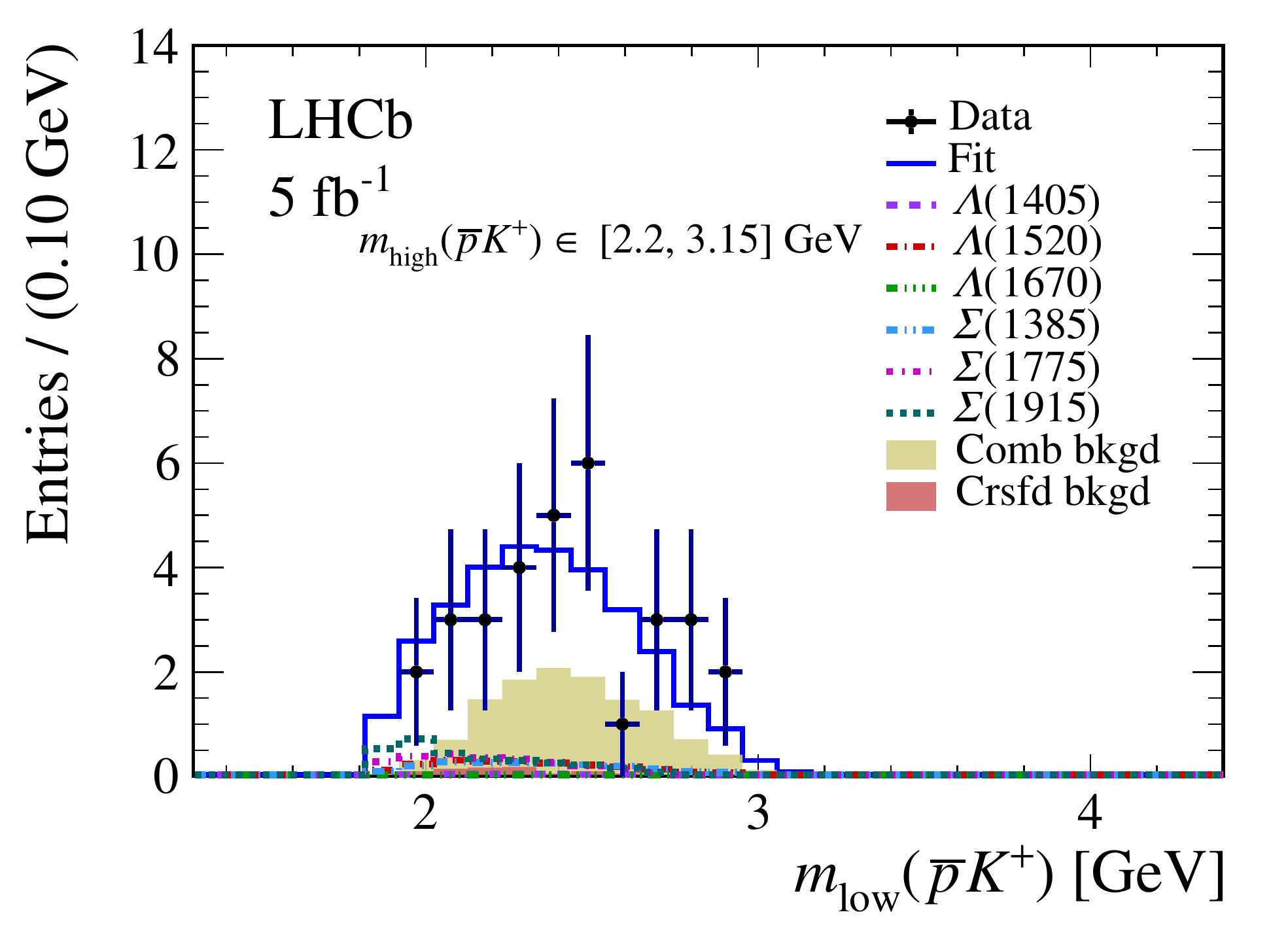} \\
\includegraphics[width=0.48\textwidth]{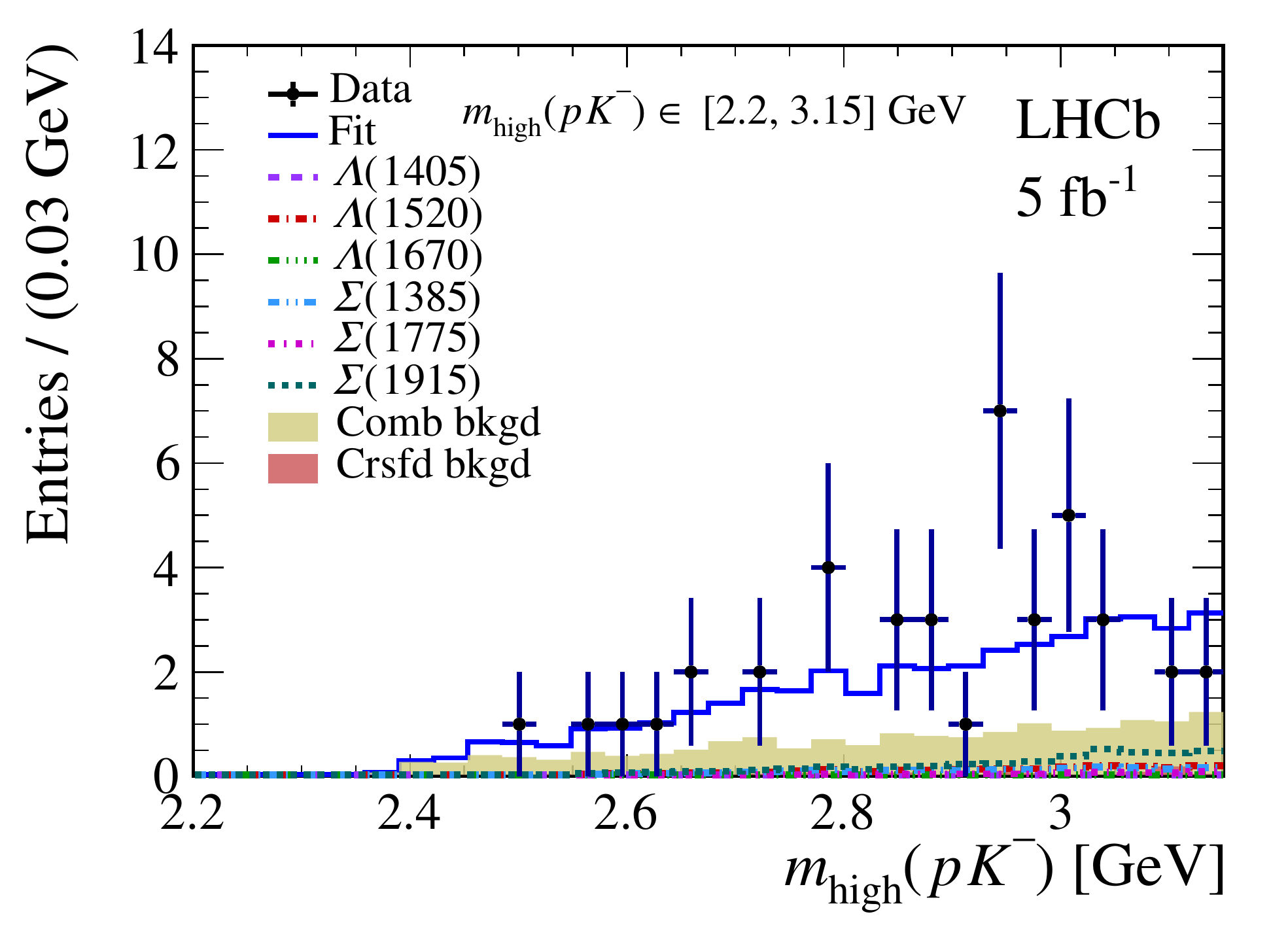}
\includegraphics[width=0.48\textwidth]{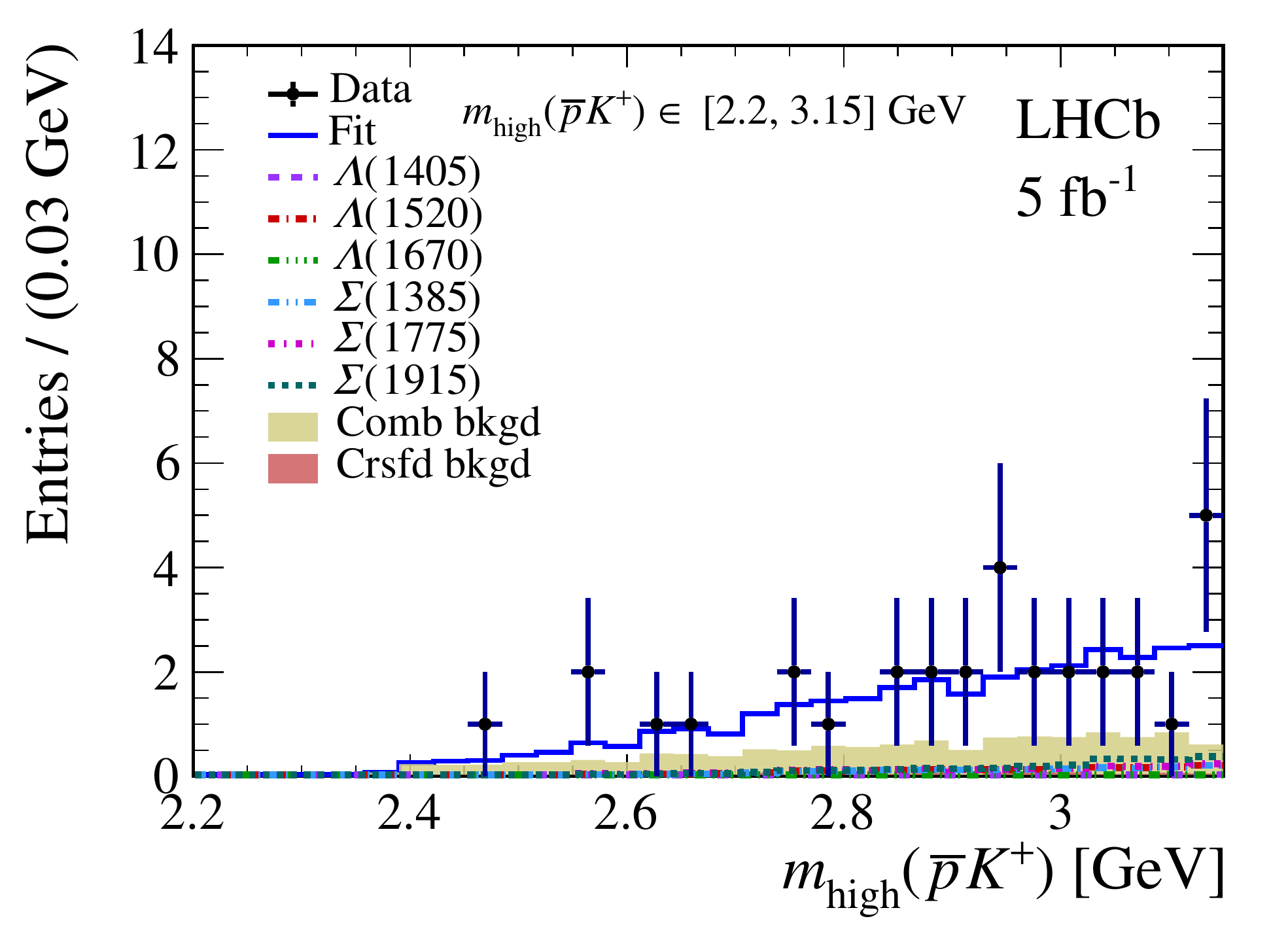} 
\caption{\small
  Distributions of (top)~$\mpklow$ and (bottom)~$\mpkhigh$, with $\mpkhigh < 3.15 \gev$ , for (left)~\Xibm\ and (right)~\Xibp\ candidates, with results of the fits superimposed.
  The total fit result is shown as the blue solid curve, with contributions from individual signal components and from combinatorial (Comb) and cross-feed (Crsfd) background shown as indicated in the legend.
}
\label{fig:mpkhigh_diffregions_low1}
\end{figure}

\begin{figure}[!tb]
\centering
\includegraphics[width=0.48\textwidth]{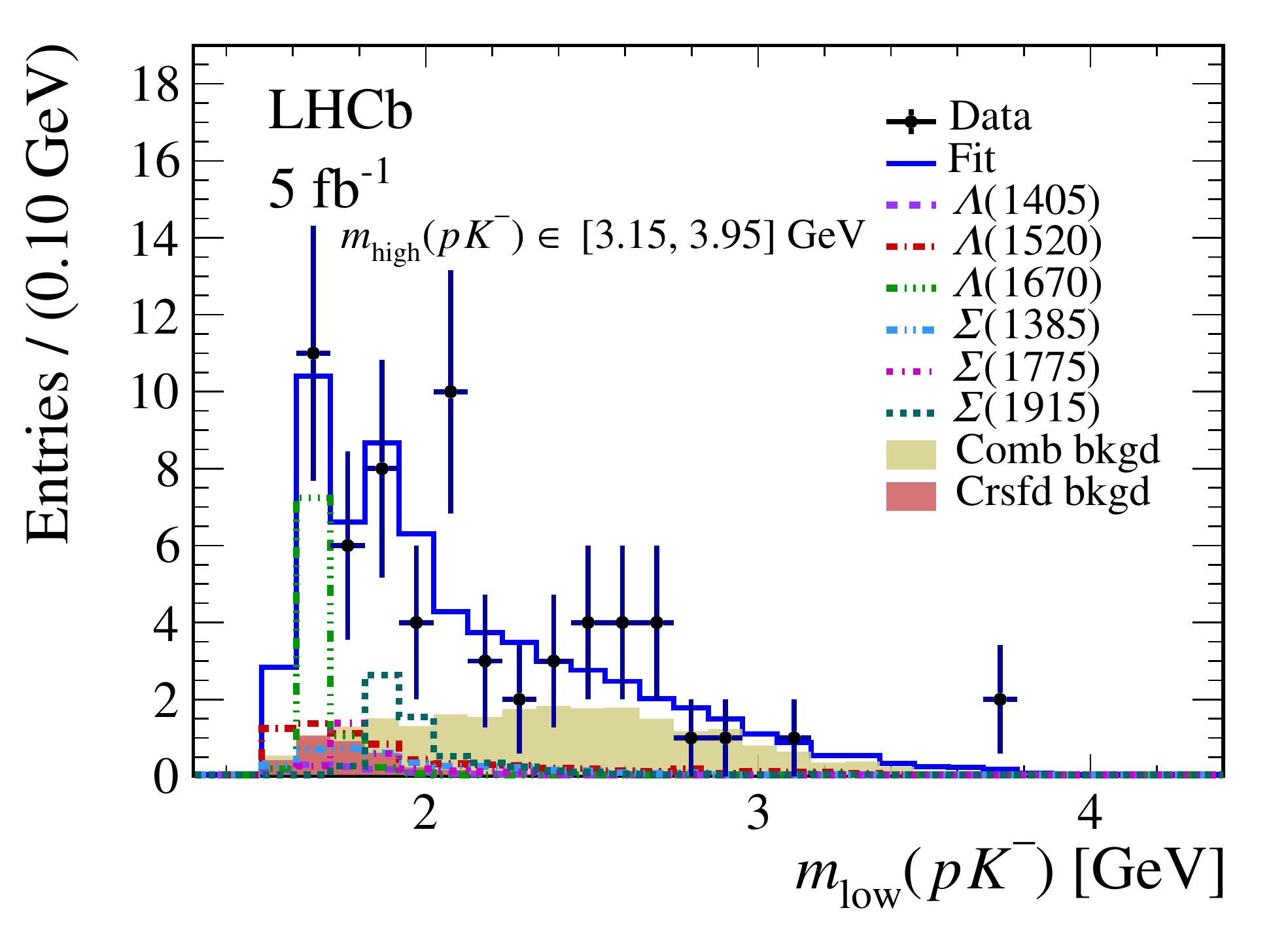} 
\includegraphics[width=0.48\textwidth]{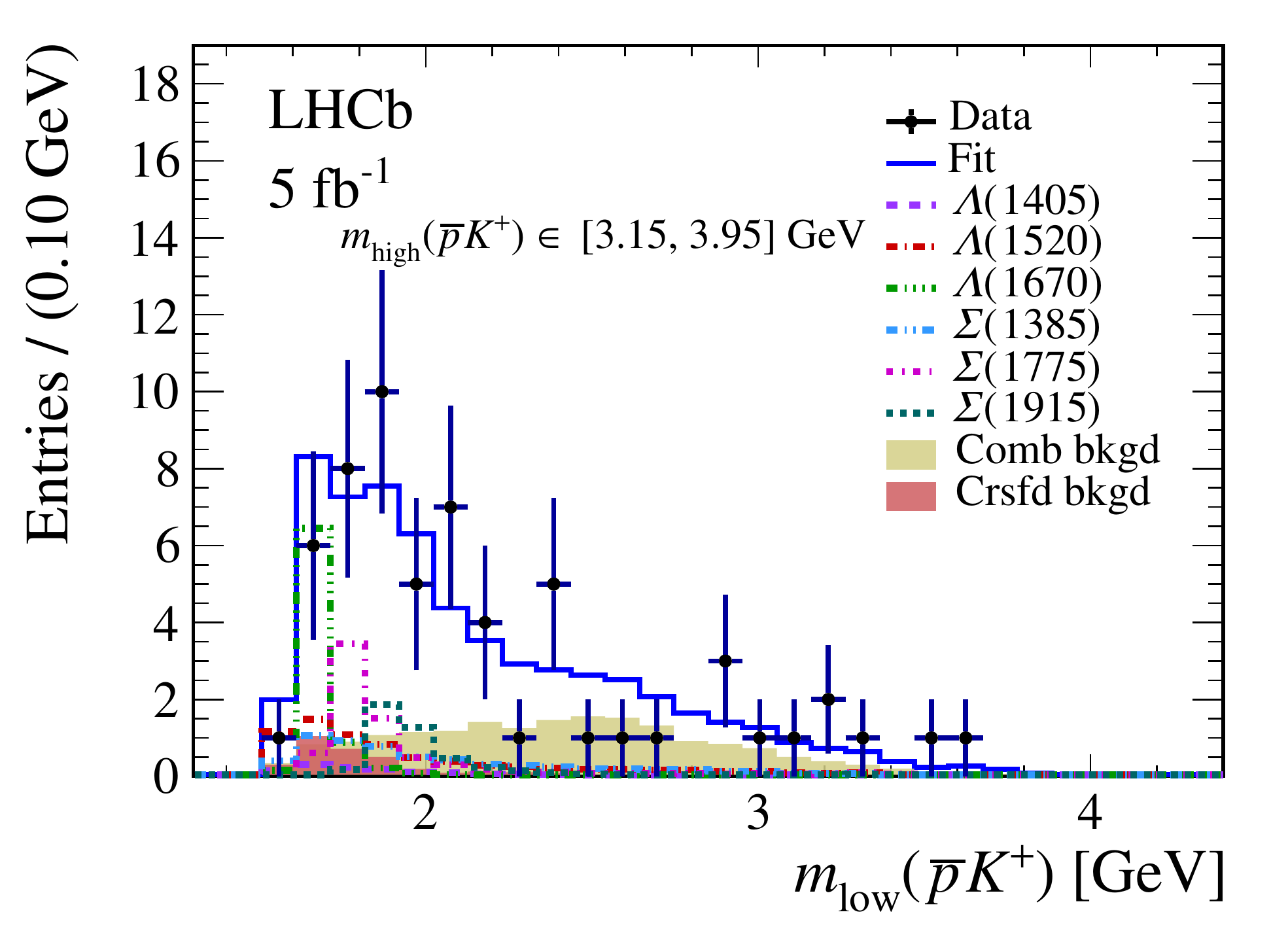} \\
\includegraphics[width=0.48\textwidth]{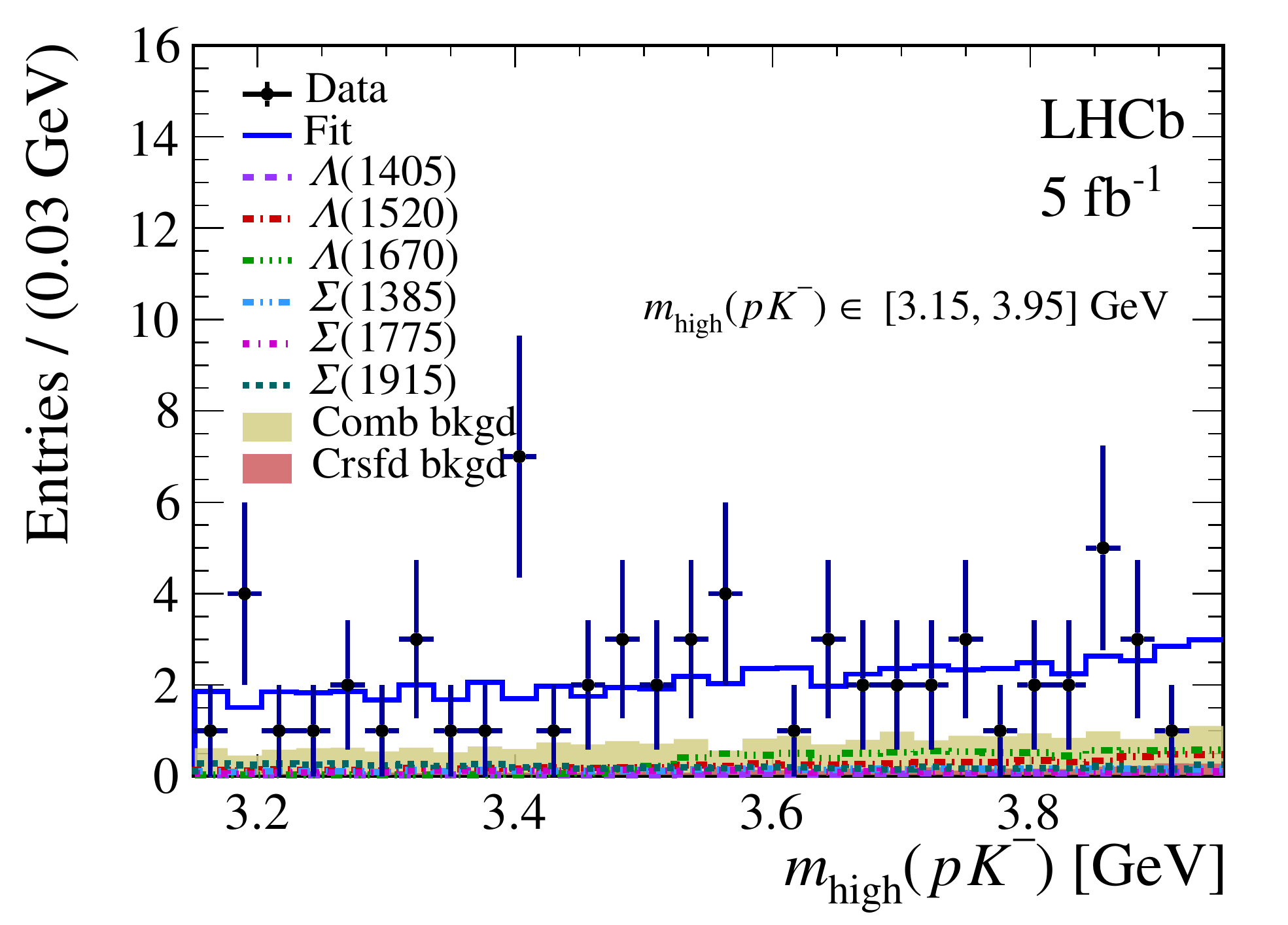} 
\includegraphics[width=0.48\textwidth]{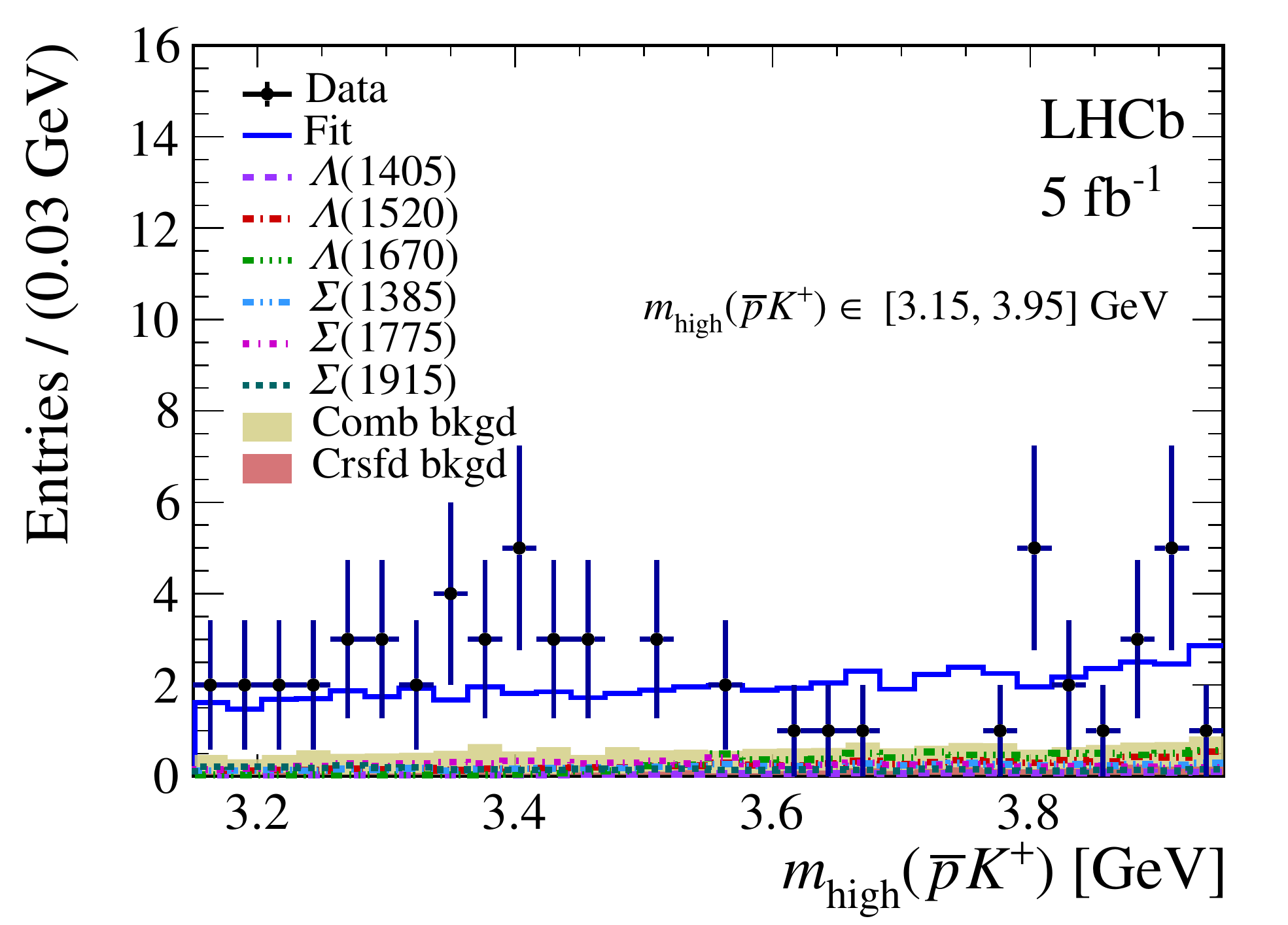} 
\caption{\small
  Distributions of (top)~$\mpklow$ and (bottom)~$\mpkhigh$, with $3.15 < \mpkhigh < 3.95 \gev$, for (left)~\Xibm\ and (right)~\Xibp\ candidates, with results of the fits superimposed.
  The total fit result is shown as the blue solid curve, with contributions from individual signal components and from combinatorial (Comb) and cross-feed (Crsfd) background shown as indicated in the legend.
}
\label{fig:mpkhigh_diffregions_low2}
\end{figure}

\begin{figure}[!tb]
\centering
\includegraphics[width=0.48\textwidth]{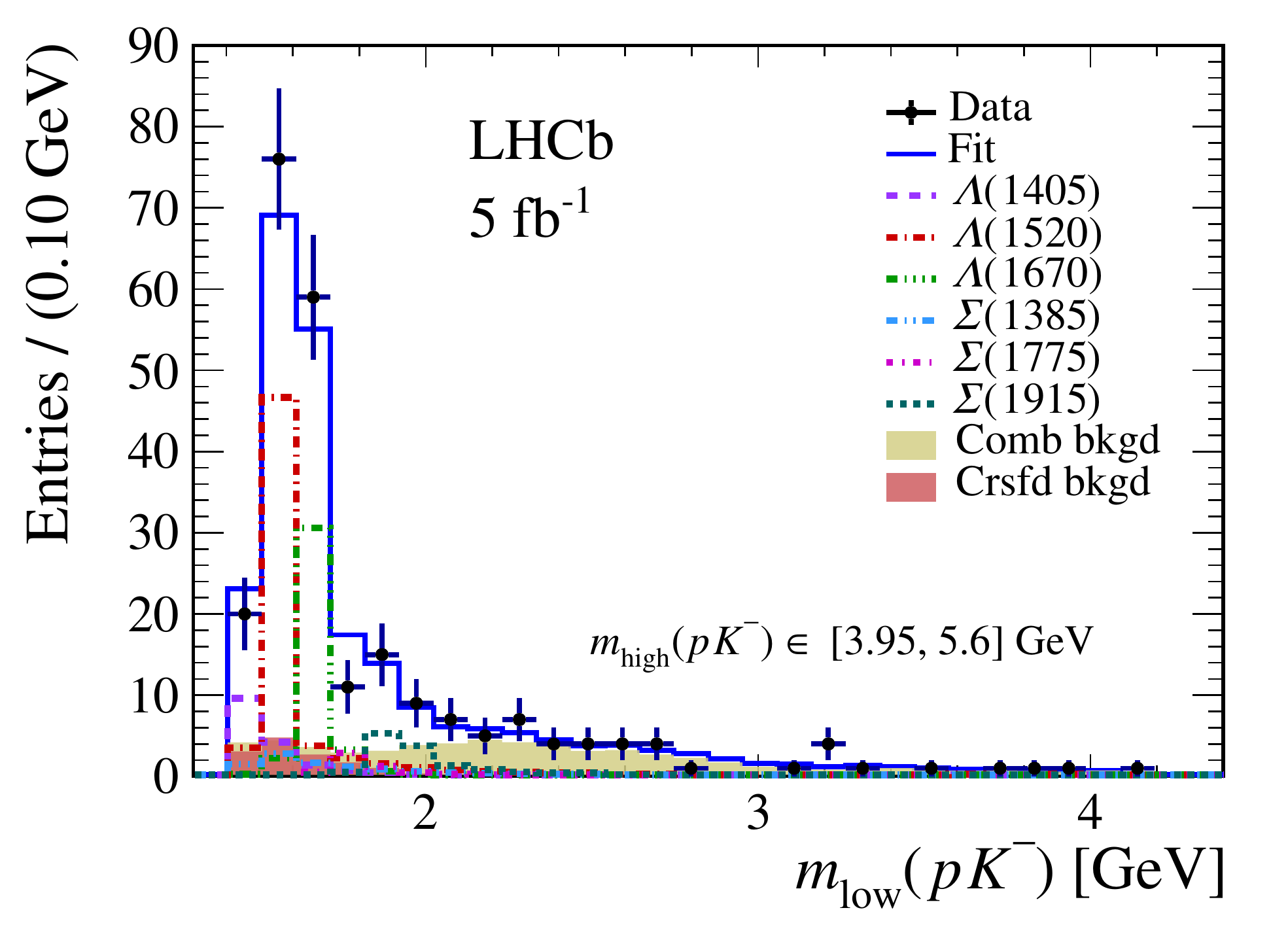} 
\includegraphics[width=0.48\textwidth]{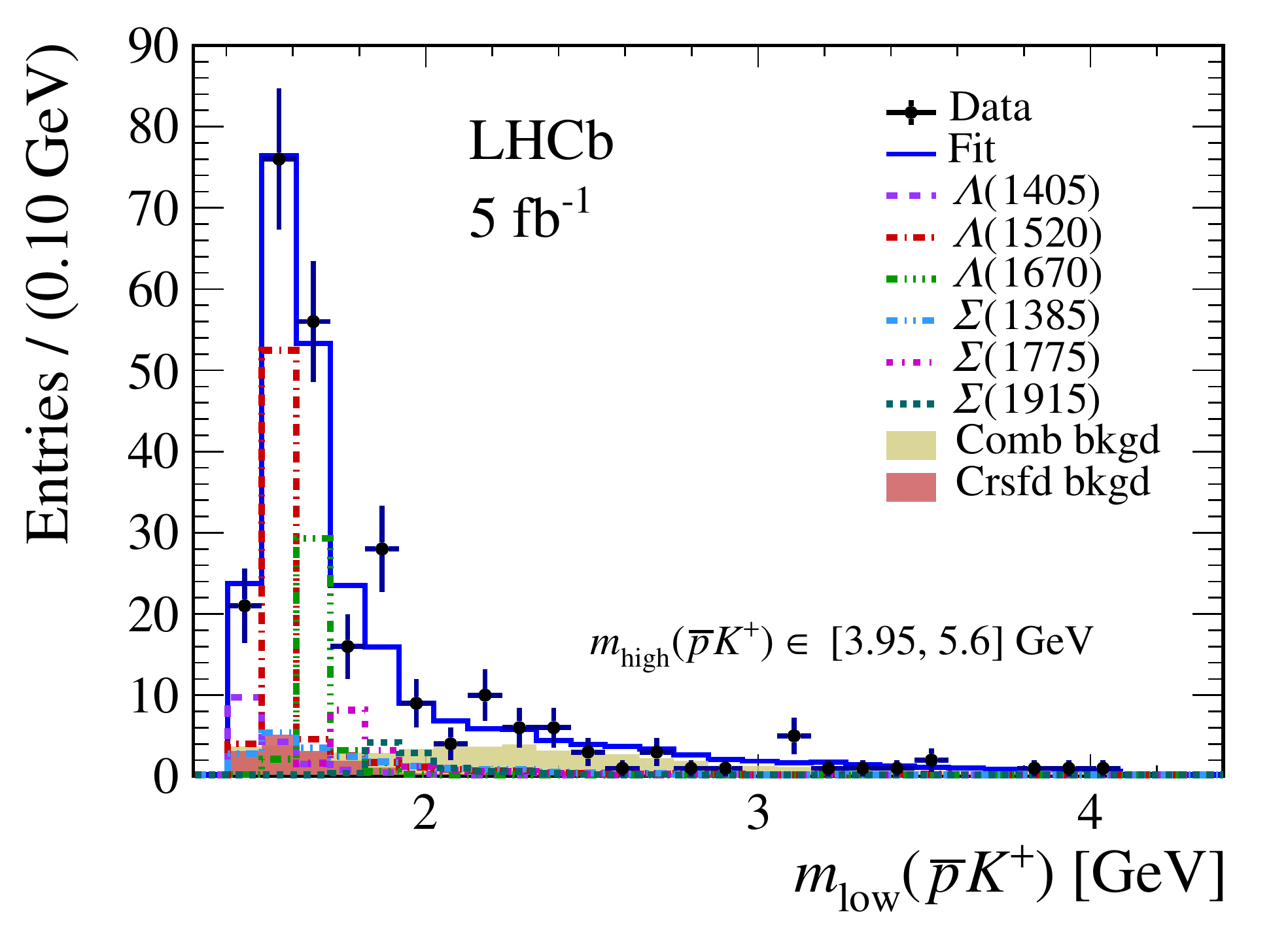} \\
\includegraphics[width=0.48\textwidth]{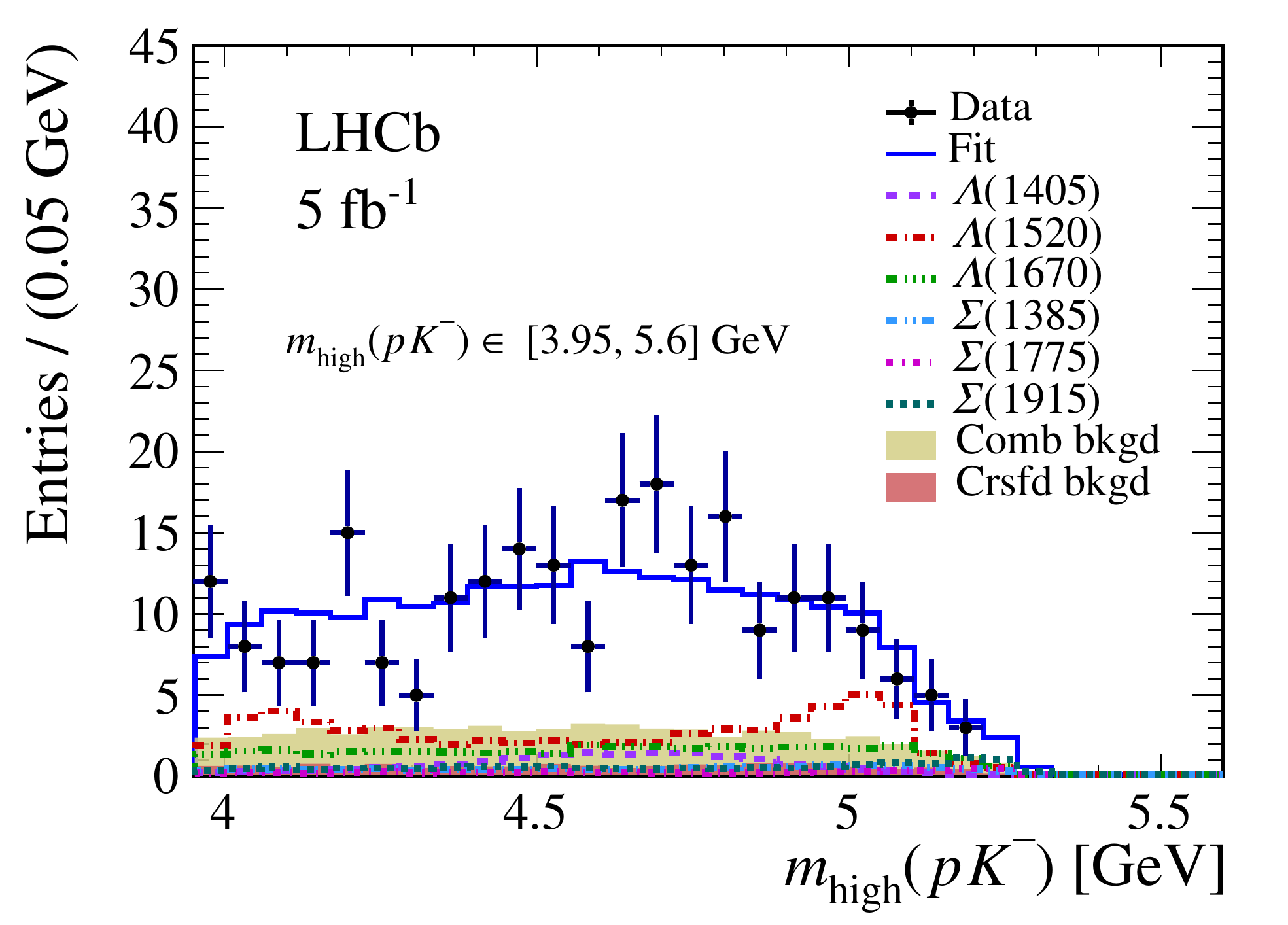} 
\includegraphics[width=0.48\textwidth]{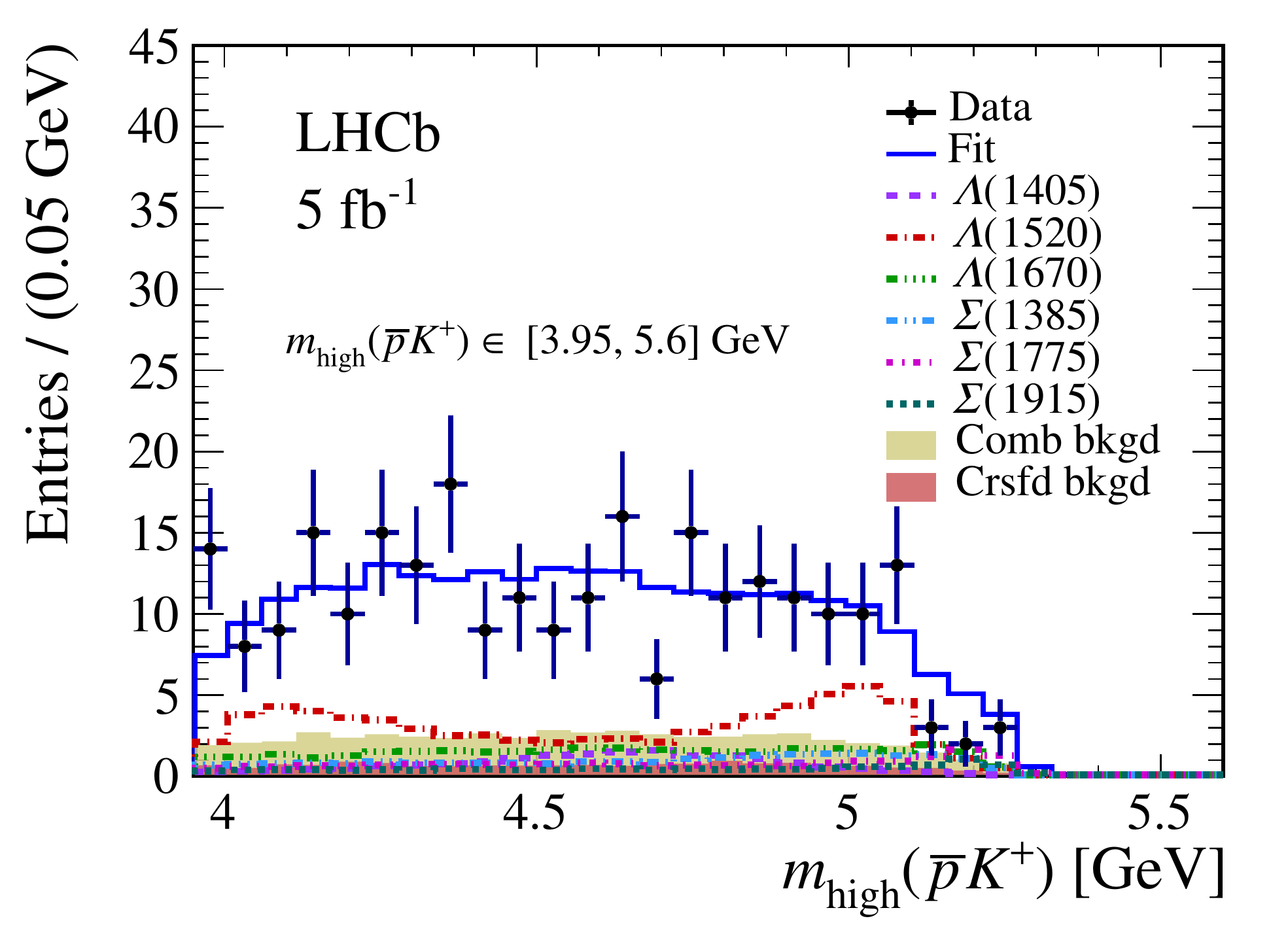} 
\caption{\small
  Distributions of (top)~$\mpklow$ and (bottom)~$\mpkhigh$, with $\mpkhigh > 3.95 \gev$, for (left)~\Xibm\ and (right)~\Xibp\ candidates, with results of the fits superimposed.
  The total fit result is shown as the blue solid curve, with contributions from individual signal components and from combinatorial (Comb) and cross-feed (Crsfd) background shown as indicated in the legend.
}
\label{fig:mpkhigh_diffregions_high}
\end{figure}

Projections of the fit result are compared to the data in slices of \mpklow\ in Figs.~\ref{fig:mpklow_diffregions_low}--\ref{fig:mpklow_diffregions_high}.
Similar projections in slices of \mpkhigh\ are shown in Figs.~\ref{fig:mpkhigh_diffregions_low1}--\ref{fig:mpkhigh_diffregions_high}.


\clearpage

\section*{Acknowledgements}
%
%
\noindent We express our gratitude to our colleagues in the CERN
accelerator departments for the excellent performance of the LHC. We
thank the technical and administrative staff at the LHCb
institutes.
We acknowledge support from CERN and from the national agencies:
CAPES, CNPq, FAPERJ and FINEP (Brazil); 
MOST and NSFC (China); 
CNRS/IN2P3 (France); 
BMBF, DFG and MPG (Germany); 
INFN (Italy); 
NWO (Netherlands); 
MNiSW and NCN (Poland); 
MEN/IFA (Romania); 
MSHE (Russia); 
MICINN (Spain); 
SNSF and SER (Switzerland); 
NASU (Ukraine); 
STFC (United Kingdom); 
DOE NP and NSF (USA).
We acknowledge the computing resources that are provided by CERN, IN2P3
(France), KIT and DESY (Germany), INFN (Italy), SURF (Netherlands),
PIC (Spain), GridPP (United Kingdom), RRCKI and Yandex
LLC (Russia), CSCS (Switzerland), IFIN-HH (Romania), CBPF (Brazil),
PL-GRID (Poland) and NERSC (USA).
We are indebted to the communities behind the multiple open-source
software packages on which we depend.
Individual groups or members have received support from
ARC and ARDC (Australia);
AvH Foundation (Germany);
EPLANET, Marie Sk\l{}odowska-Curie Actions and ERC (European Union);
A*MIDEX, ANR, IPhU and Labex P2IO, and R\'{e}gion Auvergne-Rh\^{o}ne-Alpes (France);
Key Research Program of Frontier Sciences of CAS, CAS PIFI, CAS CCEPP, 
Fundamental Research Funds for the Central Universities, 
and Sci. \& Tech. Program of Guangzhou (China);
RFBR, RSF and Yandex LLC (Russia);
GVA, XuntaGal and GENCAT (Spain);
the Leverhulme Trust, the Royal Society
 and UKRI (United Kingdom).

\addcontentsline{toc}{section}{References}
\bibliographystyle{LHCb}
\bibliography{main,standard,LHCb-PAPER,LHCb-CONF,LHCb-DP,LHCb-TDR}
 
\newpage
\centerline
{\large\bf LHCb collaboration}
\begin
{flushleft}
\small
R.~Aaij$^{32}$,
C.~Abell{\'a}n~Beteta$^{50}$,
T.~Ackernley$^{60}$,
B.~Adeva$^{46}$,
M.~Adinolfi$^{54}$,
H.~Afsharnia$^{9}$,
C.A.~Aidala$^{86}$,
S.~Aiola$^{25}$,
Z.~Ajaltouni$^{9}$,
S.~Akar$^{65}$
% ...

{\footnotesize \it

% ...
$^{9}$Universit{\'e} Clermont Auvergne, CNRS/IN2P3, LPC, Clermont-Ferrand, France\\
% ...
$^{25}$INFN Sezione di Milano, Milano, Italy\\
% ...
$^{32}$Nikhef National Institute for Subatomic Physics, Amsterdam, Netherlands\\
% ...
$^{46}$Instituto Galego de F{\'\i}sica de Altas Enerx{\'\i}as (IGFAE), Universidade de Santiago de Compostela, Santiago de Compostela, Spain\\
% ...
$^{50}$Physik-Institut, Universit{\"a}t Z{\"u}rich, Z{\"u}rich, Switzerland\\
% ...
$^{54}$H.H. Wills Physics Laboratory, University of Bristol, Bristol, United Kingdom\\
% ...
$^{60}$Oliver Lodge Laboratory, University of Liverpool, Liverpool, United Kingdom\\
% ...
$^{65}$University of Cincinnati, Cincinnati, OH, United States\\
% ...
$^{68}$Syracuse University, Syracuse, NY, United States\\
% ...
$^{86}$University of Michigan, Ann Arbor, United States, associated to $^{68}$\\
\bigskip
% ...
\medskip
}
\end{flushleft}

\end{document}